\documentclass[aps,pre,twocolumn,floatfix,showpacs, superscriptaddress,]{revtex4-2}

\usepackage{amssymb,amsmath,amsfonts}
\usepackage{graphicx,graphics}
\usepackage{epsfig}
\usepackage{epstopdf}
\usepackage{amsfonts}
\usepackage{bm,bbm}
\usepackage{amssymb,amsmath,amsfonts}
\usepackage{mathrsfs}
\usepackage{calc}
\usepackage{epsfig}
\usepackage{float}
\usepackage{color}
\usepackage{epstopdf}
\usepackage{bm,amssymb,amsmath}
\usepackage{graphicx,color}
\usepackage{ulem}
\usepackage{hyperref}

\begin{document}

\title{Probing the limits of effective temperature consistency in actively driven systems}
\author{Dima Boriskovsky}
\affiliation{Raymond \& Beverly Sackler School of Physics and Astronomy, Tel Aviv University, Tel Aviv 6997801, Israel}
\author{R{\'e}mi Goerlich}
\affiliation{Raymond \& Beverly Sackler School of Chemistry, Tel Aviv University, Tel Aviv 6997801, Israel}
\author{Benjamin Lindner}
\affiliation{Bernstein Center for Computational Neuroscience Berlin, Philippstr.~13, Haus 2, 10115 Berlin, Germany}
\affiliation{Physics Department of Humboldt University Berlin, Newtonstr.~15, 12489 Berlin, Germany}
\author{Yael Roichman}
\email{roichman@tauex.tau.ac.il}
\affiliation{Raymond \& Beverly Sackler School of Physics and Astronomy, Tel Aviv University, Tel Aviv 6997801, Israel}
\affiliation{Raymond \& Beverly Sackler School of Chemistry, Tel Aviv University, Tel Aviv 6997801, Israel}

\date{\today}
\begin{abstract}
We investigate the thermodynamic properties of a single inertial probe driven into a nonequilibrium steady state by random collisions with self-propelled active walkers. 
The probe and walkers are confined within a gravitational harmonic potential.
We evaluate the robustness of the effective temperature concept in this active system by comparing values of distinct, independently motivated definitions: a generalized fluctuation-dissipation relation, a kinetic temperature, and via a work fluctuation relation. 
Our experiments reveal that, under specific conditions, these independent measurements coincide over a wide range of system configurations, yielding a remarkably consistent effective temperature.  
Furthermore, we also identify regimes where this consistency breaks down, which delineates the fundamental limits of extending equilibrium-like thermodynamic concepts to athermal, actively driven systems.
\end{abstract}

\maketitle

\section{Introduction} 

A single colloidal particle confined within a harmonic potential and immersed in a thermal fluid offers a fundamental realization of a microscopic statistical thermometer \cite{uhlenbeck1930,wang1945}. 
In thermal equilibrium, the particle's random displacements follow a Boltzmann (Gaussian) distribution, allowing the thermodynamic temperature, $T$, of the surrounding environment to be directly determined by its mean potential energy via the principle of equipartition \cite{risken1996}. 

Near and at equilibrium, $T$ can be equivalently defined through various thermodynamic relations, which can be applied to and measured by a colloidal thermometer.
Static relations, measured in stationary states, include canonical distributions and energy equipartition, as well as configurational definitions of temperature \cite{han2004,rugh1996}. 
Dynamic relations, measured by probing the system's response properties, are given by the fluctuation-dissipation theorem \cite{einstein1905,langevin1908,callen1951,kubo1966,kubo1986}, which rigorously connects the effects of thermal noise and energy dissipation.
In addition, fluctuation theorems also relate temperature to fluctuations in thermodynamic quantities such as heat, work, and entropy \cite{gallavotti1995,gallavotti2004,jop2008,gomez2010}.

Colloidal thermometers have thus been widely employed in experiments, both to validate theoretical measures of equilibrium temperature and to identify deviations from it \cite{leptos2009,maggi2014,ciliberto2017,goerlich2022}.
Crucially, various definitions of absolute temperature must yield consistent, measurement-independent values in accordance with the zeroth law of thermodynamics.

In stark contrast, no universal definition for temperature exists for many natural and engineered systems that operate far from thermal equilibrium. 
Nonetheless, the notion of \textit{effective} temperature has frequently arisen as a heuristic extension of equilibrium statistical mechanics to diverse systems operating far from equilibrium, including glassy materials \cite{cugliandolo1997,ono2002,berthier2002,joubaud2009,bouchbinder2011}, biological suspensions \cite{wu2000,leptos2009,maggi2014,argun2016,maggi2017,wiese2024,di2024}, driven granular media \cite{puglisi1999,poschel2001,van2004,ojha2004,villamaina2008,villamaina2009,shokef2006,bunin2008,puglisi2002,feitosa2004,abate2008,joubaud2012,gnoli2014,chastaing2017,zeng2022}, and active matter \cite{palacci2010,martinez2013,levis2015,dieterich2015,ginot2015,petrelli2020,flenner2020,fodor2016,han2017,solon2022,ye2020,park2020,boudet2022,shea2022,goerlich2022,geva2025}. 

In particular, under well-separated timescales, an equilibriumlike fluctuation-dissipation relation (FDR) can become applicable, leading to a definition of an effective temperature, $T_{\text{eff}}$.
While the FDR-based $T_{\text{eff}}$ has proven useful in various fields of physics\cite{marconi2008,baiesi2009,ciliberto2017,maes2020,baldovin2022review}, its thermodynamic interpretation and range of applicability remain subjects of ongoing research \cite{martens2009,cugliandolo2011,puglisi2017temperature,holubec2020,sorkin2024}. 

\begin{figure*}[t] 
    \includegraphics[width=0.575\textwidth]{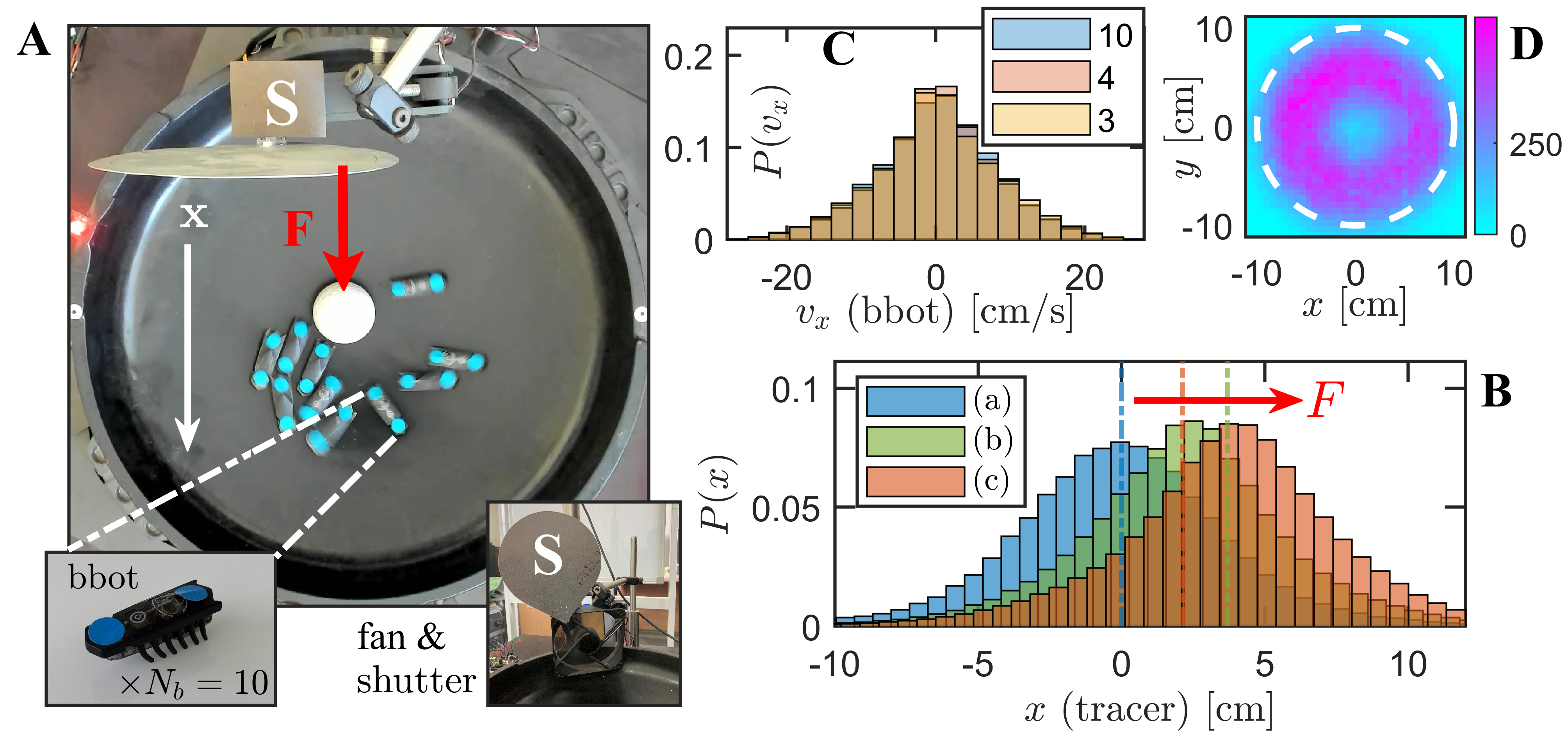}
    \includegraphics[width=0.405\textwidth]{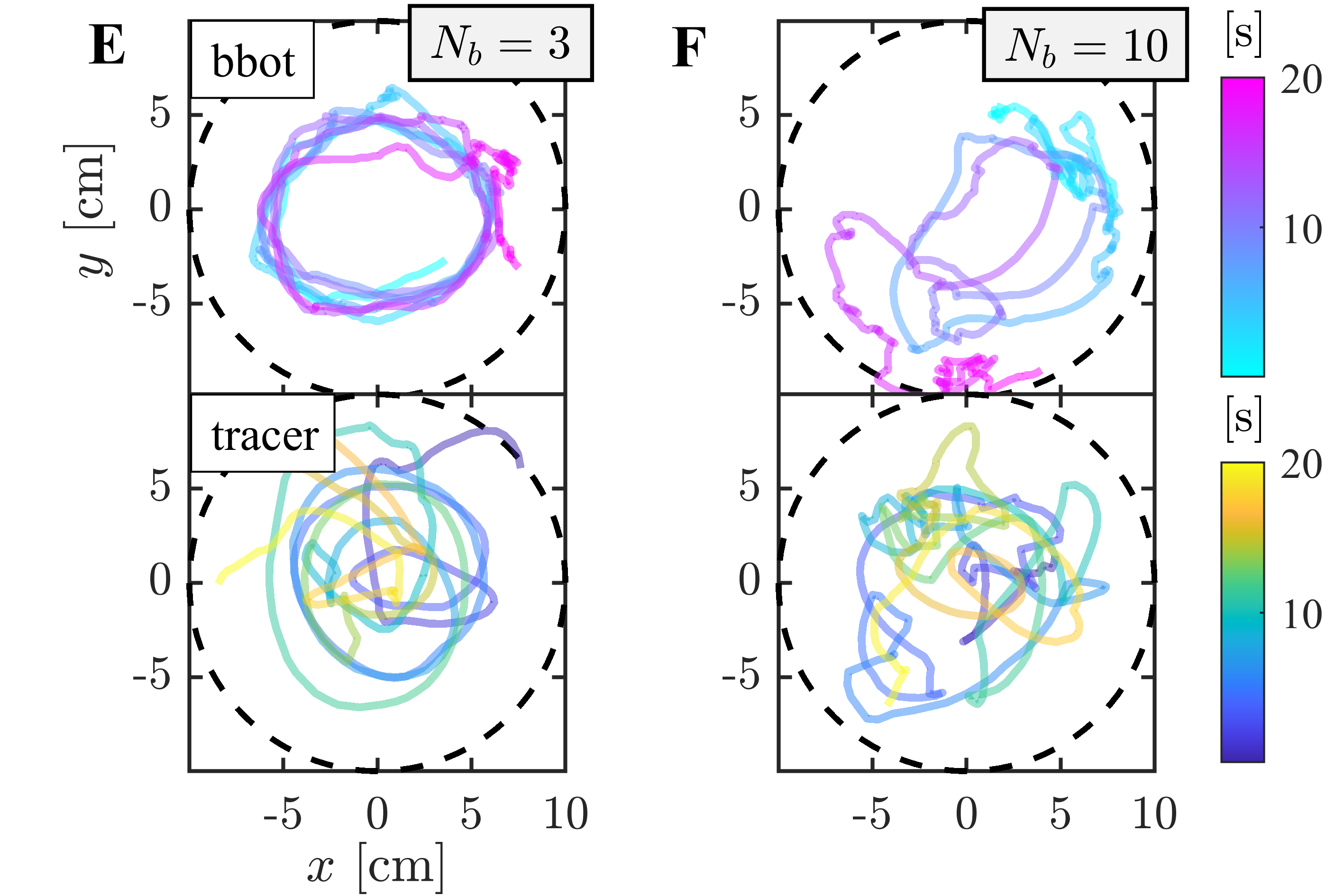}
    \caption{
    \textbf{The system:} \ 
    \textbf{A.} Experimental setup: A Styrofoam ball (diameter $\sim4$~cm, $1$~g) is trapped in a gravitational harmonic potential, a plastic bowl (diameter 38cm, depth 5cm), and subjected to collisions with $N_b=10$ self-propelled bbots (inset: standard bbot, $4\times1$~cm, $7.1$~g). The ball is repeatedly perturbed with a uniform air stream created by an external fan along the $x$-axis (white arrow) to test a fluctuation-response relation. To enforce an abrupt onset and release of the perturbation, a mechanical shutter is used (denoted by 'S'). \
    \textbf{B.} Exemplary results (with $N_b=10$) for three independent tracer's steady states: 
    (a) an unperturbed state; 
    (b) a weakly perturbed state ($10$~V fan operating voltage, $F_0=\kappa\Delta x_\epsilon\approx 62$$\mu$N); 
    and (c) a strongly perturbed state ($13.5$~V, $F_0\approx 107.8$$\mu$N). 
    These stationary position distributions were obtained from combined time and ensemble averages, using 375 trajectories of 1 minute length. \ 
    \textbf{C.} Velocity distributions of bbot assemblies with different $N_b=3,4,$ and 10.
    Instantaneous velocities were extracted from recordings of bbot trajectories tracked over 40~minutes using a frame rate of 30 frames per second (fps). \ 
    \textbf{D.} Spatial distribution of 10 bbots within the harmonic trap, with a soft boundary (steep curvature) indicated by a dashed circle of radius $10$~cm. \
    \textbf{E, F.} Typical $20$~second trajectories of a single bbot in systems with $N_b=3$ (E) and $N_b=10$ (F) (upper panels), alongside the corresponding tracer trajectories (lower panels).
    }
    \label{fig:1}
\end{figure*}

A remaining key challenge lies in identifying under which conditions distinct theoretical definitions of effective temperatures yield mutually consistent values far from equilibrium \cite{jou2000, hecht2024}.
Further examples of nonequilibrium definitions that capture meaningful physical behavior in their statistical formulation include equipartition-based approaches, such as granular temperatures \cite{poschel2001,puglisi1999}, and those derived from fluctuation-relations (FR), applicable to both long-time nonequilibrium steady states (NESS) and nonequilibrium transitions \cite{gallavotti2004,feitosa2004,puglisiFR}.
Such definitions typically require stable steady states, the coexistence of stochastic and deterministic dynamics, and access to distinct NESS observables - conditions that are often difficult to achieve experimentally.
Although no theoretical principle ensures the coincidence of these definitions far from equilibrium, our experiments indicate a range of conditions in which a consistent effective temperature can emerge in NESS, driven by active or athermal fluctuations (cf. \cite{ono2002,maggi2014,chastaing2017}). 

\begin{figure}[t] 
    \centering
    \includegraphics[width=0.39\textwidth]{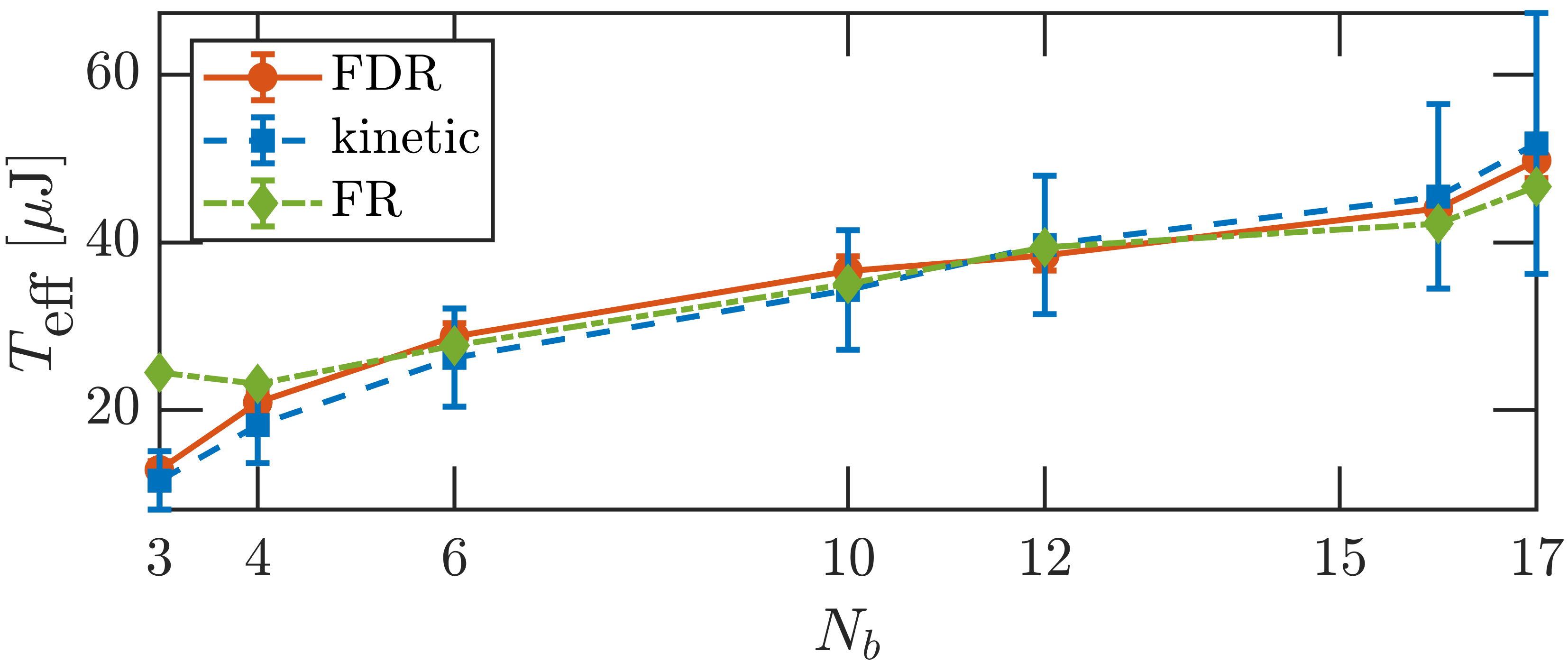}
    \caption{
    \textbf{Thermometer consistency:} \ 
    $N_b=\{ 3,4,6,10,12,15,17 \}$ bbots.
    Effective temperature measurements obtained using three independent methods: the potential temperature $T_{\text{pot}}=T_{\text{eff}}$ (circles, Eq. \ref{eq:temp}), the modified kinetic temperature $\tilde{T}_{\text{kin}}$ (squares, Eq. \ref{eq:temp2}), and a constant temperature $T_{\text{FR}}$ derived from a work FR (diamonds, Eq. \ref{eq:FR}).
    Notably, these \textit{static} temperatures validate the FDR of Eq. \ref{eq:FDR1} (for $N_b>3$) and define a consistent effective temperature $T_{\text{eff}}$. 
    }
    \label{fig:6}
\end{figure}

In this work, a statistical thermometer is realized experimentally, based on a macroscopic tracer particle confined in a harmonic trap and driven into a NESS by random inelastic collisions with self-propelled walkers (see Section 2 and Fig. \ref{fig:1}).
Previously \cite{boriskovsky2024}, it was shown that, with a large enough number of walkers, this system satisfies a linear (dynamic) FDR. 
Here, we assess the consistency of the FDR-based $T_{\text{eff}}$ by comparing it to three independent measures: the tracer's potential energy and modified kinetic energy (in an unperturbed NESS), and a temperature based on a steady state work FR, in a NESS under a strong external perturbation.
Our main result is displayed in Fig.~\ref{fig:6}: all three independent measurements of $T_{\text{eff}}$ agree across a range of active bath parameters. 
We further compare our results with theoretical predictions and discuss conditions under which the consistency of $T_{\text{eff}}$ breaks down.

\section{Experimental setup and basic characterization} 

Our experimental setup is illustrated in Fig. \ref{fig:1}\textbf{A}: A lightweight styrofoam ball (the tracer) is confined within a parabolic plastic arena and is subjected to random collisions with an assembly of vibration-driven self-propelled particles (bristlebots, or \textit{bbots}, specifically Hexbug$^{\text{TM}}$).
The system is imaged from above using a standard webcam (Brio 4k, Logitech), and particle trajectories in the horizontal plane (XY) are extracted using a custom image analysis algorithm. The confining potential is approximately harmonic, with constant stiffness $\kappa=mga=28.2\pm3 \text{gs}^{-2}$, set by the substrate curvature $a$, the gravitational acceleration $g$, and the tracer mass $m=1\pm0.1$g.

An external airflow can be applied via a fan (in the $x$ direction) and abruptly switched-off by a physical shutter, affecting primarily the tracer particle.
Stationary probability densities of the tracer's position along the $x$-axis for different intensities of air streams are shown in Fig. 1\textbf{B}. The main effect of the air stream is a shift of the mean value of the histograms in the potential well, i.e., an unperturbed state (a), a weakly perturbed state (b), and a strongly perturbed state (c). 
This mechanism is used to test the validity of both FDRs and FRs, as detailed in Sections 3.1 and 3.3 respectively (see SI$^{\dag}$ figures 1-2 for further details).
We assume independent statistics along all axes and focus on analyzing the tracer’s position and velocity components projected onto the $x$-axis, where the perturbation occurs.

\textbf{Nonequilibrium properties of the active bath.-} \ 
In this setting, a single bbot with sufficient inertia typically performs circular motion around the trap center \cite{dauchot2019}, with a preferred chirality (clockwise for Hexbugs).
As the number of bbots $N_b$ increases, frequent collisions with other bbots (and the tracer) randomize their propulsion direction and result in an active gas-like state (see SM movie 1).
In particular, the bbot velocity distribution remains independent of $N_b$ whereas its position distribution deviates from the original circular path with large $N_b$, as seen in Fig.~\ref{fig:1}\textbf{C,D}. 

Typical bbot trajectories are shown in Fig. \ref{fig:1}\textbf{E,F}. For $N_b=3$, collisions are rare, and bbots follow long-lived circular paths; for $N_b=10$, frequent collisions yield erratic, stochastic motion. 
In general, these self-propelled particles exhibit rich dynamics, including alignment with boundaries and emergent collective motion \cite{giomi2013}. 
Here, the passive tracer primarily experiences random inelastic collisions with the active bbots (lower panels).
These interactions introduce both noise and dissipation, which result in a collision-induced stochastic motion reaching a steady state within the harmonic potential. 
In this work, we focus on the properties of the tracer particle, as detailed below.

\begin{figure}[t!] 
    \centering
    \includegraphics[width=0.23\textwidth]{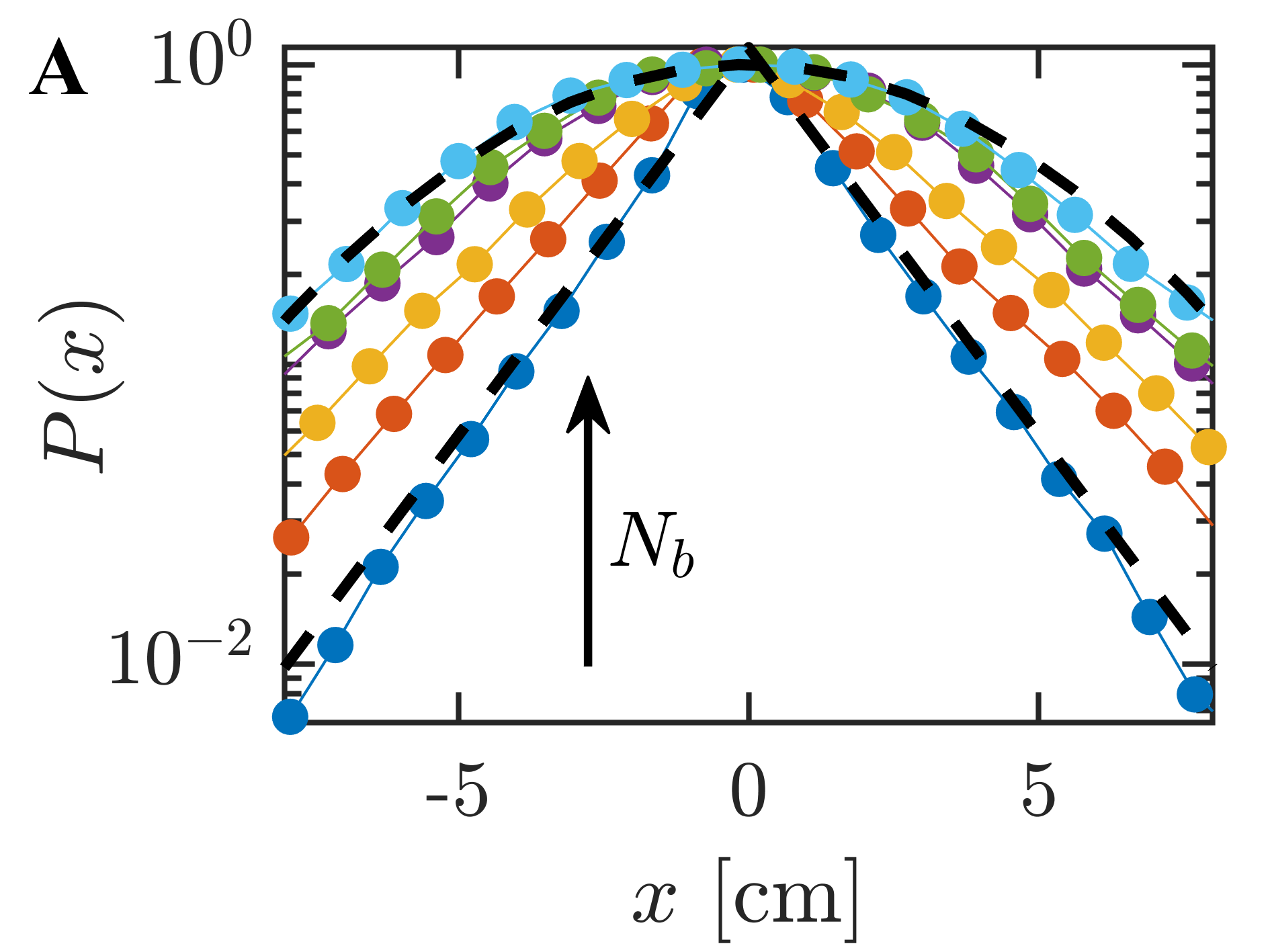}
    \includegraphics[width=0.23\textwidth]{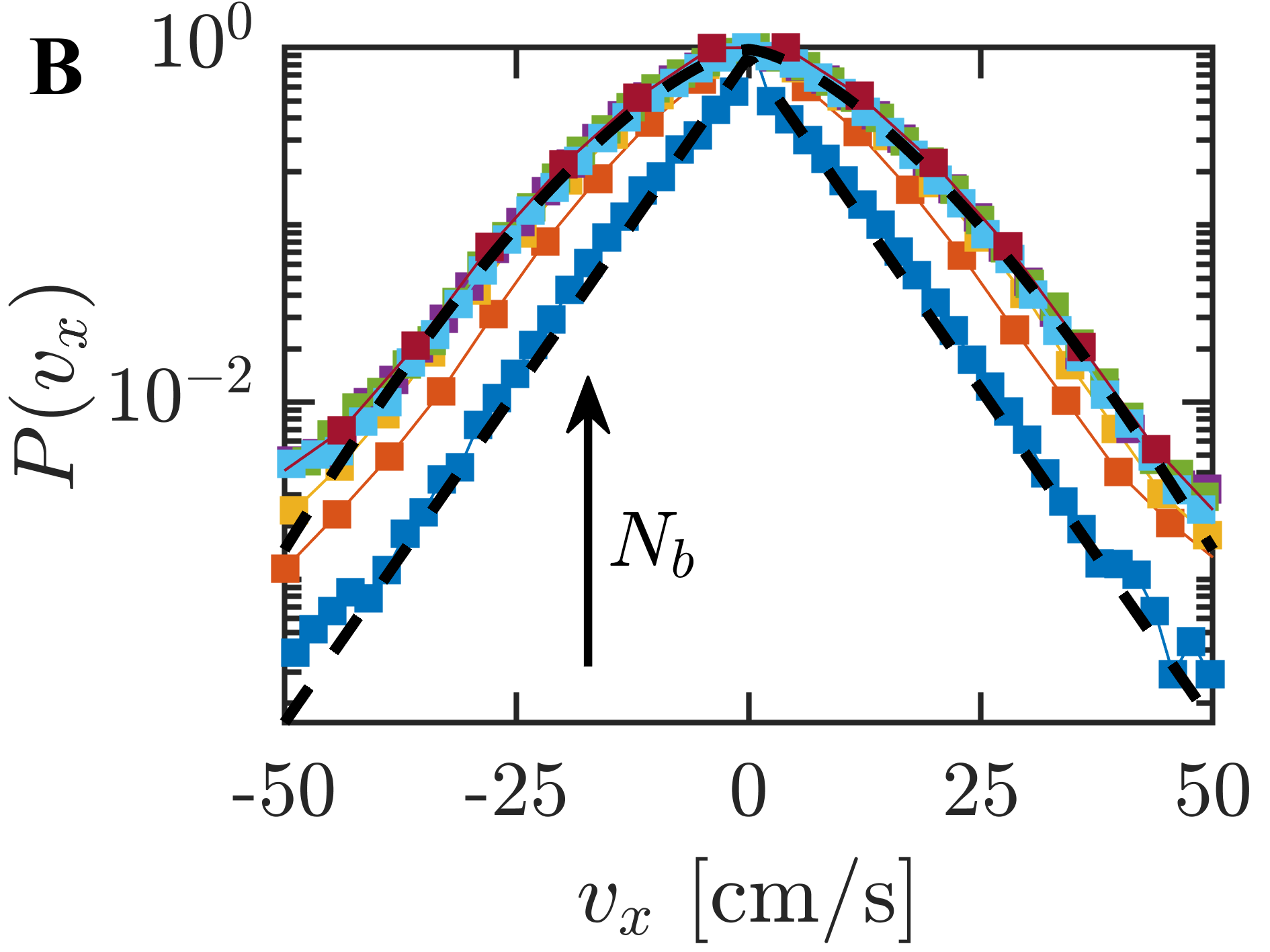}
    \includegraphics[width=0.23\textwidth]{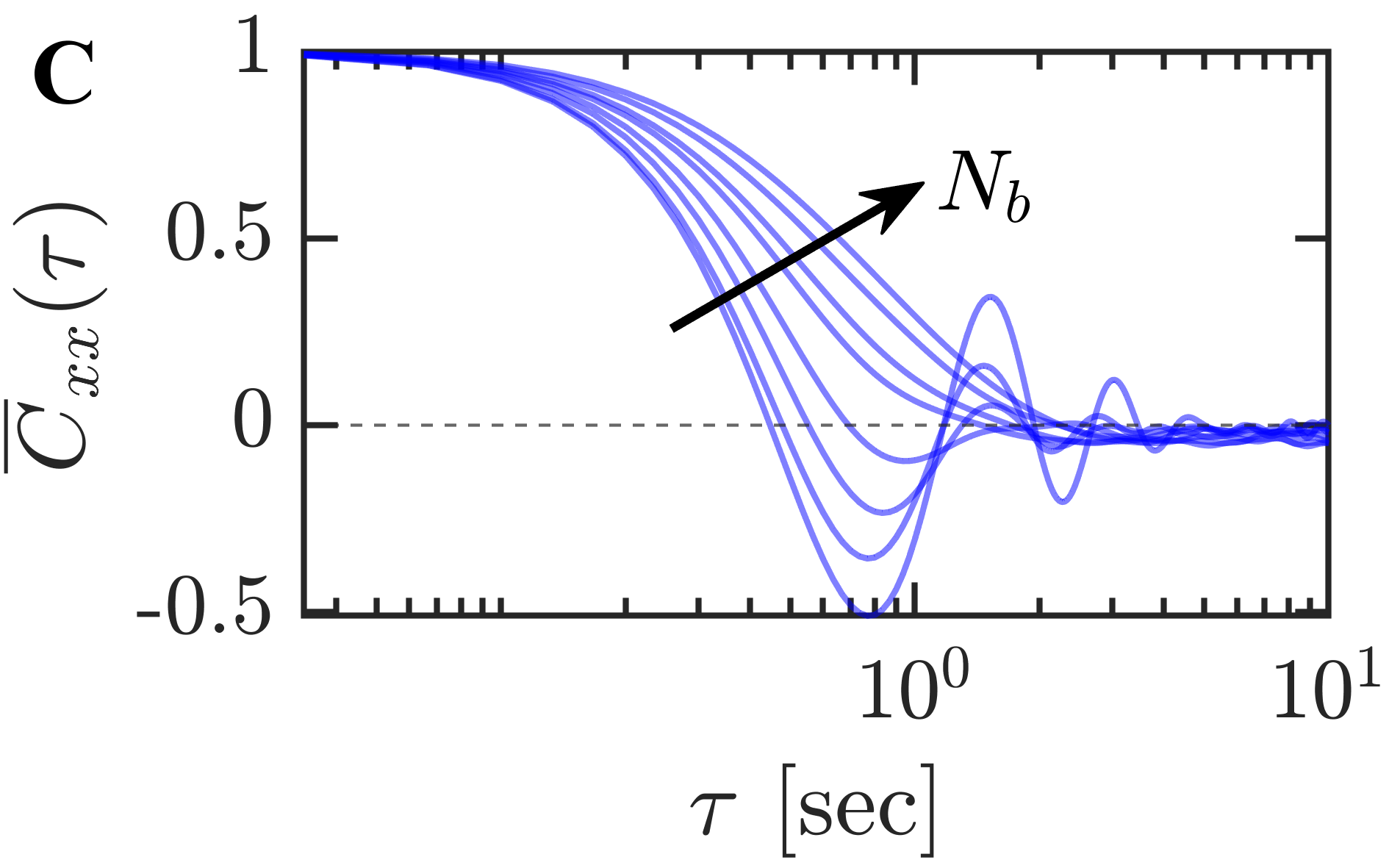}
    \includegraphics[width=0.23\textwidth]{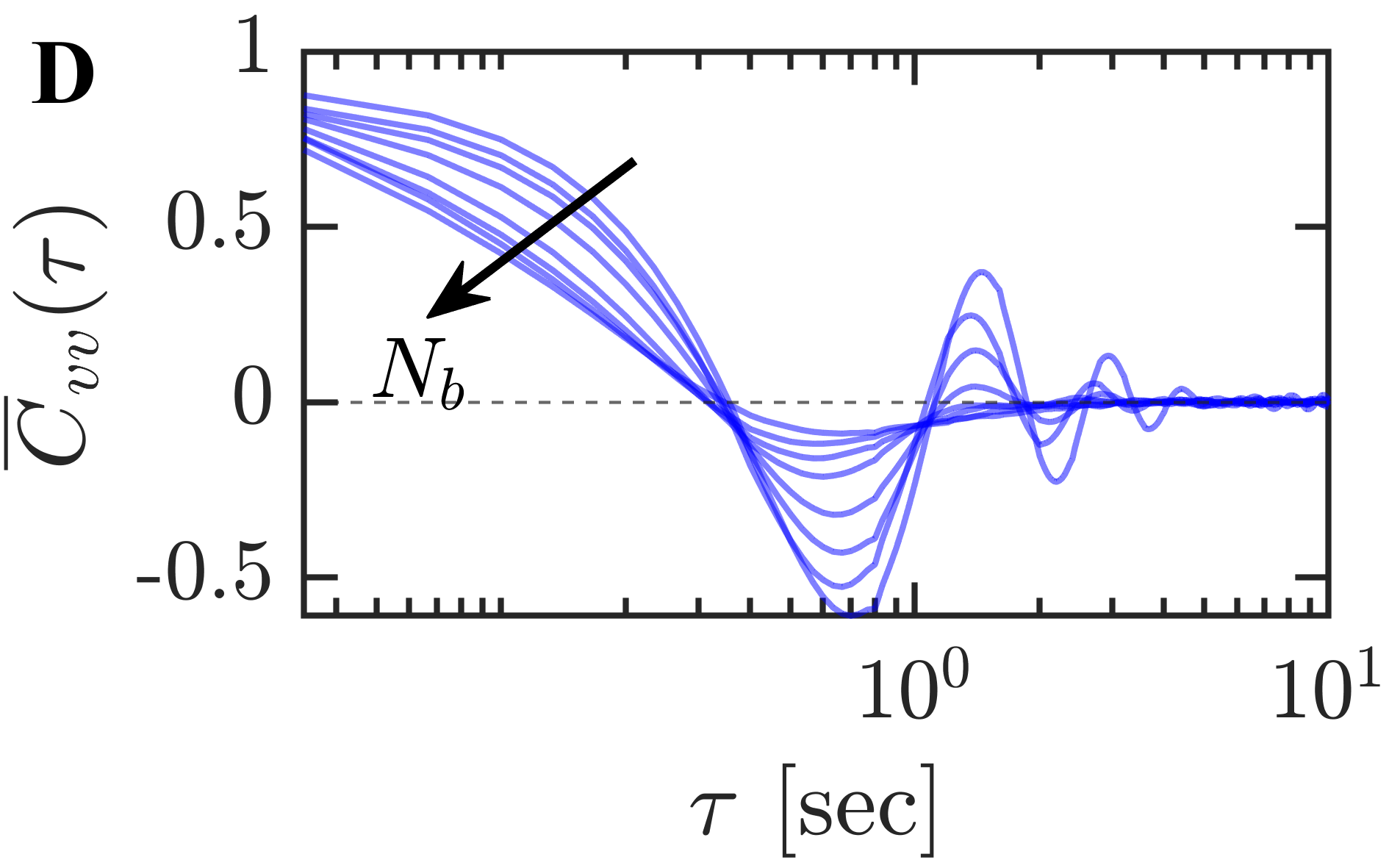}
    \includegraphics[width=0.23\textwidth]{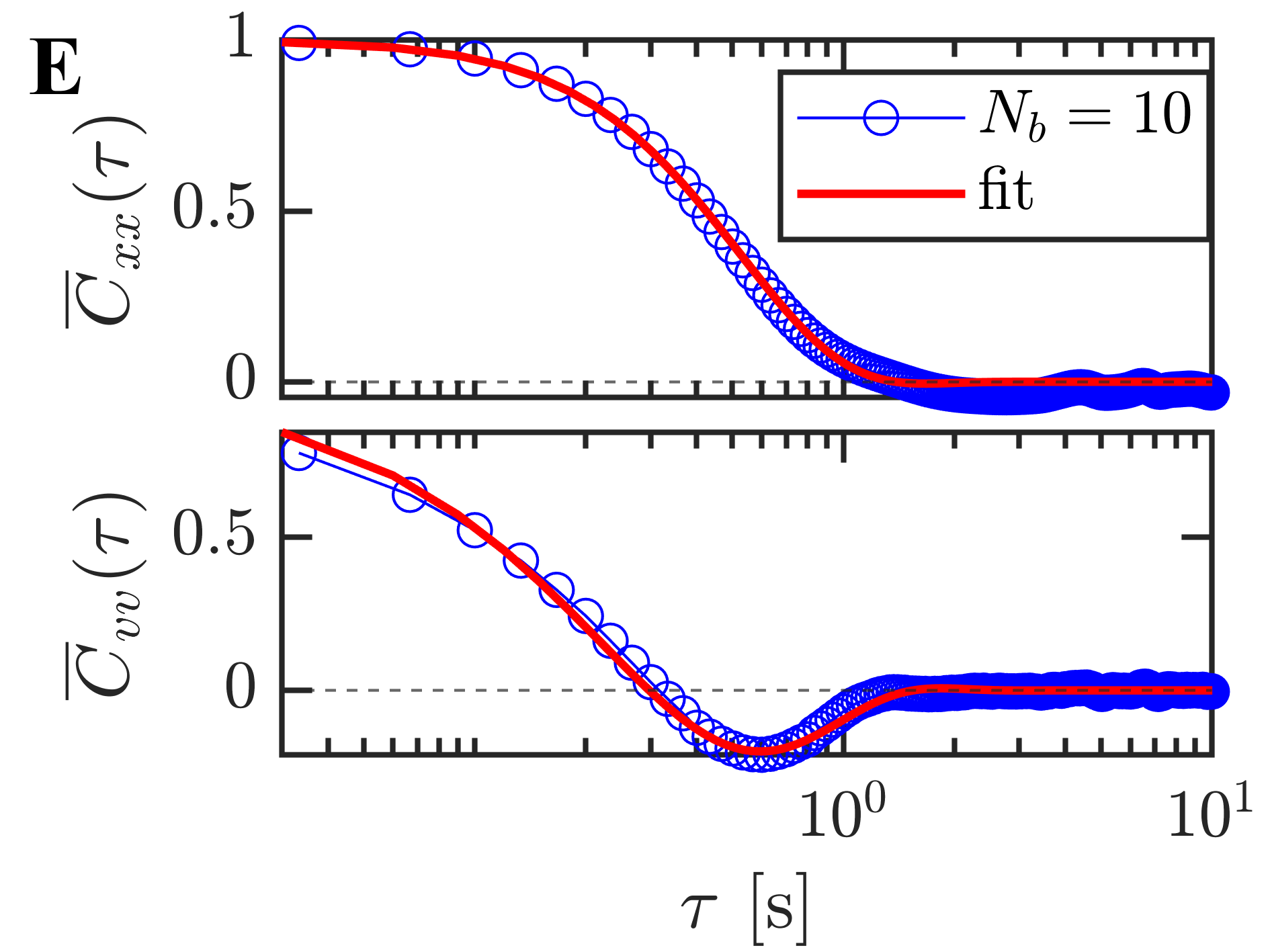}
    \includegraphics[width=0.23\textwidth]{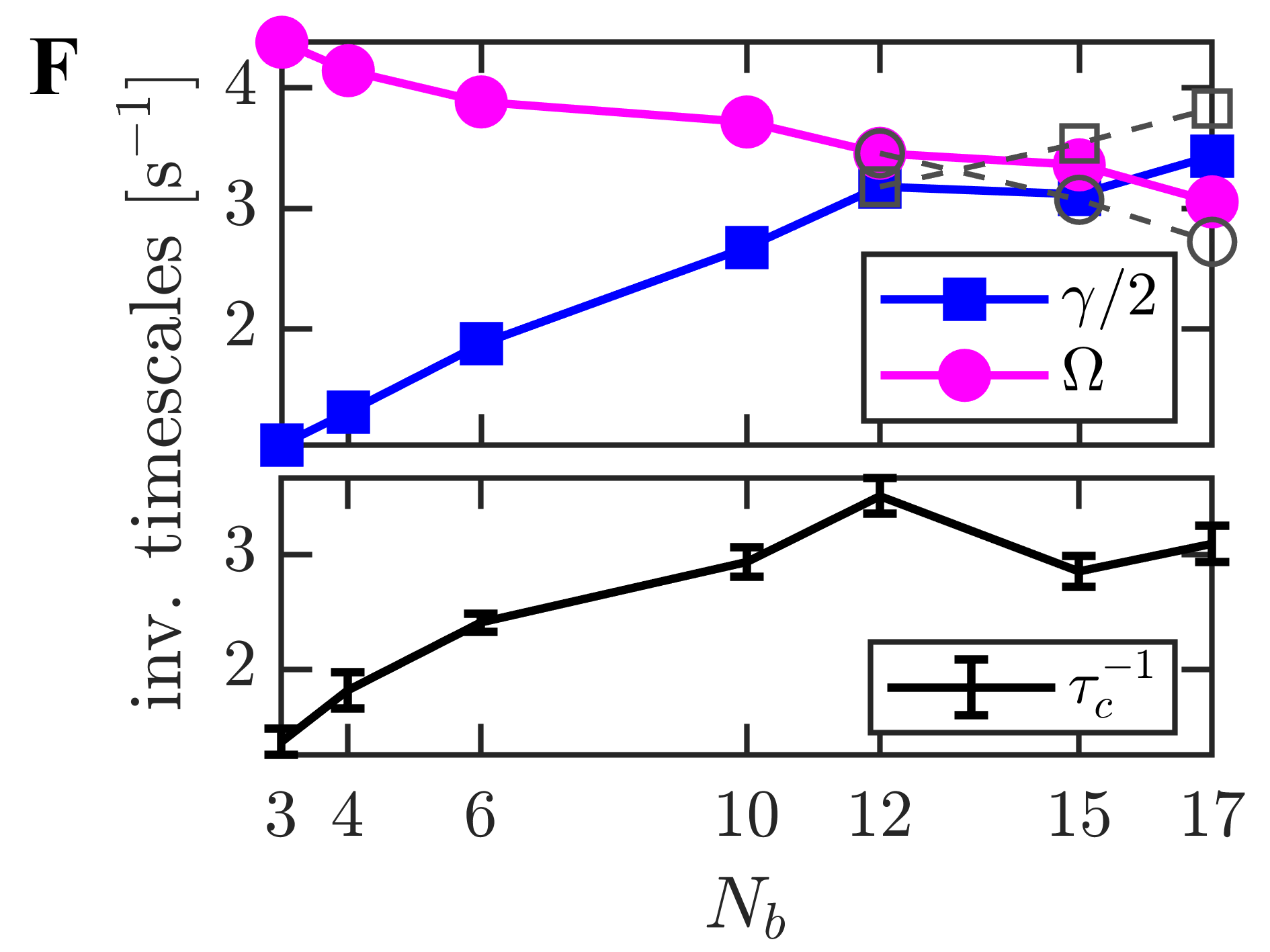}
    \caption{
    \textbf{NESS dynamics and statistics:}
    Data were obtained for active bath configurations with $N_b = \{3,4,6,10,12,15,17\}$ bbots, using a tracer of mass $m = 1 \pm 0.1$~g confined in a potential of stiffness $\kappa = 28.2 \pm 3$~g/s$^2$. Results are obtained from combined time and ensemble averages over an ensemble of 375 one-minute-long tracer trajectories recorded at 30~fps for each $N_b$. \
    \textbf{A.} Rescaled position ($x$) distributions. Dashed line show fits to exponential ($N_b=3$) and Gaussian ($N_b=15$) functions. \
    \textbf{B.} Rescaled velocity ($v_x$) distributions. An exponential fit is shown for $N_b=3$ and a stretched exponential fit for $N_b=15$. \
    \textbf{C,D.} Position (C) and velocity (D) autocorrelation functions (ACFs), showing increasingly damped dynamics with higher $N_b$. \
    \textbf{E.} Position and velocity ACFs for $N_b=10$, fitted with Eq. \ref{eq:Langevin2} using $\gamma$ and $\Omega$ as fitting parameters, yielding consistent descriptions for both $C_{xx}$ and $C_{vv}$. \
    \textbf{F.} Extracted relaxation rate ($\gamma$) and trapping frequency ($\Omega$) as functions of $N_b$ (upper panel).
    The impact of under-sampling on the observed dynamics is pronounced for large $N_b$ ($15$ and $17$). Hollow markers (dashed lines) show measurements at $30$~fps, while colored markers correspond to $60$~fps. The lower frame rate leads to apparent overdamped behavior, whereas the higher frame rate reveals dynamics consistent with a critically-damped regime. The collision frequency ($\tau_c^{-1}$) is plotted versus $N_b$ (lower panel), where $\tau_c$ is the measured mean-free time between tracer-bbot collisions. 
    }
    \label{fig:2}
\end{figure}

\textbf{Nonequilibrium properties of the tracer.-} \ 
The NESS statistics and the dynamics of the tracer particle (in unperturbed conditions) are shown in Fig. \ref{fig:2}, with the number of bbots $N_b$ used as a control parameter. 
These results were obtained by averaging 375 particle trajectories of 1-minute duration and a time interval of $\Delta t=1/30$~s.

Both the stationary position ($x$) and velocity ($v_x$) distributions of the tracer generally deviate from Gaussian behavior.
As shown in Fig. \ref{fig:2}\textbf{A}, the position distribution evolves with increasing $N_b$, transitioning from an exponential form toward a Gaussian-like profile, particularly within the harmonic trap boundary $|x|<10$\thinspace{cm} (for $|x|>10$~cm the arena possesses a steeper curvature, i.e., soft boundary for the trapped tracer and bbots).
The velocity distributions (Fig. \ref{fig:2}\textbf{B}) exhibit an exponential decay for small  $N_b=3$, with reliable statistics within $|v_x|<50$\thinspace{cm/s}. 
At larger $N_b$, the velocity distributions converge to a distinct non-Gaussian shape that is well described by a stretched exponential function.

The stationary dynamics of the tracer are further characterized in Figs. \ref{fig:2}\textbf{C,D} via the position and velocity autocorrelation functions (ACF). 
An increase in the rate of dissipative collisions upon adding more bbots, leads to more strongly damped tracer motion. 
This is reflected in a suppression of oscillations in the relaxation dynamics, and in the reduction of the relaxation time.

The position and velocity ACFs are well fit by the generic solution of a noisy damped harmonic oscillator (Fig. \ref{fig:2}\textbf{E}).
We focus on the velocity ACF that is given by \cite{wang1945,yaghoubi2017},
\begin{align}
    \overline{C}_{vv}(t) =\frac{C_{vv}(t)}{\langle v_x^2\rangle_0}= e^{-t \gamma/2}\left( \cos{\omega t}-\gamma\frac{\sin{\omega t}}{2\omega} \right), \label{eq:Langevin2} 
\end{align}
\noindent 
where the bar indicates normalization, $\langle v_x^2\rangle_0$ is the (stationary) velocity variance, with the angular brackets representing an ensemble averaged, and $\omega^2\equiv\Omega^2-\gamma^2/4$.

Fitting the ACF to Eq.~\ref{eq:Langevin2} provides system-specific values of the effective inverse timescales: the damping rate, $\gamma$, and the harmonic frequency, $\Omega$.

These inverse timescales are displayed in Fig. \ref{fig:2}\textbf{F} as a function of $N_b$, showing a transition between underdamped ($\Omega<\gamma/2$) towards critically-damped ($\Omega\approx\gamma/2$) dynamics as $N_b$ increases.
A reduction in the relaxation time, $2\gamma^{-1}$, corresponds to an increase in collision frequency, evaluated by image analysis as the frequency of tracer-bbot collisions, $\tau_c^{-1}$ (black line).

At high densities ($N_b=15$ and $17$), fast sequential collisions require a higher temporal resolution to capture the tracer's inertial dynamics. 
In these cases, we perform and use additional recordings with $\Delta t=1/60$~s to extract the effective inverse timescales (solid lines).
At this sampling rate the fit parameters are consistent with the critical regime, whereas $\Delta t=1/30$~s results in overdamped behavior (dashed lines).
These measurements of velocity fluctuations and their corresponding system-specific inverse timescales are further discussed in the context of the kinetic temperature in Sec.~3.2.

\section{Effective temperatures} 

In the following sections, we aim to assess whether three independent definitions of effective temperatures can yield mutually consistent values across different conditions and observables. 
Specifically, we examine the FDR for position and velocity observables under a step-perturbation (3.1), \ effective equipartition between kinetic and potential energies (3.2), \ and a work FR in a perturbed NESS (3.3).

\subsection{FDR temperature} 

By subjecting the tracer to a small step-perturbation (an abrupt arrest of a constant force $F_0$ at $t=0$), a linear FDR can be expressed for both position and velocity observables as \cite{haga2015,yaghoubi2017} (setting $k_B=1$):
\begin{align}
    R_x(t) = \frac{F_0}{T_{\text{eff}}}C_{xx}(t), \label{eq:FDR1} \\
    R_v(t) = \frac{F_0}{T_{\text{eff}}}C_{vx}(t), \label{eq:FDR2}
\end{align}
\noindent 
where $R_x$ and $R_v$ are the mean response functions of the position and velocity observables, respectively.
The correlation functions in the (long-time) unperturbed NESS are given by a position ACF, $C_{xx}=\langle x(t)x(0)\rangle_0$, and a position-velocity cross-correlation function, $C_{vx}=\langle v_x(t)x(0)\rangle_0$. 
In this context, an effective temperature $T_{\text{eff}}$ is defined as a proportionality constant.

Notably, a static FDR can be recovered in Eq. \ref{eq:FDR1} at $t=0$, with $C_{xx}(0)\rightarrow\langle x^2\rangle _0$ and $R_x(0)\rightarrow\Delta x_\epsilon$, where $\langle x^2\rangle_0$ is the position variance in the unperturbed NESS, and $\Delta x_\epsilon=\langle x\rangle_\epsilon-\langle x\rangle_0$ is the mean displacement between the average position in the two steady states, perturbed (sub-index $\epsilon$) and unperturbed (sub-index $0$).
Considering that $F_0=\kappa\Delta x_\epsilon$, the effective temperature $T_{\text{eff}}$ coincides with the definition of potential energy equipartition,
\begin{align}
    T_{\text{pot}}=\kappa\langle x^2\rangle_0 . \label{eq:temp}
\end{align}
\noindent
Thus, $T_{\text{eff}}=T_{\text{pot}}$ is always fulfilled when the system obeys the dynamic FDR Eq. \ref{eq:FDR1}, even in the presence of non-Gaussian steady state distributions.

\begin{figure}[t] 
    \centering
    \includegraphics[width=0.23\textwidth]{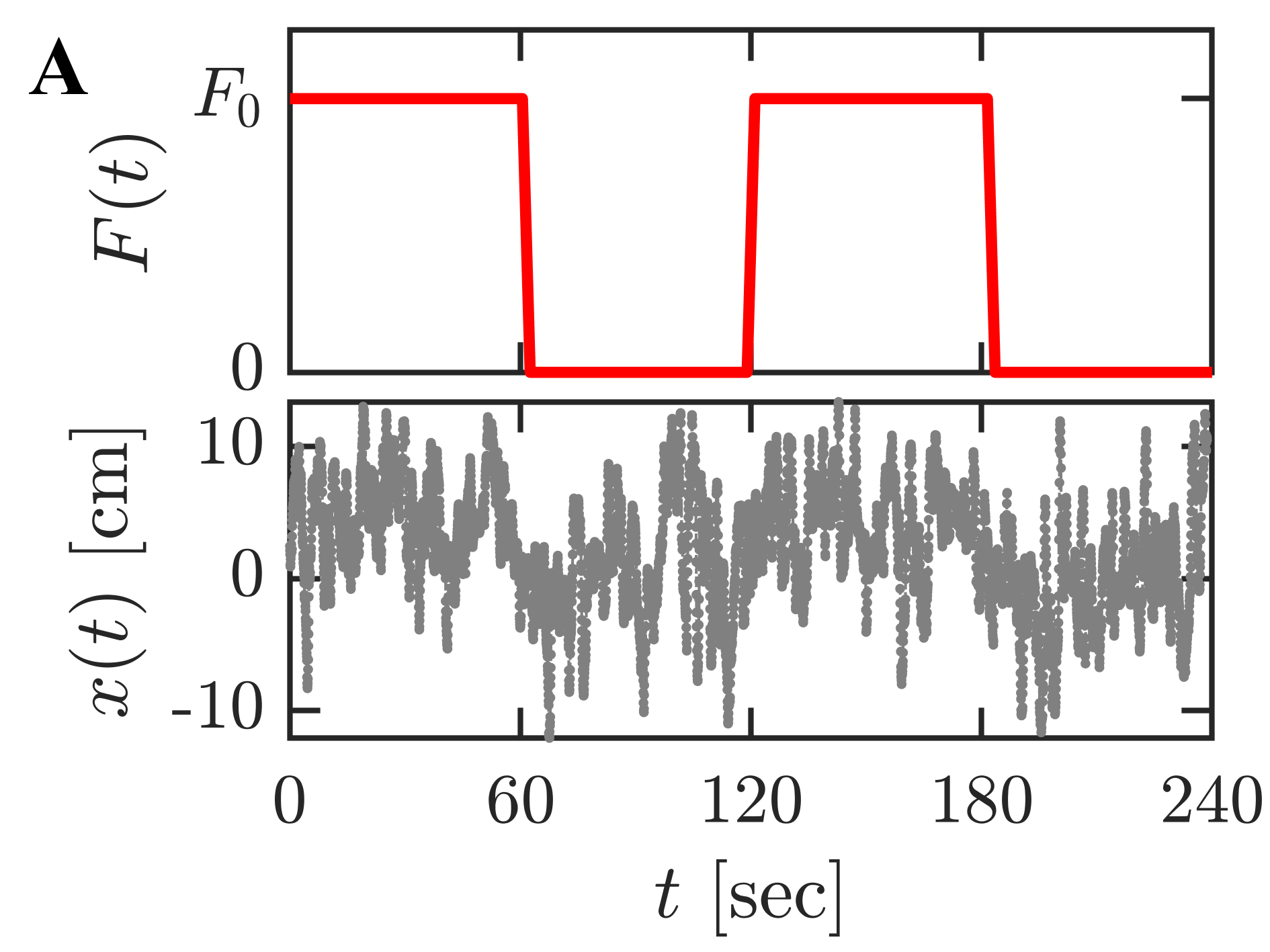}
    \includegraphics[width=0.23\textwidth]{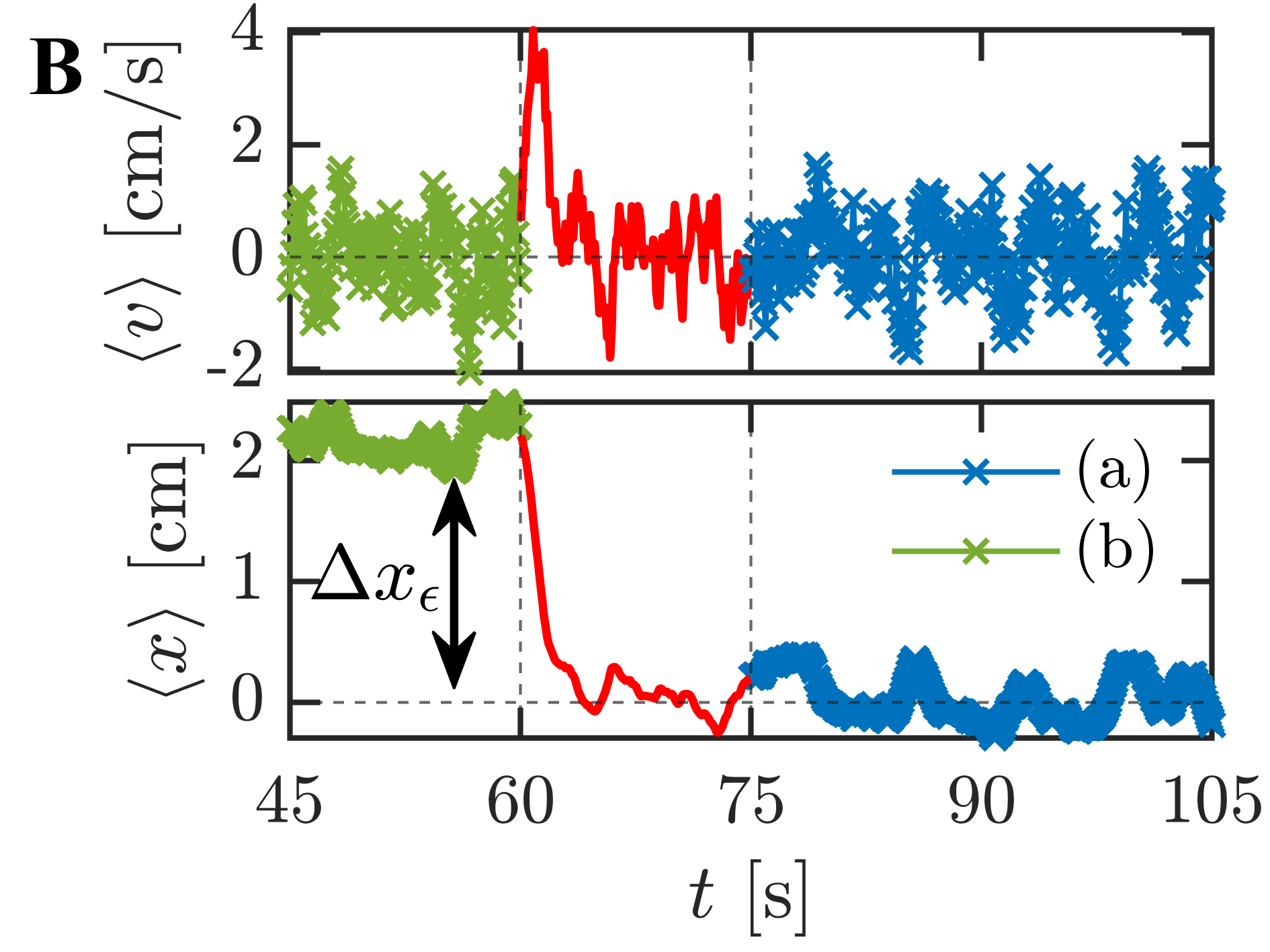}
    \includegraphics[width=0.23\textwidth]{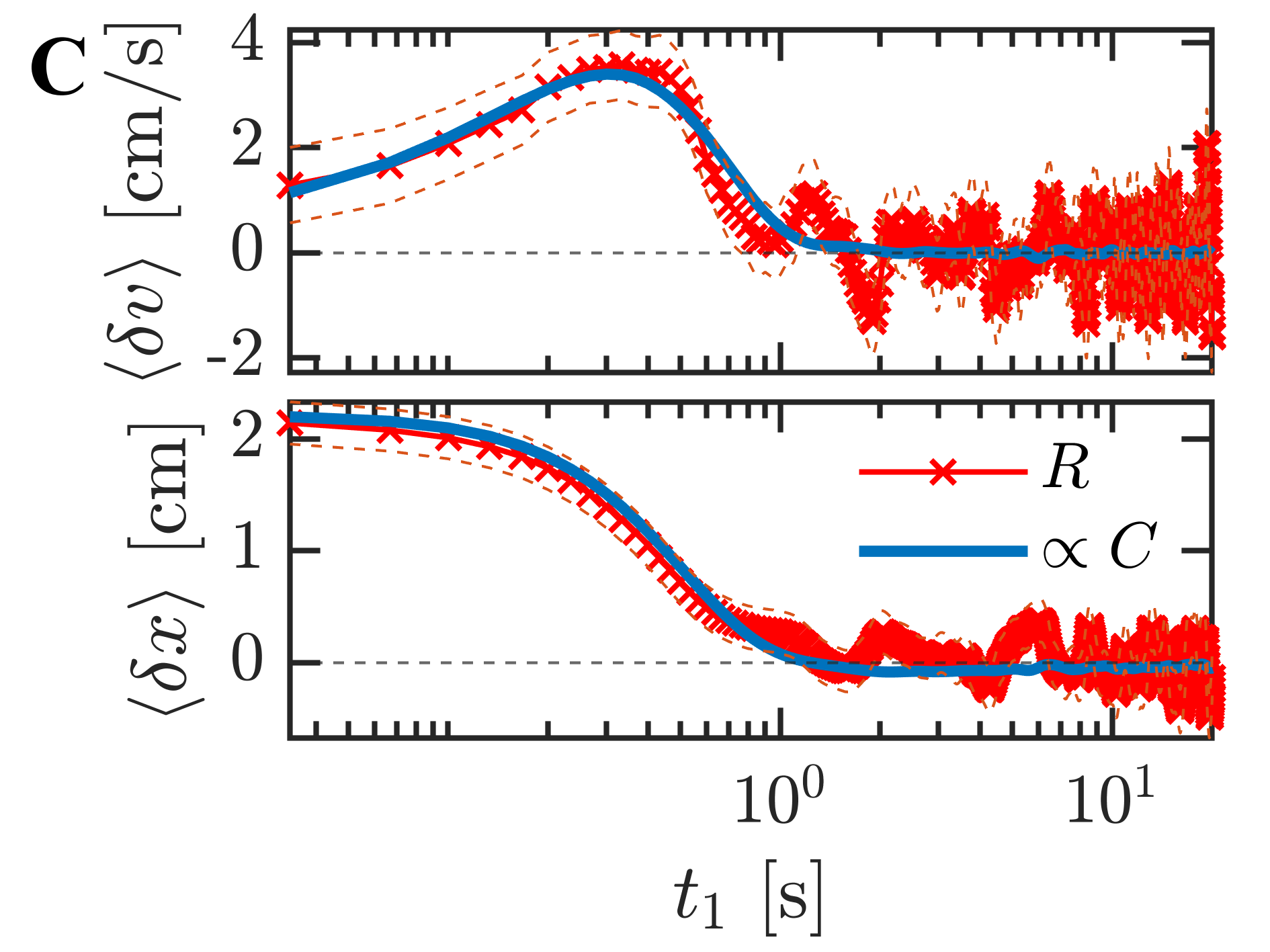}
    \includegraphics[width=0.23\textwidth]{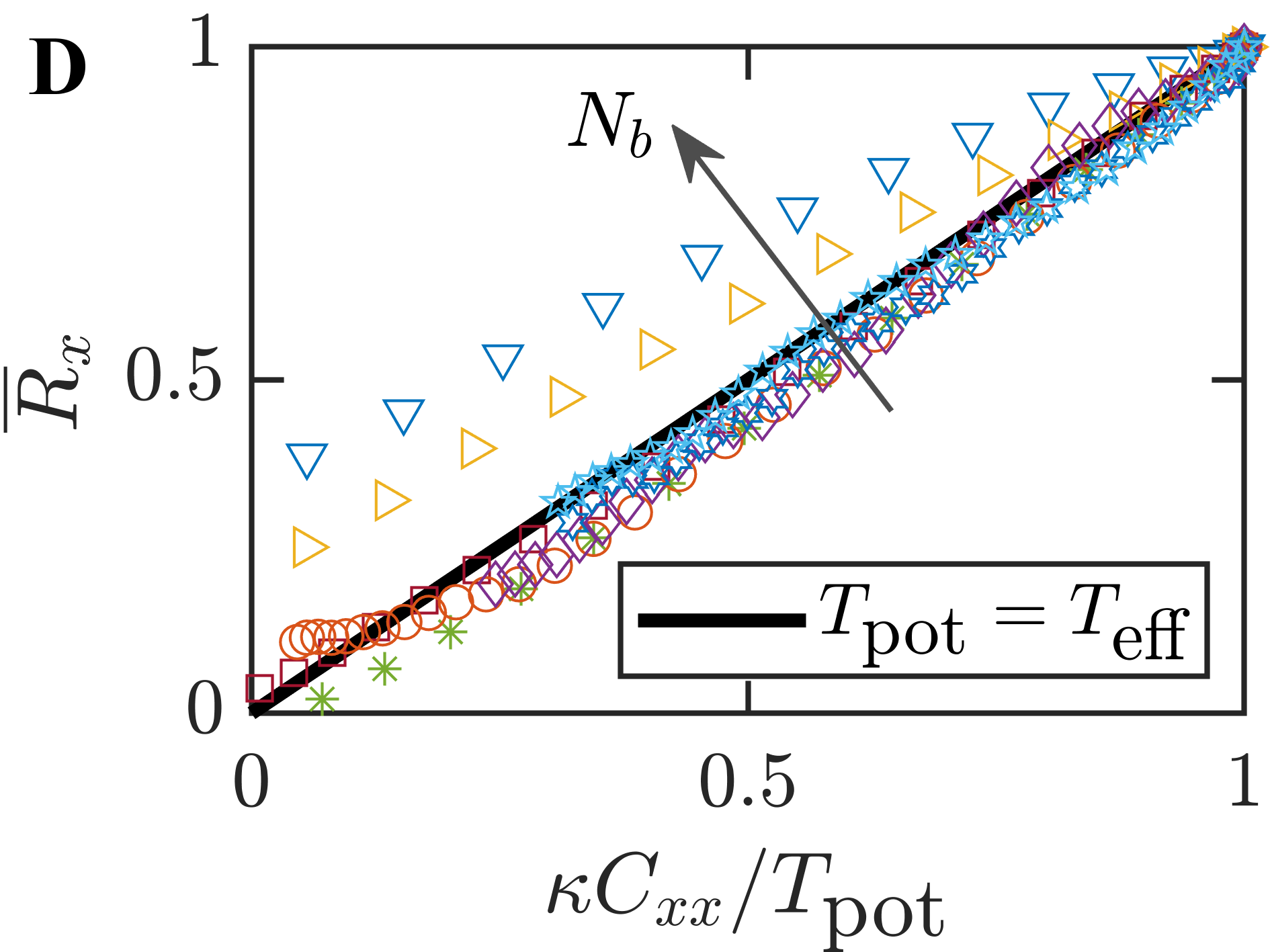}
    \caption{
    \textbf{FDR test of the full-response to a step-perturbation:} \ 
    Results for mean values are obtained from pure ensemble averages over 375 two-minute step-perturbation sequences (at $t_0=60$~s), under weak external airflow applied by a fan at an operating voltage $10$~V; correlation functions are computed in the steady state by a combined time and ensemble average.
    \textbf{A.} Example of a typical perturbation sequence $x(t)$, where the fan is turned on for one minute and abruptly turned off for the subsequent minute. \
    \textbf{B.} 
    Time-dependent mean velocity $\langle v_x(t) \rangle$ and mean position $\langle x(t) \rangle$. The shown interval captures the transient response following perturbation arrest (red line) and the unperturbed steady state (a), from which correlation functions are computed. The perturbed steady state is displaced from the trap center by $\Delta x_{\epsilon} = 2.2$ cm (b). \ 
    \textbf{C.} FDR analysis using Eq. \ref{eq:FDR1} (lower panel) and Eq. \ref{eq:FDR2} (upper panel). These results were obtained with $N_b=10$. \
    \textbf{D.} Parametric plot of position FDRs for $N_b=\{2,3,4,6,10,12,15,17\}$ realizations, where the effective temperature $T_{\text{eff}}$ determines the linear slope.
    In equilibrium-like behavior, all FDR data collapse onto a single line with slope $T_{\text{eff}} \sim \kappa \langle x^2 \rangle_0$. Clear deviations from this equality are observed for the lowest densities ($N_b=2$ and 3, triangles), indicating FDR violations.
    }
    \label{fig:3}
\end{figure}

Fig. \ref{fig:3}\textbf{A} illustrates the experimental protocol designed to measure the system's linear response (see also Refs. \cite{boriskovsky2024,engbring2023}). 
The tracer is subjected to a small mechanical perturbation (external fan, $10$V operating voltage, $F_0\approx62\mu$N on average).
During an experiment, every two minutes, the fan is turned on for a minute and abruptly turned off at $t_0=60$s for the following minute.
In Fig. \ref{fig:3}\textbf{B} we show the time-dependent ensemble averages $\langle x(t)\rangle$ and $\langle v_x(t)\rangle$, obtained with $N_b=10$.
These trajectories capture the relaxation dynamics towards a steady state following the force removal (a), as well as a perturbed steady state (b), characterized by a shifted mean $\Delta x_\epsilon$ (see also Fig. \ref{fig:1}\textbf{B}).

Fig. \ref{fig:3}\textbf{C} presents the corresponding FDR tests for both position and velocity observables.
The system's full response to the abrupt force arrest at time $t_1=t-t_0$ is evaluated as $R_x(t_1)=\langle x(t_1)\rangle-\langle x\rangle_0$ and $R_{v}(t_1)=\langle v_x(t_1)\rangle-\langle v_x\rangle_0$, and compared with the unperturbed correlation functions, $C_{xx}$ and $C_{xv}$. 
Within measurement error, these results validate the FDRs given in Eqs. \ref{eq:FDR1} and \ref{eq:FDR2}, with an effective temperature determined by Eq. \ref{eq:temp}. 
Thereby, in this system the tracer's potential temperature satisfies a dynamic FDR in both position and velocity observables.
We note that the signal-to-noise ratio in $R_v$ diminishes with increasing $N_b$, rendering the velocity response experimentally inaccessible for larger values ($N_b>10$).
The individual FDR tests of the experimental setups are further provided in the SI$^{\dag}$ (see figures 3-4 therein).

Fig. \ref{fig:3}\textbf{D} presents a parametric plot of the position FDR, showing normalized fluctuation and response quantities across different $N_b$ configurations.
Specifically, the solid linear line represents validation of Eq. \ref{eq:FDR1} with $T_{\text{eff}}=T_{\text{pot}}$.
Violations of Eq.~\ref{eq:FDR1} are mainly observed with the lowest $N_b=2$ and $3$, indicating a breakdown of the effective temperature description (as detailed in Ref. \cite{boriskovsky2024}).
In contrast, with larger $N_b$, the tracer's potential temperature $T_{\text{pot}}$ satisfies a dynamic FDR and is therefore consistent with $T_{\text{eff}}$. 

\begin{figure}[t] 
    \centering
    \includegraphics[width=0.235\textwidth]{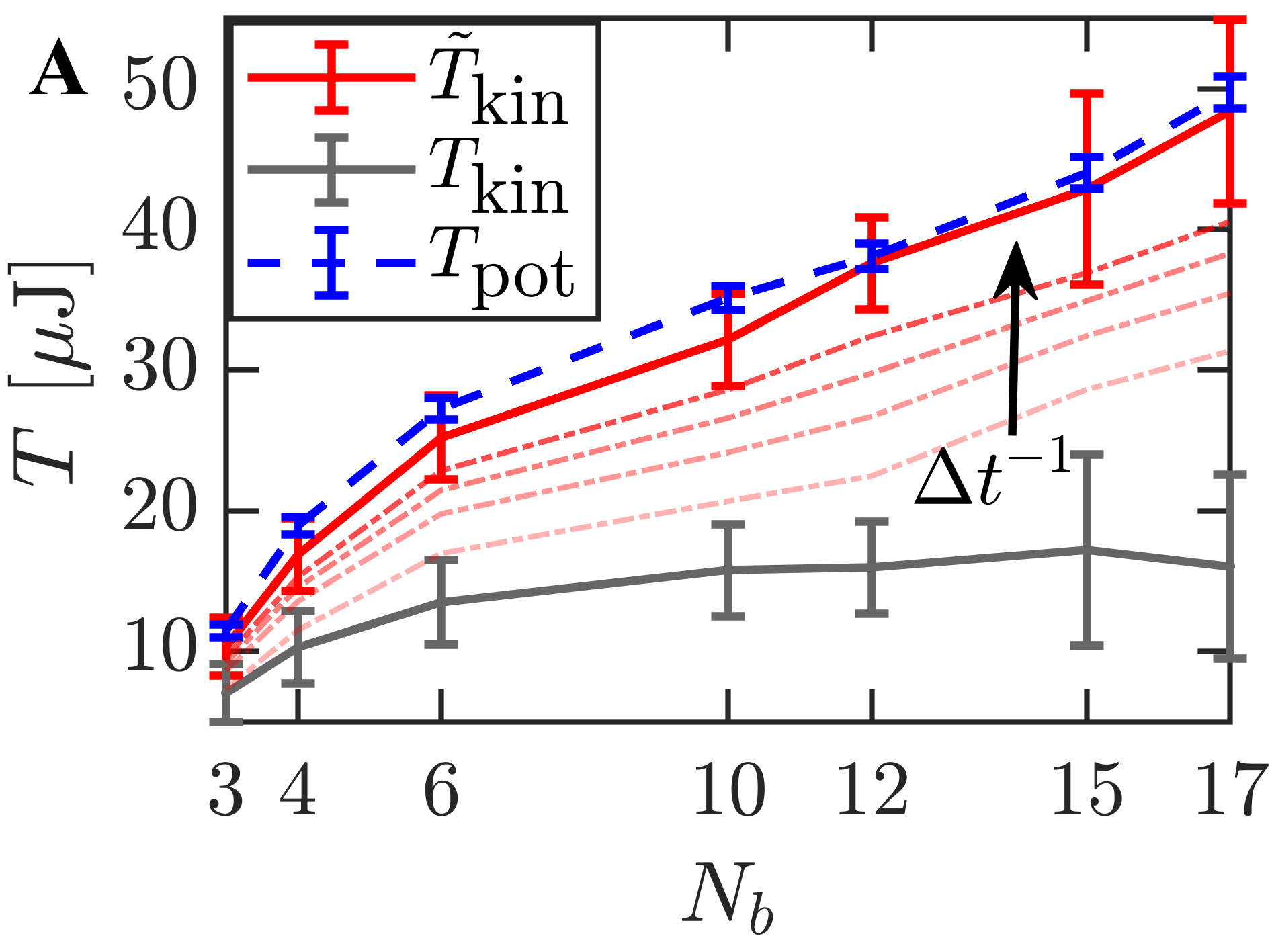}
    \includegraphics[width=0.235\textwidth]{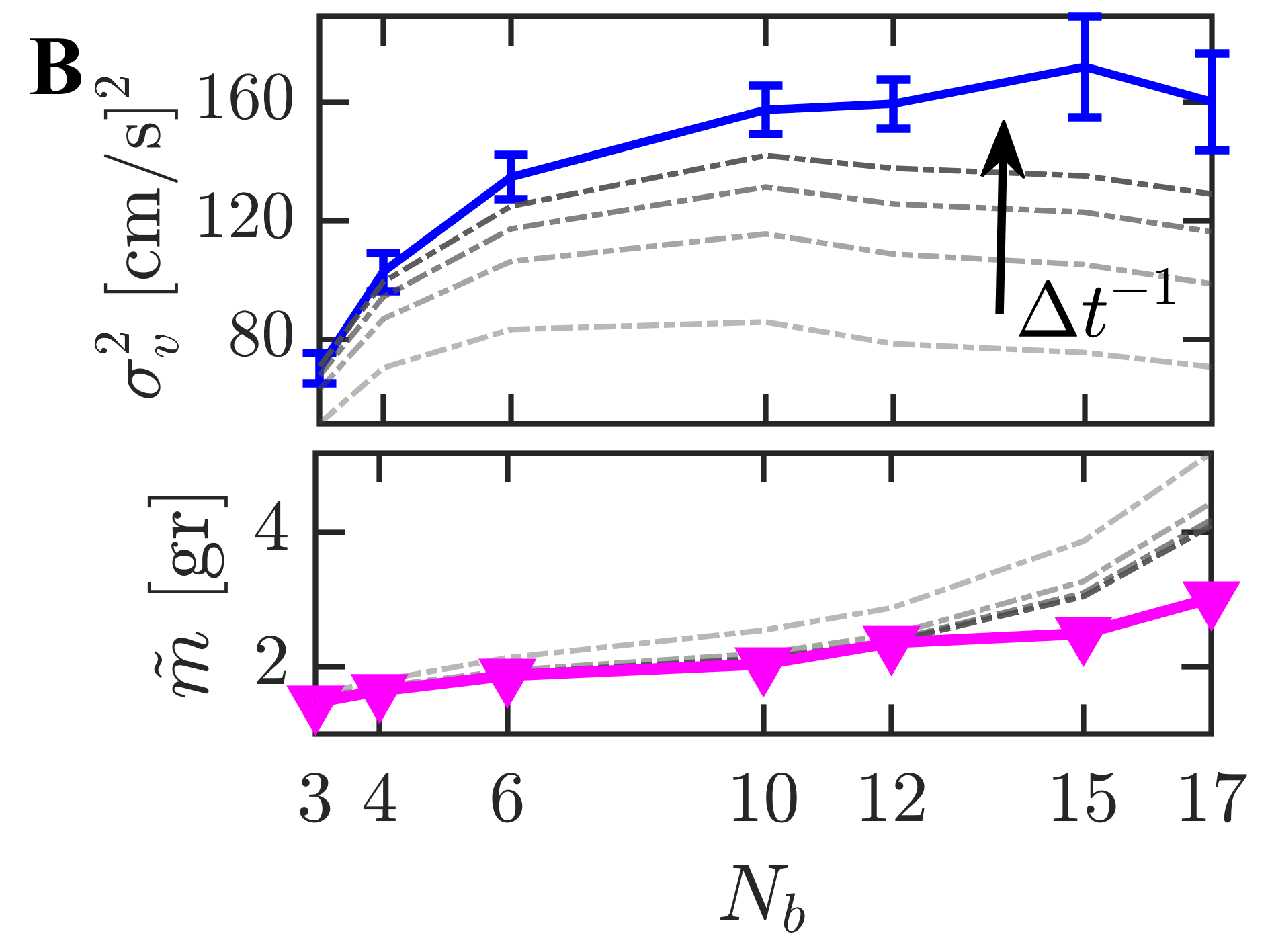}
    \caption{
    \textbf{The modified kinetic temperature:} \ 
    Data are presented for configurations with $N_b = \{3,4,6,10,12,15,17\}$ bbots, averaged over 375 one-minute tracer trajectories recorded in unperturbed conditions. \
    \textbf{A.} Comparison of the standard kinetic temperature $T_{\text{kin}}$ (solid gray line), calculated using the tracer mass $m \approx 1$~g (Eq. \ref{eq:kinetic}), the modified kinetic temperature $\tilde{T}_{\text{kin}}$ (solid red line), evaluated using Eq. \ref{eq:temp2}, and the potential temperature $T_{\text{pot}}$ (dashed blue line). Error bars indicate standard deviations. Dot-dashed red lines show deviations in $\tilde{T}_{\text{kin}}$ arising from increasing the measurement time interval $\Delta t$. \
    \textbf{B.} Separate contributions to $\tilde{T}_{\text{kin}}$: (upper panel) the velocity variance $\sigma_v^2 = \langle v_x^2 \rangle_0$, and (lower panel) the effective mass $\tilde{m}$, evaluated from stationary ACF dynamics using Eqs. \ref{eq:Langevin2} and \ref{eq:temp2}. Gray dot-dashed lines indicate deviations due to increasing $\Delta t$ in both panels.
    }
    \label{fig:4}
\end{figure}

\subsection{Kinetic and potential energy partition} 

To probe deviations from equilibrium behavior, we define the tracer's kinetic temperature as,
\begin{align}
T_{\text{kin}}=m\langle v^2_x\rangle _0 . \label{eq:kinetic}
\end{align}
\noindent
As seen in  Fig.~\ref{fig:4}, this definition seems to result in a contradiction between the kinetic and potential temperatures, since deviations from the potential temperature $T_{\text{pot}}$ (Eq.~\ref{eq:temp}) are observed for all $N_b$ configurations (solid gray line in Fig.~\ref{fig:4}\textbf{A}). 
Namely, $T_{\text{kin}}$ saturates and becomes largely independent of $N_b$ in the high-density limit.
This clear mismatch is further supported by direct measurements of the effective trapping frequencies $\Omega$, extracted from the tracer's velocity ACFs (see Fig.~\ref{fig:2}\textbf{F}).
Across all system setups, $\Omega$ yields values lower than the \textit{natural} frequency $\sqrt{\kappa/m}\approx5.3~\rm{s^{-1}}$, even in an underdamped dynamic regime. 

\begin{figure}[t] 
    \centering
    \includegraphics[width=0.235\textwidth]{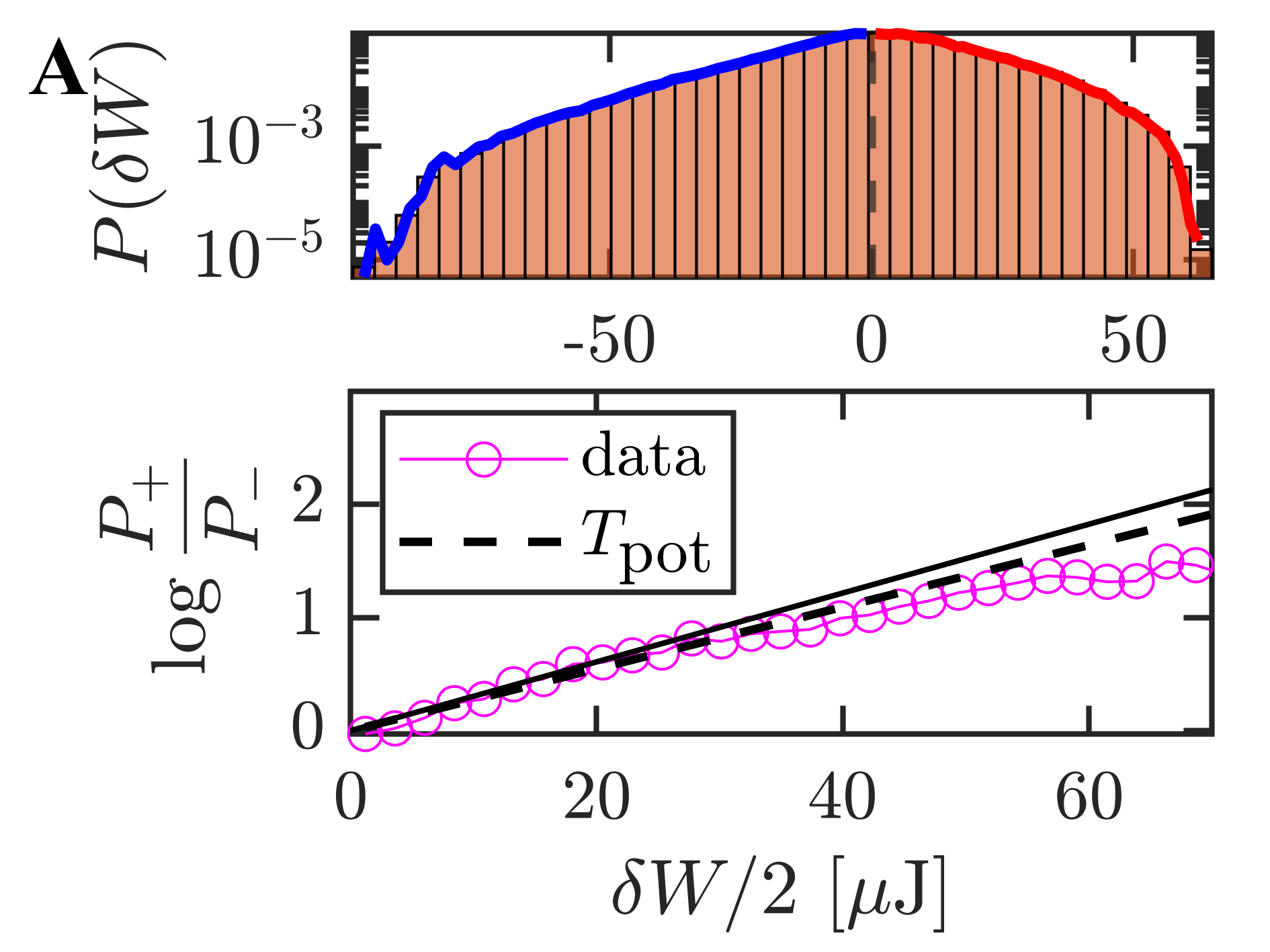}
    \includegraphics[width=0.235\textwidth]{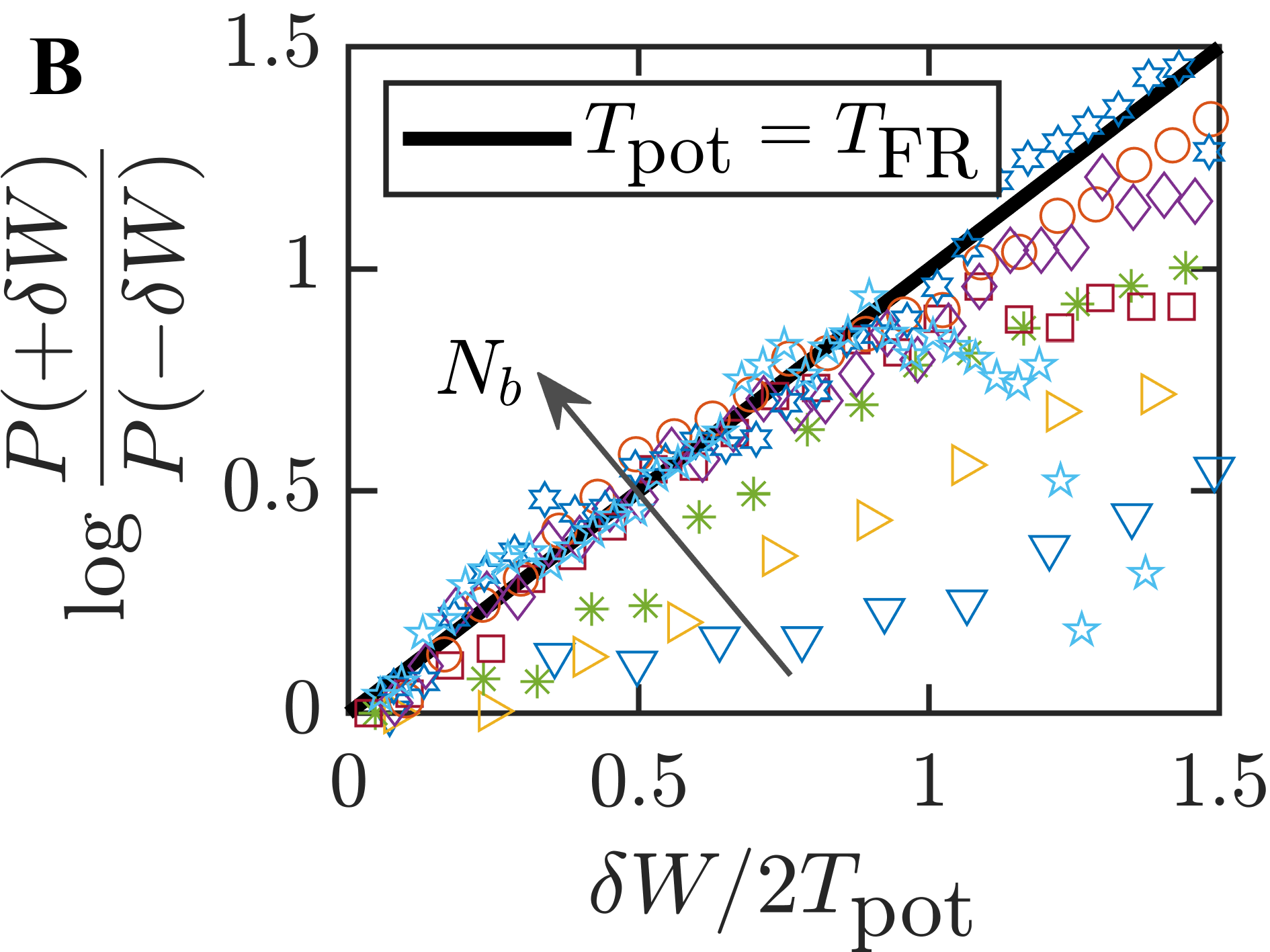}
    \caption{
    \textbf{Steady state work FR test:} \ 
    Results present averages over 375 one-minute perturbed tracer trajectories, under a strong external airflow applied by a fan at an operating voltage $13.5$~V.
    \textbf{A.} Work FR test for $N_b=10$. 
    The upper panel shows an asymmetric work ($\delta W$) distribution in a NESS under a strong continuous perturbation (airflow). 
    The temperature measurement of $T_{\text{FR}}$, extracted from a fit to Eq. \ref{eq:FR} (small $\delta W$, solid line), is consistent with the potential temperature ($T_{\text{pot}}$, dashed line) determined from the unperturbed definition in Eq. \ref{eq:temp}. \
    \textbf{B.} Normalized plot of work FR tests across various $N_b=\{2,3,4,6,10,12,15,17\}$ configurations. 
    The solid (black) line indicates an ideal agreement between the effective temperatures obtained from the work FR and from the equipartition definition ($T_{\text{pot}}=T_{\text{FR}}$), demonstrating thermodynamic-like consistency. 
    }
    \label{fig:5}
\end{figure}

We note, however, that for tracer particles embedded in active or driven granular media, an effective mass emerges as a consequence of persistent athermal fluctuations and memory effects \cite{maes2020fluctuating,seguin2022,giona2024}. Accordingly, we define an effective mass, $\tilde{m}=\kappa\Omega^{-2}$, derived from the velocity correlations.
Mapping the dynamics onto a linear Langevin equation, consistent with the ACF solution in Eq.~\ref{eq:Langevin2}, relies on short persistence times and typically necessitates a modification of system parameters \cite{wisniewski2024,shea2024}.
Consequently, the definition of the kinetic temperature is modified to,
\begin{align}
    \tilde{T}_{\text{kin}}=\tilde{m} \langle v_x^2\rangle _0 , \label{eq:temp2} 
\end{align}
\noindent
where $\tilde{m}$ is defined as a $N_b$-specific effective mass, under a constant potential stiffness $\kappa$.

Using the effective mass to compute the kinetic temperature yields agreement between the two effective temperatures (solid red line in Fig. \ref{fig:4}\textbf{A}).
We note that the effective mass correction is derived from the correlation function of the unperturbed NESS, independent of both the step-response measurement and the FDR result (previous section).
Deviations of $\tilde{T}_{\text{kin}}$ measurements resulting from low measurement rates are shown as red dashed lines in Fig. \ref{fig:4}\textbf{A}.

Fig.~\ref{fig:4}\textbf{B} shows the separate contributions to $\tilde{T}_{\text{kin}}$: the velocity variance $\langle v_x^2 \rangle_0$ (upper panel) and the effective mass $\tilde{m}$ (lower panel), across different $N_b$. 
We note that $\tilde{m}$ deviates from the tracer's actual mass ($m = 1$ g) and results in values that monotonically increase with $N_b$.
Both measurements show a dependence on the sampling rate ($\Delta t^{-1}$), particularly for large $N_b>10$ systems where $\langle v_x^2 \rangle_0$ saturates.
Nevertheless, under well-resolved velocity fluctuations, the associated modified kinetic temperature $\tilde{T}_{\text{kin}}$ is consistent with the potential temperature $T_{\text{pot}}$, linking tracer dynamics with its linear response through $T_{\text{eff}}=T_{\text{pot}}$.

\subsection{Effective temperature based on work FR} 

Here we investigate a work FR in a NESS subjected to a constant force $F_0=\kappa \Delta x_\epsilon $ (applied at time $t\rightarrow-\infty$).
An asymmetric work distribution is obtained by the energetic fluctuation around the steady state, defined as $\delta W(t) = F_0 \left[x_\epsilon(t)-\langle x_\epsilon\rangle\right]$.
A steady state work FR can then be expressed as \cite{wang2002FR,narayan2003,gomez2010},
\begin{align}
    \log{\frac{P_\epsilon(+\delta W)}{P_\epsilon(-\delta W)}} = \frac{\delta W}{2T_{\text{FR}}}, \label{eq:FR}
\end{align}
\noindent
where $P_\epsilon(\pm \delta W)$ are negative and positive stationary work distributions, and $T_{\text{FR}}$ is defined as an FR-based effective temperature.

Fig. \ref{fig:5}\textbf{A} presents a work FR test conducted with $N_b=10$.
The upper panel shows the work distribution in a strongly perturbed steady state (state (c) in Fig. \ref{fig:1}\textbf{B}).
The red and blue lines correspond to positive and negative work fluctuations, respectively.
In the lower panel, the left-hand side of Eq. \ref{eq:FR} is plotted as a function of $\delta W$. For small enough values of $\delta W<80$~$\mu$J, the probabilities are sufficiently sampled, and $T_{\text{FR}}$ can be extracted from the slope of the linear relation (solid line).
For large $\delta W$, deviations from linearity stem from limited sampling of these rare fluctuations.

The dashed linear line represents $T_{\text{FR}}=T_{\text{pot}}$, where $T_{\text{pot}}$ is obtained independently in the unperturbed state (a).
Namely, the two temperature measures agree, providing an experimental validation of the work FR (Eq. \ref{eq:FR}). This result can be extended to different $N_b$, as shown in Fig.~\ref{fig:5}\textbf{B}. Here, the solid linear line represents $T_{\text{FR}}=T_{\text{pot}}$.
Deviations for large values of the work fluctuations are expected due to under-sampling. 
Clear violations of Eq.~\ref{eq:FR} are mainly observed for $N_b=2$ - $4$. For larger $N_b$ systems, $T_{\text{FR}}$ is also consistent with the FDR-based measure $T_{\text{eff}}$.
Notably, these temperatures were measured in independent measurements under different perturbations. 
This result shows that, for a broad range of parameters ($N_b=6$ - $17$), the same effective temperature that governs static (equipartition) and dynamic (FDR) properties of the system also rules the irreversibility of energy exchanges.


\begin{figure}[t] 
    \centering
    \includegraphics[width=0.22\textwidth]{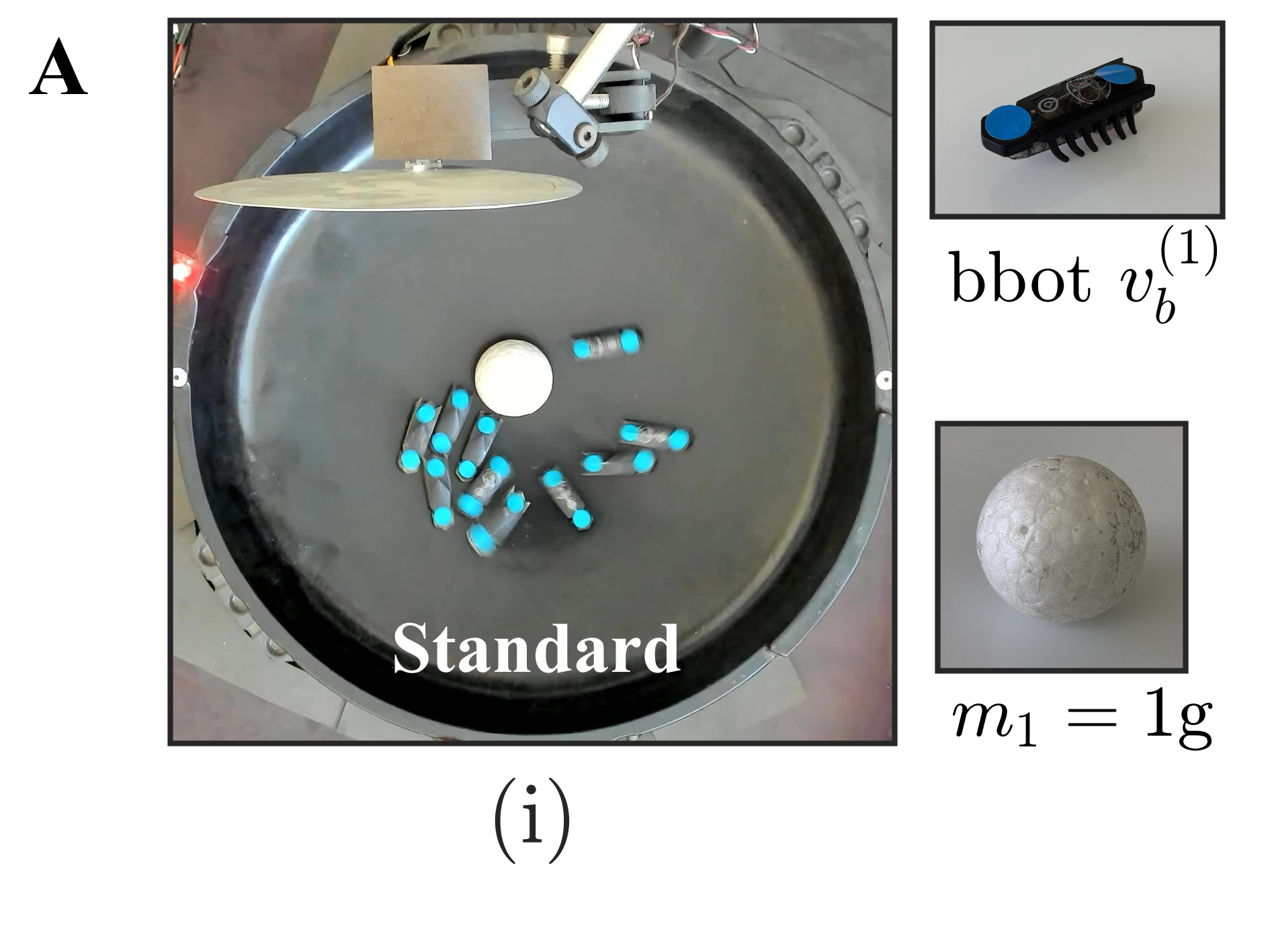}
    \includegraphics[width=0.22\textwidth]{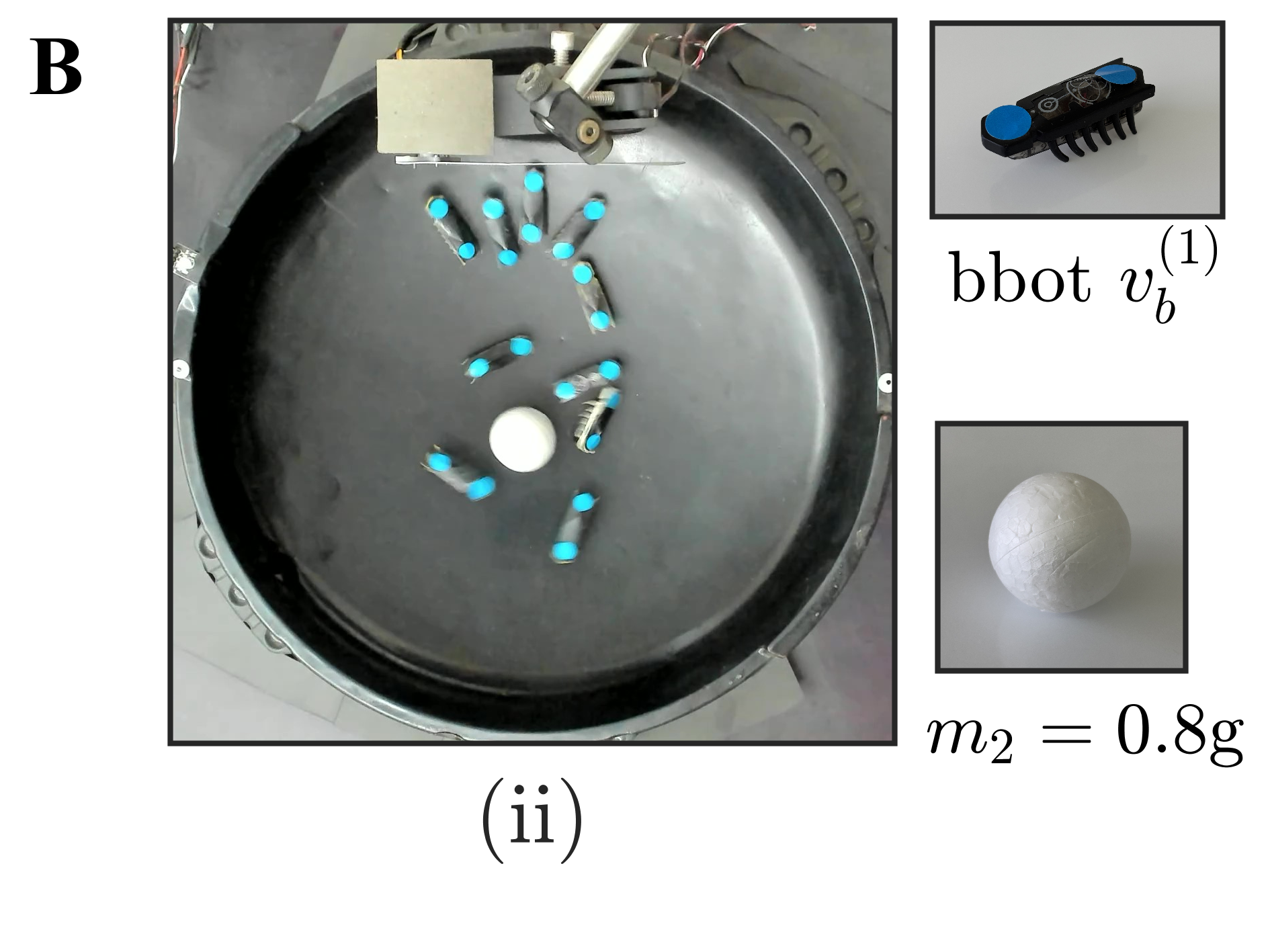}
    \includegraphics[width=0.22\textwidth]{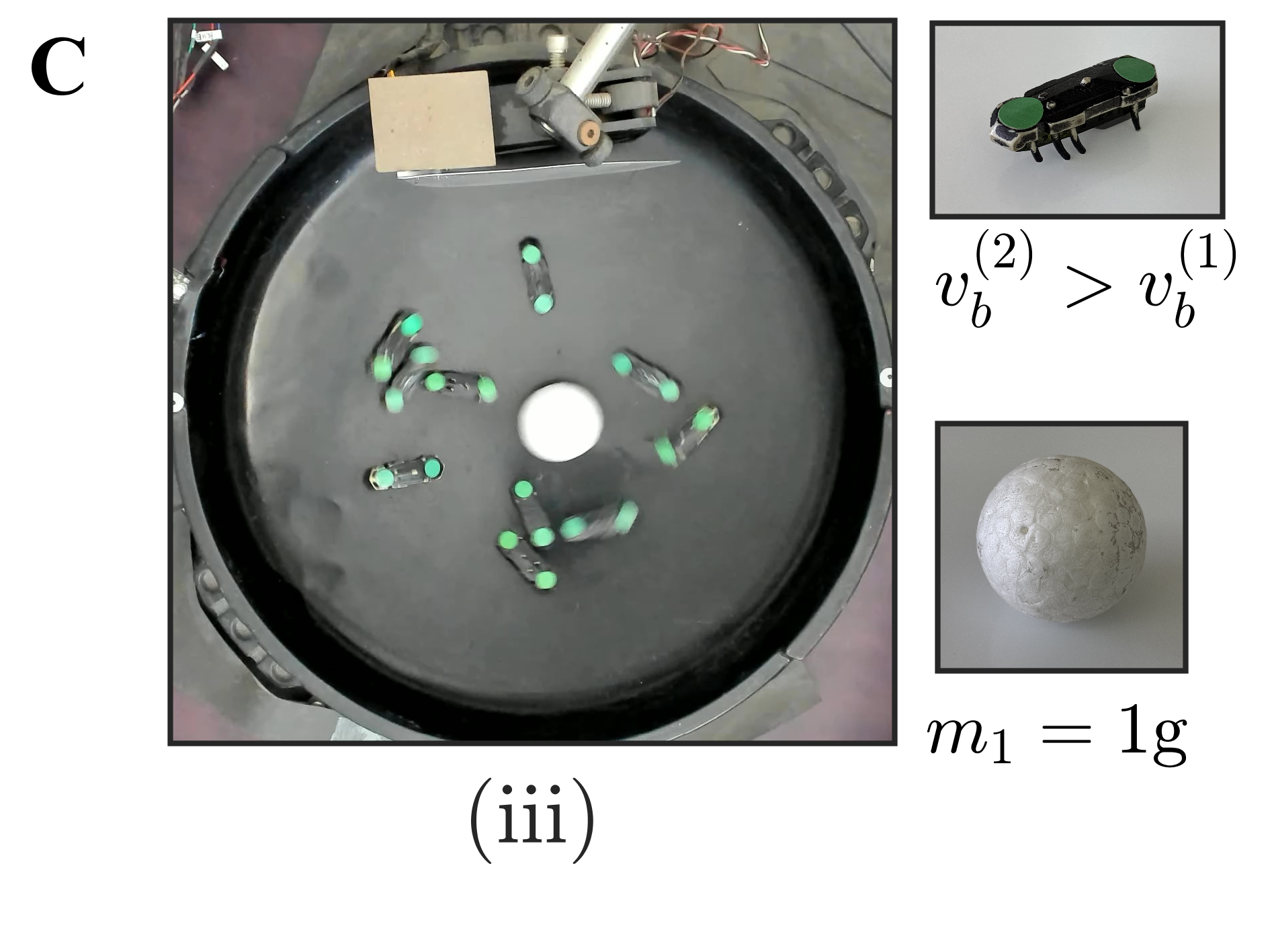}
    \includegraphics[width=0.22\textwidth]{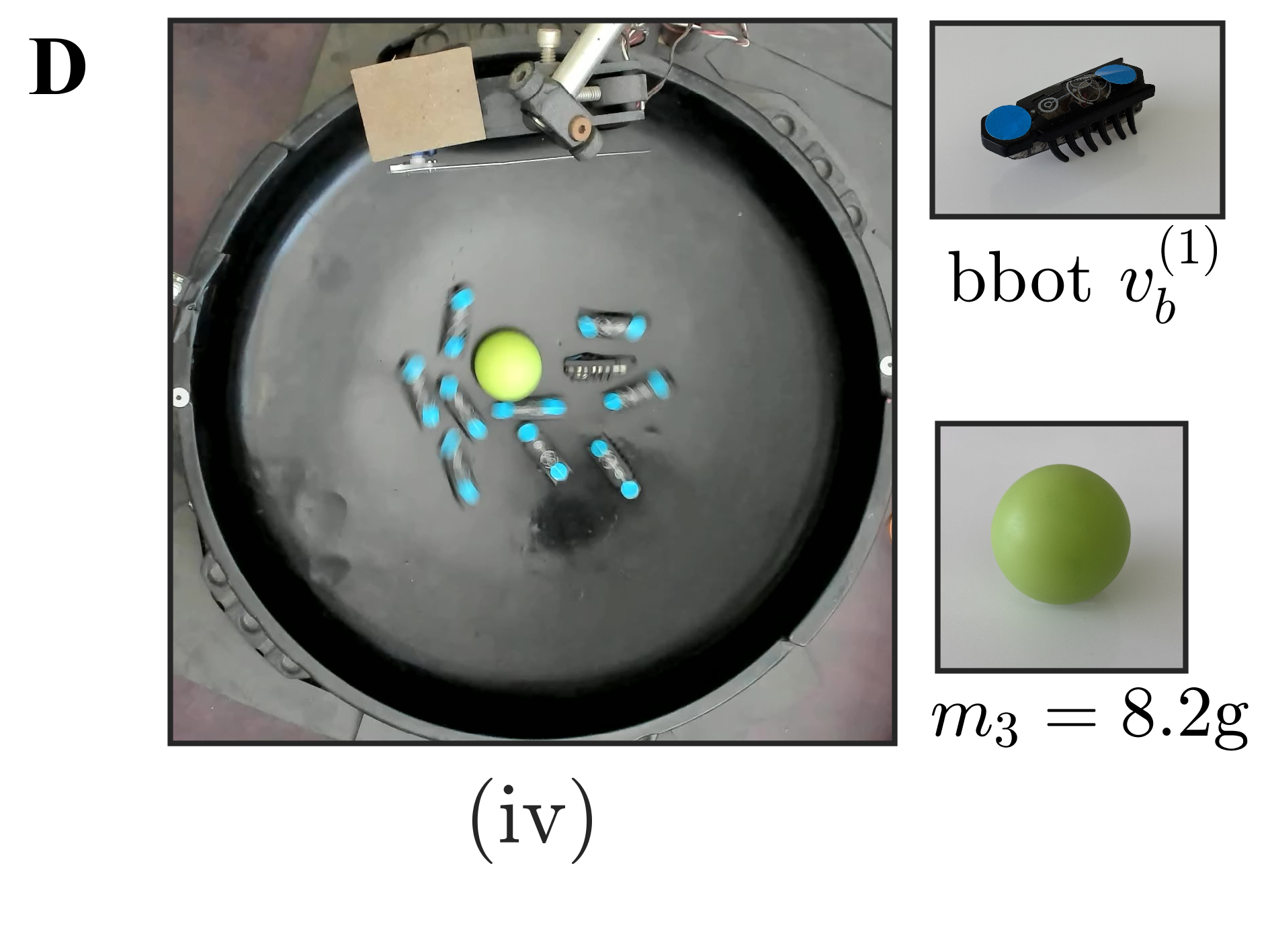}
    
    \includegraphics[width=0.23\textwidth]{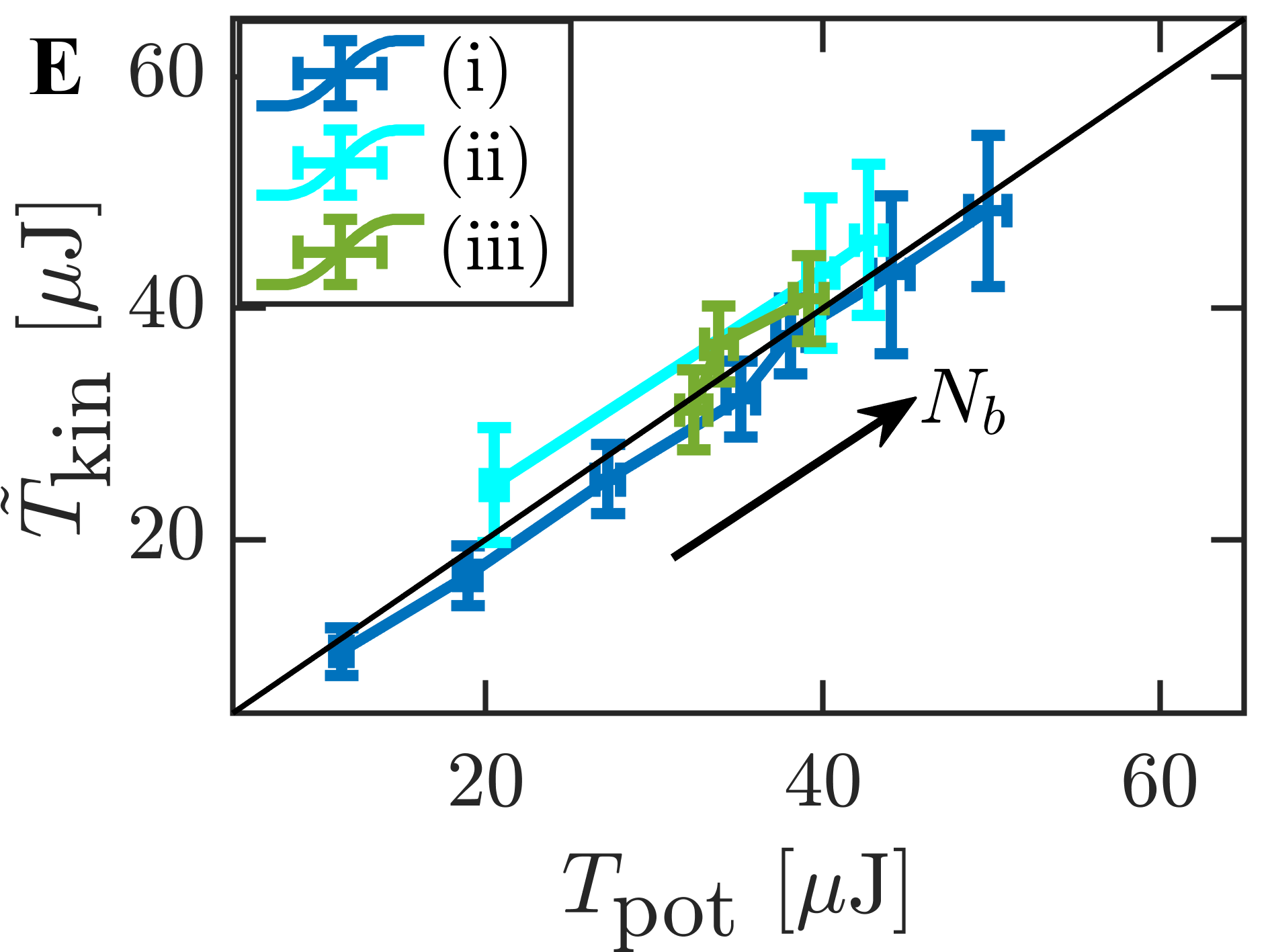}
    \includegraphics[width=0.23\textwidth]{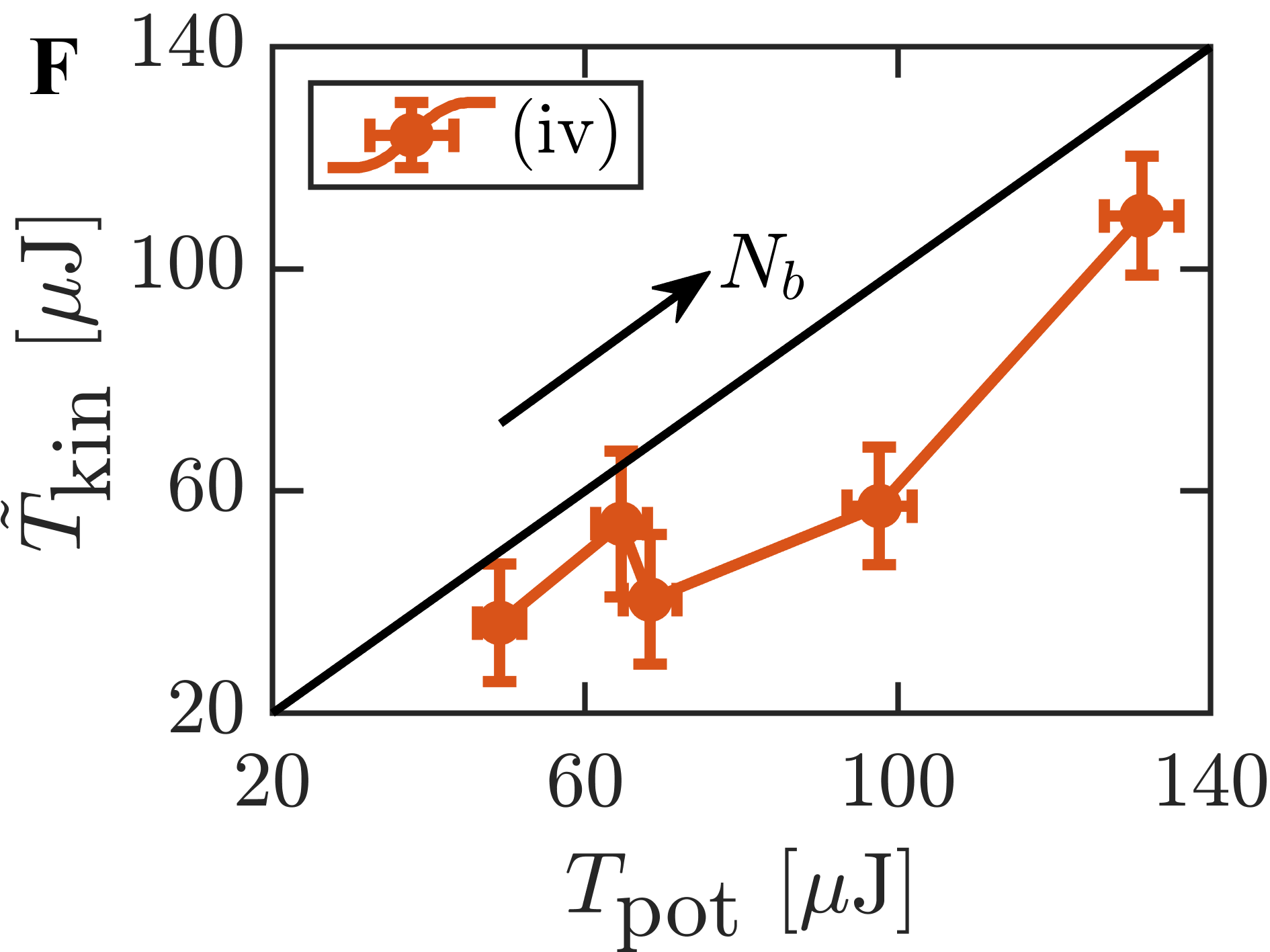}
    \caption{
    \textbf{Experimental applicability of effective temperature consistency.} \
    \textbf{A.} Standard experimental setup as presented in Fig. \ref{fig:1} and detailed in Sections 2 and 3. \
    \textbf{B.} Configuration using a smaller Styrofoam tracer particle ($m = 0.8$~g, diameter 3.5~cm). \
    \textbf{C.} Configuration with faster self-propelled bbots (see SI$^{\dag}$ for details). \
    \textbf{D.} Configuration employing a heavy tracer particle ($m = 8.2$~g, diameter 3.8~cm). \
    \textbf{E.} Comparison of energy partitioning between the modified kinetic temperature $\tilde{T}_{\text{kin}}$ (Eq. \ref{eq:kinetic}) and potential temperature $T_{\text{pot}}$ (Eq. \ref{eq:temp}) as functions of $N_b$ for configurations (i), (ii), and (iii). \
    \textbf{F.} Systematic violations of effective equipartition observed for configuration (D) with the heavy tracer, indicating the breakdown of thermodynamic-like consistency in this regime.
    }
    \label{fig:7}
\end{figure}

\section{Limitations of a consistent effective temperature} 

To further assess the range of consistency of effective temperature measures, we repeated the experiments from Section 3 for other system configurations.
The \textit{standard} experimental setup (detailed in Fig. \ref{fig:1} and shown in Fig. \ref{fig:7}\textbf{A}) was modified by varying both the passive tracer and the surrounding active particles. 
Specifically, these modifications included: using a smaller styrofoam tracer particle (Fig. \ref{fig:7}\textbf{B}); employing faster active bbots (Fig. \ref{fig:7}\textbf{C}); and introducing a heavy tracer particle (Fig. \ref{fig:7}\textbf{D}), i.e., heavier than the bbots ($m_{\text{bot}}=7.1$~g). 
Further system details are provided in the SI$^{\dag}$ (see figures 6-8 therein).

Figs. \ref{fig:7}\textbf{E,F} compare the tracer's modified kinetic temperature ($\tilde{T}_{\text{kin}}$) and potential temperature ($T_{\text{pot}}$) as a function of the number of bbots ($N_b$) for these configurations. The solid linear line represents the condition: $\tilde{T}_{\text{kin}}=T_{\text{pot}}$. 
In the alternative experimental setups that include a light tracer (Figs. \ref{fig:7}\textbf{B,C}), $\tilde{T}_{\text{kin}}$ and $T_{\text{pot}}$ coincide within measurement error and increase with $N_b$, consistent with our previous findings (Fig. \ref{fig:4}). 
These systems can also yield consistent $T_{\text{eff}}$ measurements whether via FDR (Eq. \ref{eq:FDR1}) or work FR (Eq. \ref{eq:FR}) tests (see details in SI$^{\dag}$).
However, when replacing the lightweight tracers with a heavy tracer, we observe differences between the measured $T_{\text{pot}}$ and $\tilde{T}_{\text{kin}}$ (see Fig.~\ref{fig:7}\textbf{F}).

Ultra-high-speed videos (SM movie 2) reveal the mechanical origin of this deviation. For lightweight tracers, interactions with the self-propelled bbots occur as discrete impacts, yielding a measurable ballistic regime between collisions. 
In contrast, heavy tracers can experience rapid recurrent collisions with the bbots, occurring on timescales shorter than the experimental sampling interval ($\Delta t$).
Upon strong collisions the tracer visibly changes the bbot propulsion dynamics. 
In addition, the heavy tracer remains embedded among the bbots, whereas a lighter tracer may be kicked upwards, escaping entrapment. 

These effects modify the coupling between tracer and environment and may thus lead to the observed discrepancy.
We note that the external air-driven perturbation is generally too weak to induce a significant perturbation to the heavy tracer.
Due to both effects, the FDR and FR tests fail for all considered $N_b$ systems.

\section{Summary and discussion} 

In this work, we have experimentally demonstrated that a passive granular tracer, confined in a harmonic potential and driven by an active fluctuating medium, can sustain a NESS that is well-characterized by a single effective temperature $T_{\text{eff}}$.
This effective temperature serves as a unifying parameter that encapsulates both the system's dynamic response properties and its stationary fluctuations (as detailed in Fig. \ref{fig:6}).
We find that introducing an effective mass is essential to properly define the tracer’s kinetic temperature, ensuring its consistency with the effective temperature of the NESS.

We identify three key features that appear essential to this equilibriumlike behavior of effective temperature consistency.
First, both the spontaneous fluctuations of the tracer and its response to weak external perturbations are governed by the same physical mechanism, i.e. random collisions with the active medium \cite{boriskovsky2024}. 
This is manifested in the validity of a linear FDR \cite{kubo1986,baldovin2022review}. \ 
Second, the light tracer is only weakly coupled to the environment, which minimizes the chance of physical entrapment or caging by bath particles, an effect known to induce long-term memory in the tracer's dynamics \cite{kurchan2013,caprini2023}. \ 
Finally, the bath dynamics (bbot motion) and the fluctuations they induce are largely unaffected by the external driving (airflow) that acts mainly on the tracer particle \cite{gnoli2014,puglisi2017temperature,hecht2024}. 

The agreement between different definitions of effective temperature observed here is unexpected for two reasons.
First, the noise is a result of non-exponential collision times with the bath particles that show temporal correlation due to active self-propulsion. 
Under these conditions, the system exhibits non-trivial steady state distributions, in variance with the Poisson shot noise conditions leading to Boltzmann-like steady states under which such agreement has been predicted \cite{dybiec2017,di2024,kanazawa2015minimal}.
Second, the use of an inertial tracer introduces memory effects in active matter, often breaking Markovian dynamics and precluding a single consistent effective temperature \cite{caprini2022, caprini2023, sprenger2023, hecht2024}. That this agreement persists despite these deviations from the idealized theoretical framework suggests a broader robustness of effective temperature concepts than previously recognized.


\section*{Conflicts of interest}
There are no conflicts to declare.

\section*{Acknowledgements}
DB, RG and YR acknowledge support from the European Research Council (ERC) under the European Union’s Horizon 2020 research and innovation program (Grant agreement No. 101002392). 
BL would like to thank Deutsche Forschungsgemeinschaft for their support (DFG grant LI 1046/10-1).

\bibliography{REFS} 

\clearpage
\newpage

\appendix 
\setcounter{figure}{0}

\section*{Supplementary Information}

\subsection*{Further results and measurements}
\begin{figure}[t] 
    \centering
    \includegraphics[width=0.19\textwidth]{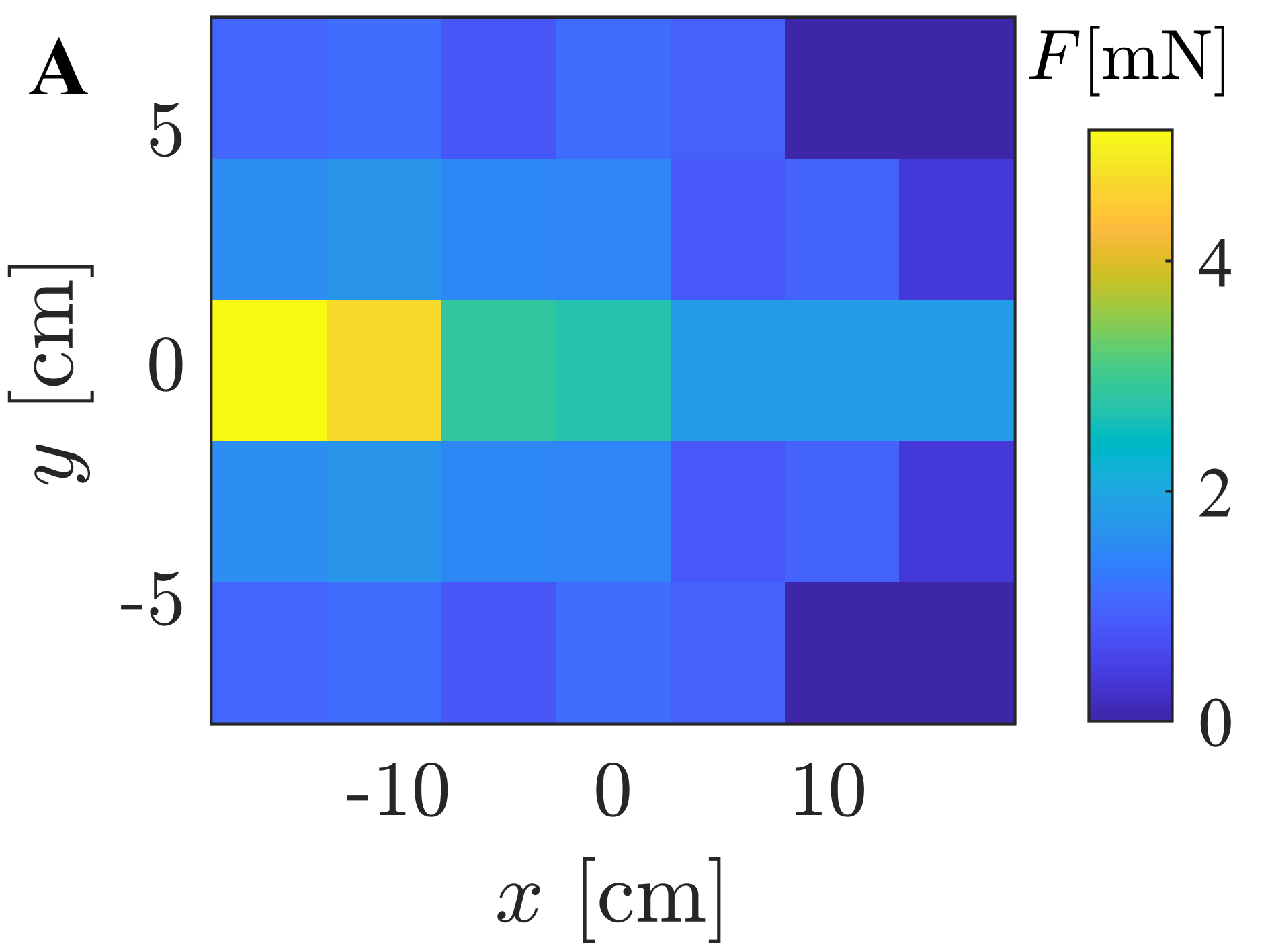}
    \includegraphics[width=0.19\textwidth]{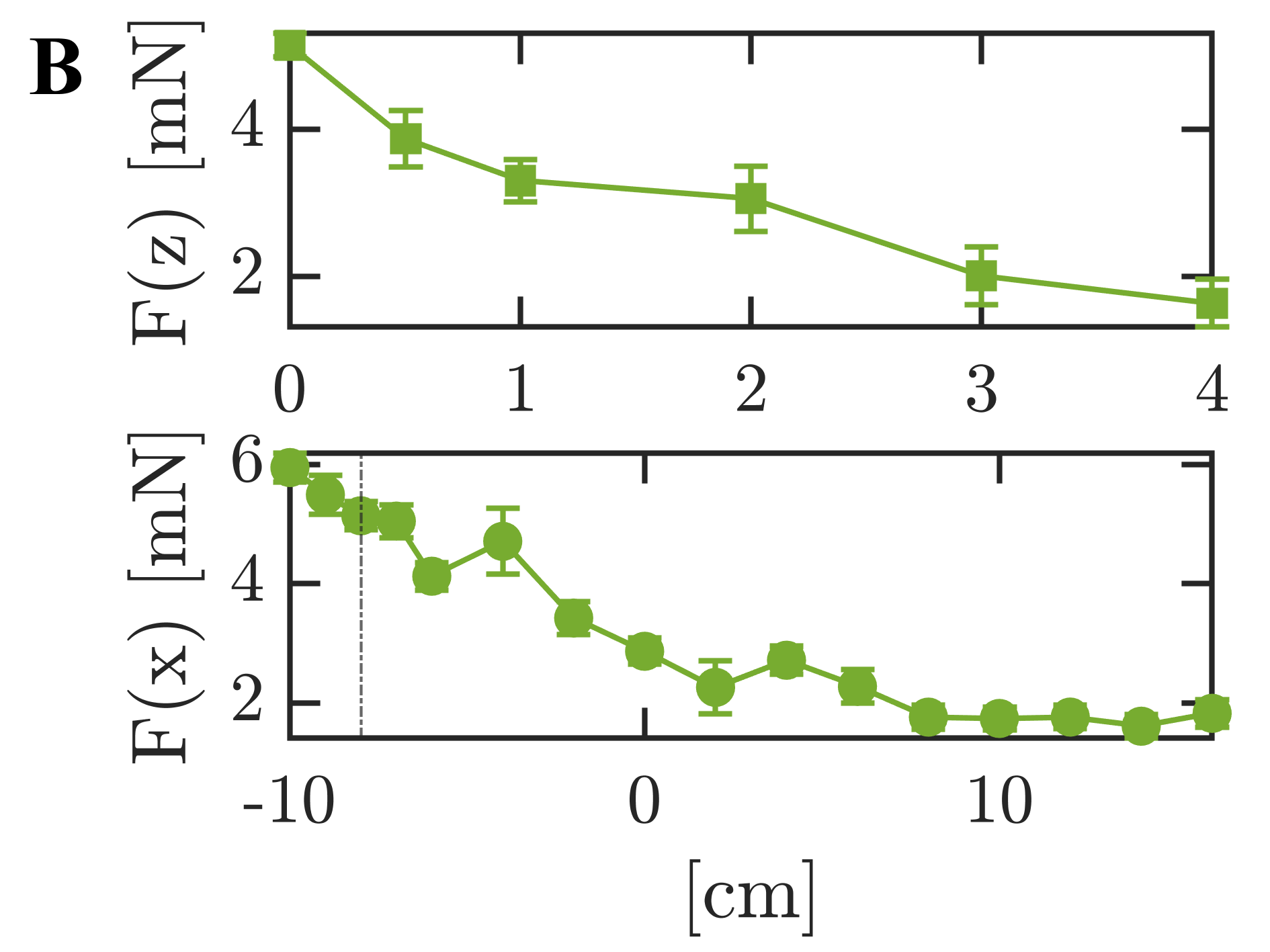}
    \includegraphics[width=0.19\textwidth]{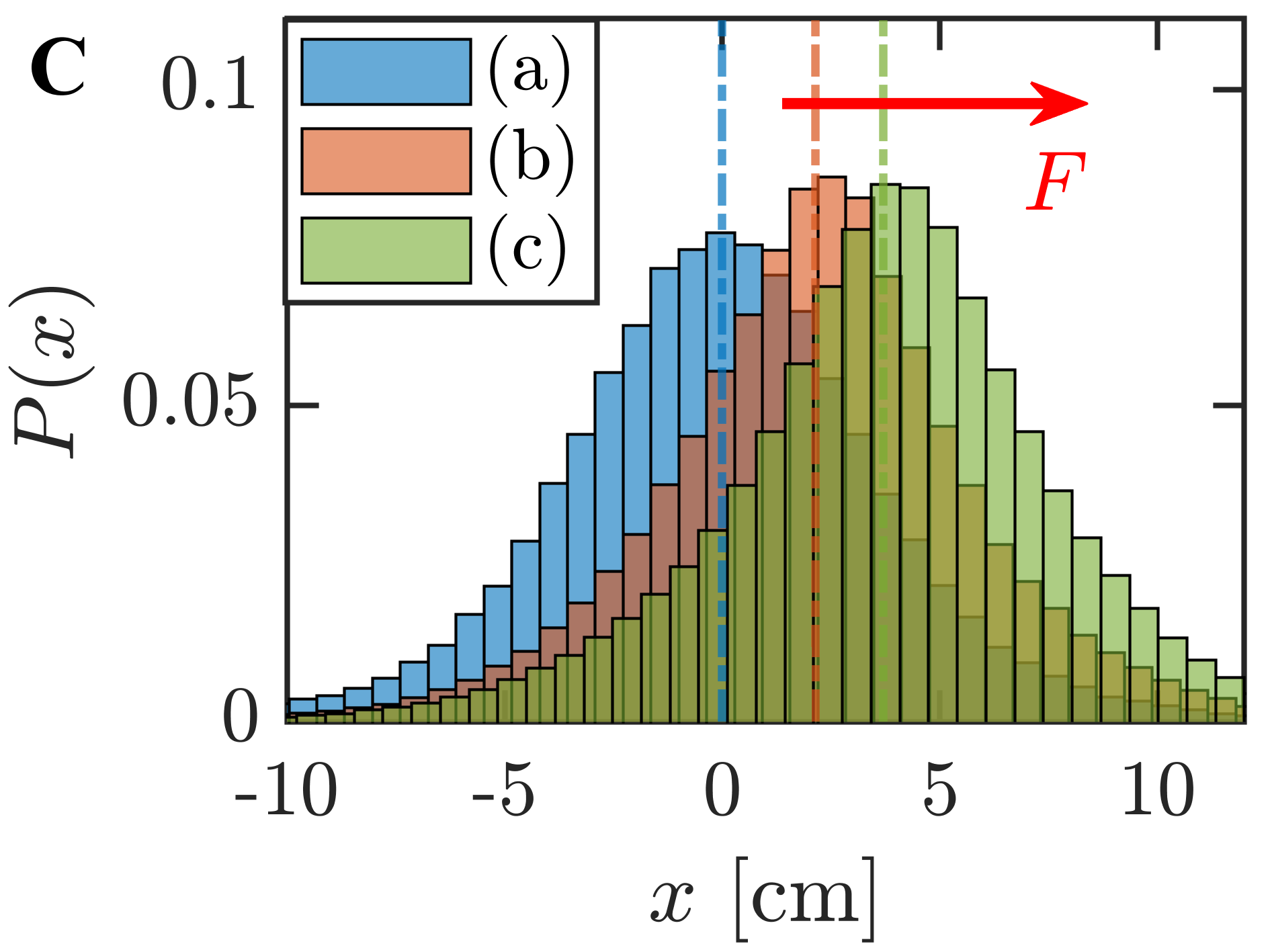}    
    \includegraphics[width=0.19\textwidth]{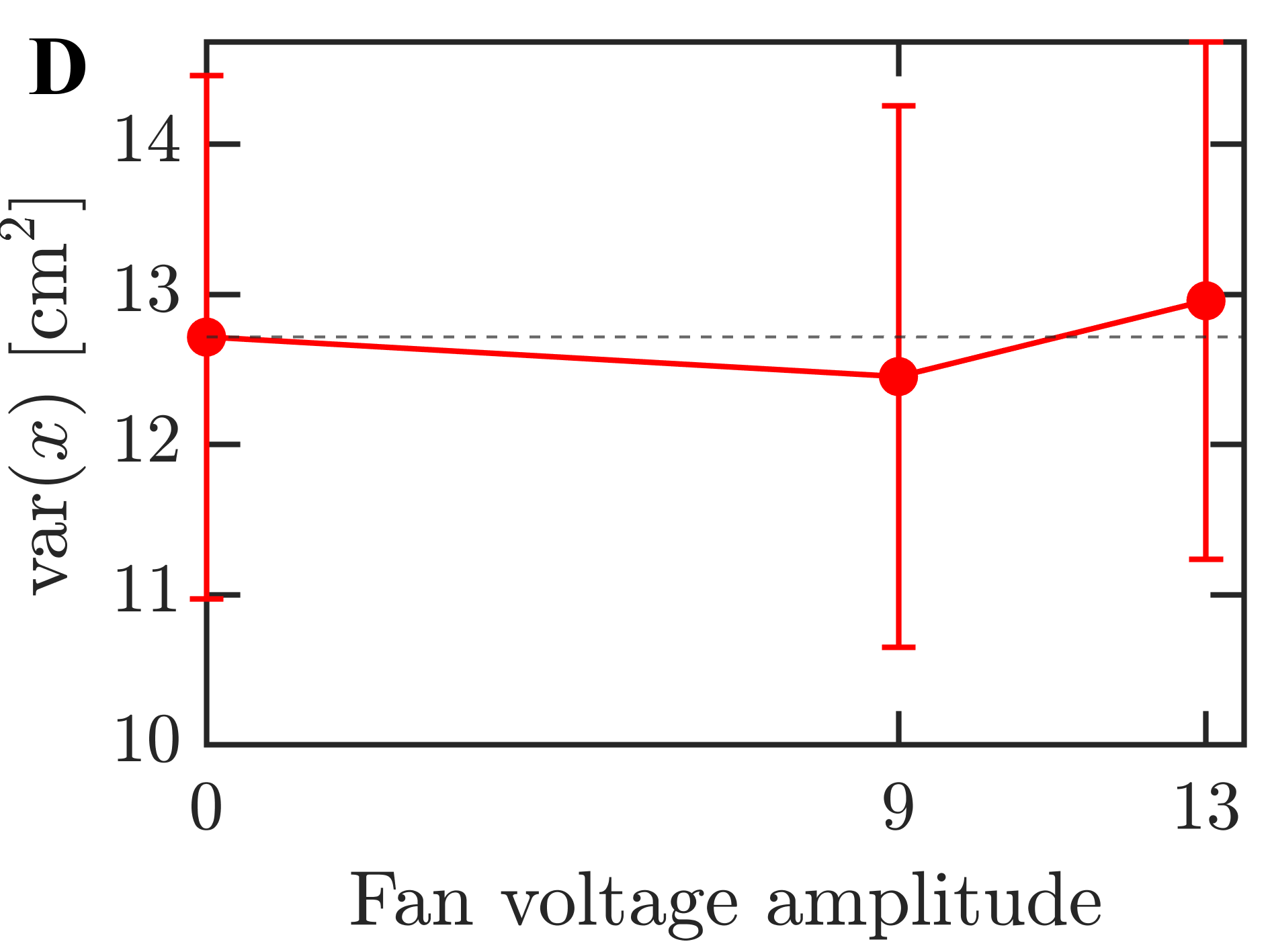}    
    \caption{ 
    \textbf{Forces and stationary statistics under external air stream.} \ 
    \textbf{A, B.} Average forces acting on the tracer in the absence of bbots, on a flat surface.
    Measurements were obtained using a force sensor (Mark-10 Force Gauge) with $10$V operating fan voltage:    
    (A) as a function of $x-y$ position; (B) along the main air stream axis $x$ (lower panel) - and as a function of height $z$ (upper panel), at a distance $x=8$cm marked by the vertical dashed line.   
    Error bars are standard deviations of $50$ measurements. \ 
    \textbf{C.} Stationary probability distributions of the tracer’s position under different air stream intensities in the harmonic trap (unperturbed, $10$V, and $13.5$V fan operating voltage), with $N_b=10$. \ 
    \textbf{D.} Corresponding position variances. 
    Results were obtained from time and ensemble averages over $M = 375$ realizations of one-minute tracer trajectories. Error bars indicate standard deviations.
    }
    \label{fig:SIfig11}
\end{figure}

\begin{figure}[t] 
    \centering
    \includegraphics[width=0.19\textwidth]{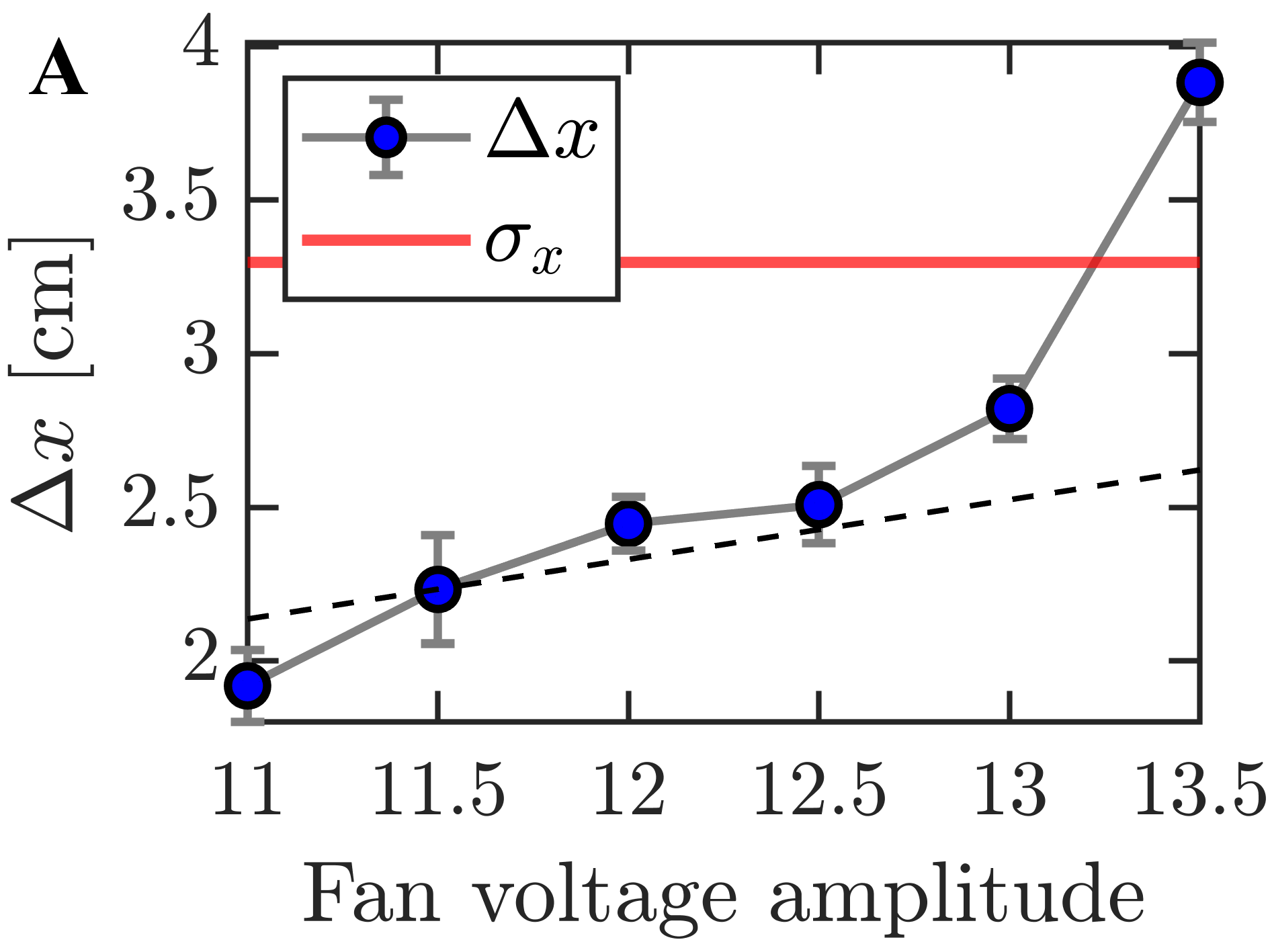}    
    \includegraphics[width=0.19\textwidth]{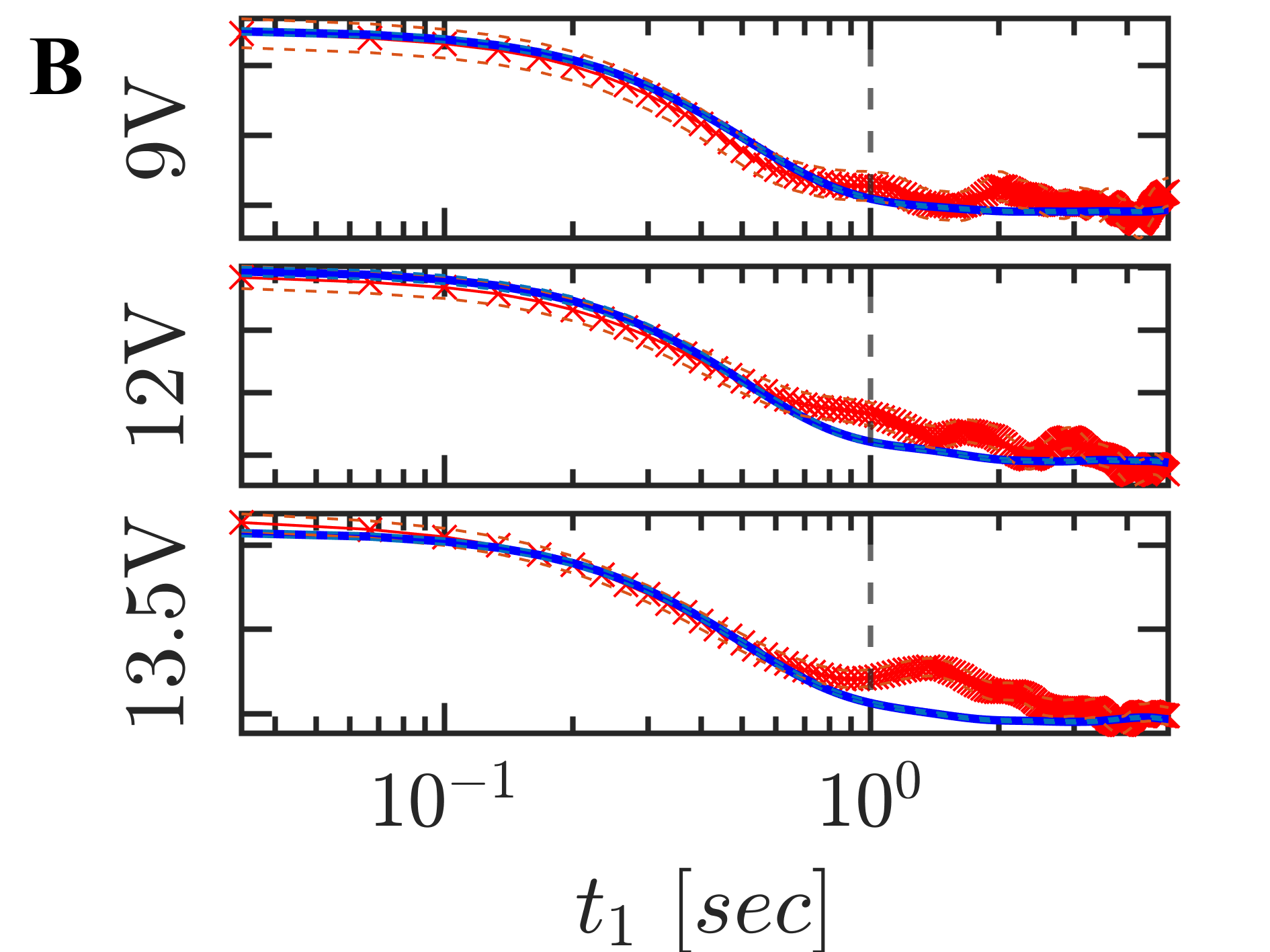}    
    \caption{ 
    \textbf{The FDR test under different perturbation amplitudes.} \ 
    \textbf{A.} The mean displacement $\Delta x$ under a perturbation of a tracer in a $N_b=10$ bbot bath, as a function of fan operating voltage $V$ (error bars are standard errors). 
    The applied force $F_0=k\Delta x$ increases above $V=11$V. The tracer's position standard deviation, $\sigma_x$, is used to define the range of small perturbations (solid line).
    The dashed line is a linear fit of the first 4 data points, with $\Delta x = 0.19\cdot V$, whereas $V=13$V and $13.5$V are out of the linear regime. \ 
    \textbf{B.} \ The generalized FDRs with $T_{\text{eff}}\sim\langle \Delta x^2\rangle_0$ are plotted for $V=9, \ 12, \ 13.5$~V, with $N_b=10$. The results present an average over $M=375$ perturbation sequences. 
    }
    \label{fig:SIfig3}
\end{figure}

\begin{figure}[t] 
    \centering
    \includegraphics[width=0.23\textwidth]{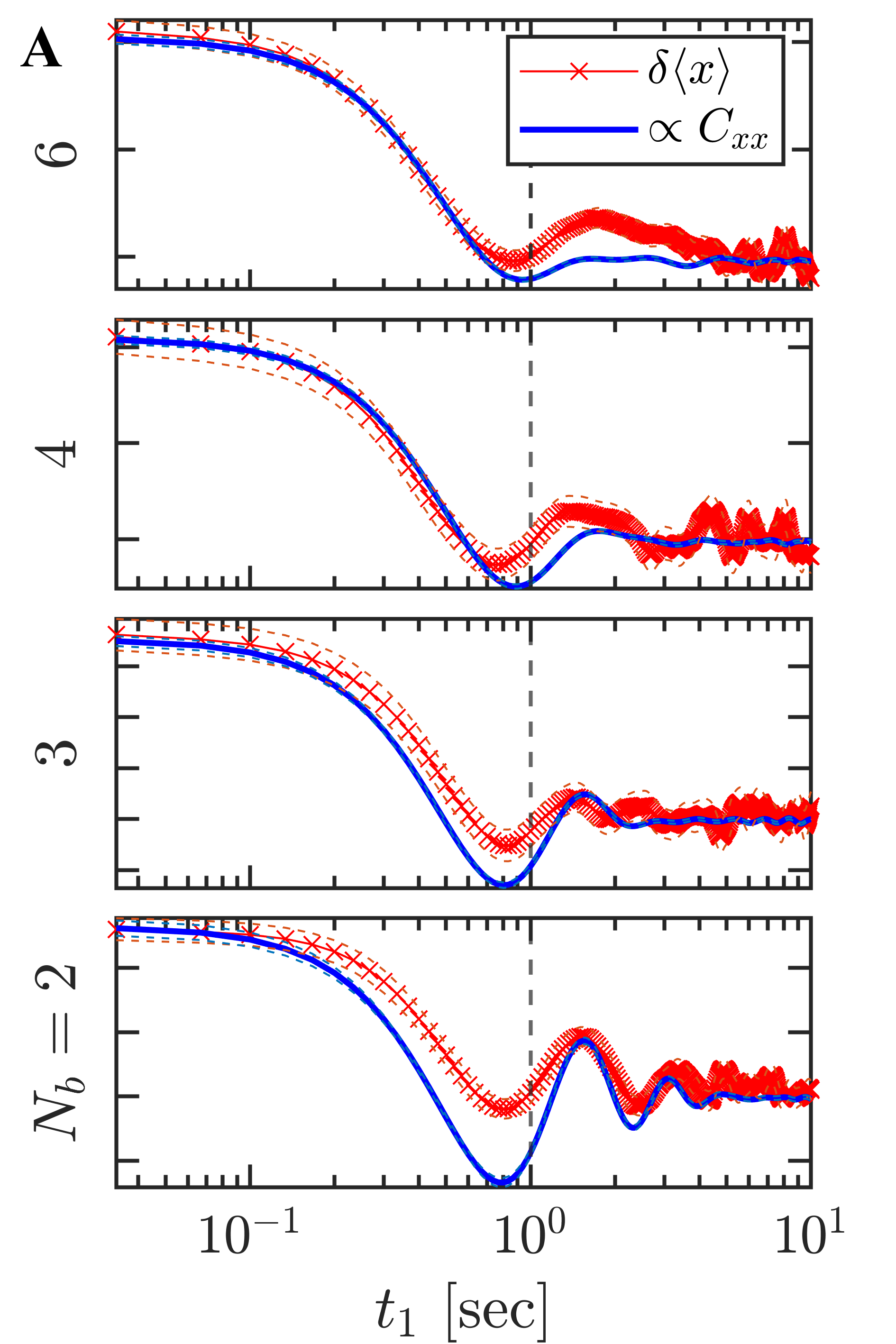}
    \includegraphics[width=0.23\textwidth]{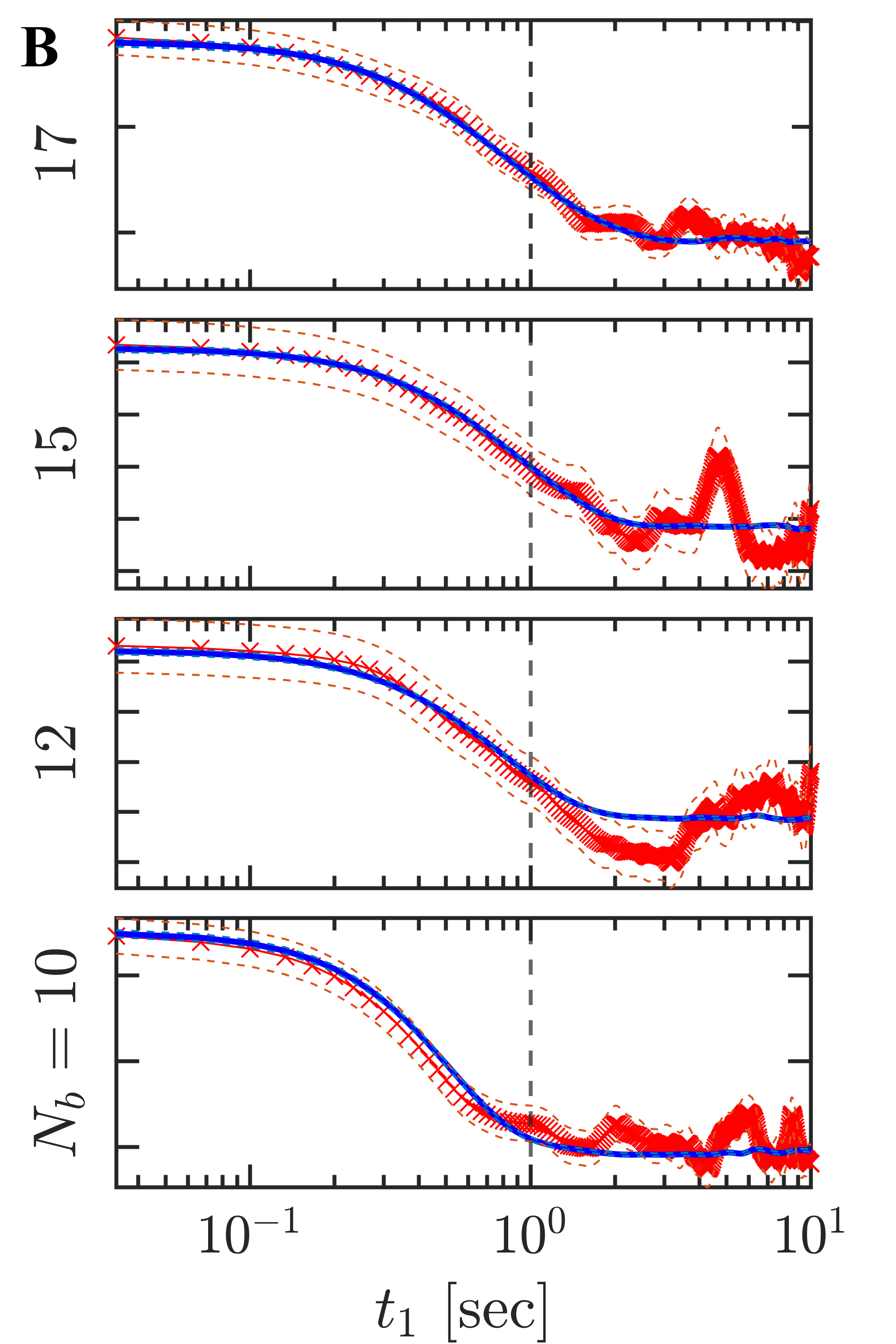}
    \caption{
    \textbf{Position ($x$) FDR results with different numbers of bbots.} 
    \ \textbf{A,B.} FDRs with $T_{\text{eff}}\sim\langle \Delta x^2\rangle_0$ for $N_b = \{2,3,4,6,10,12,15,17\}$ bbot baths.
    The results were obtained for an ensemble average of $M=375$ perturbation sequences, with the same tracer $m\approx 1$~g, gravitational stiffness $k\approx 28.2~\text{g/s}^{-2}$, and fan operating voltage ($9$~V) with $\Delta x_{\epsilon}<\sigma_x$.
    The vertical dashed lines are $T_c=1$~s.
    }
    \label{fig:SIfig4}
\end{figure}

\begin{figure}[t] 
    \centering
    \includegraphics[width=0.23\textwidth]{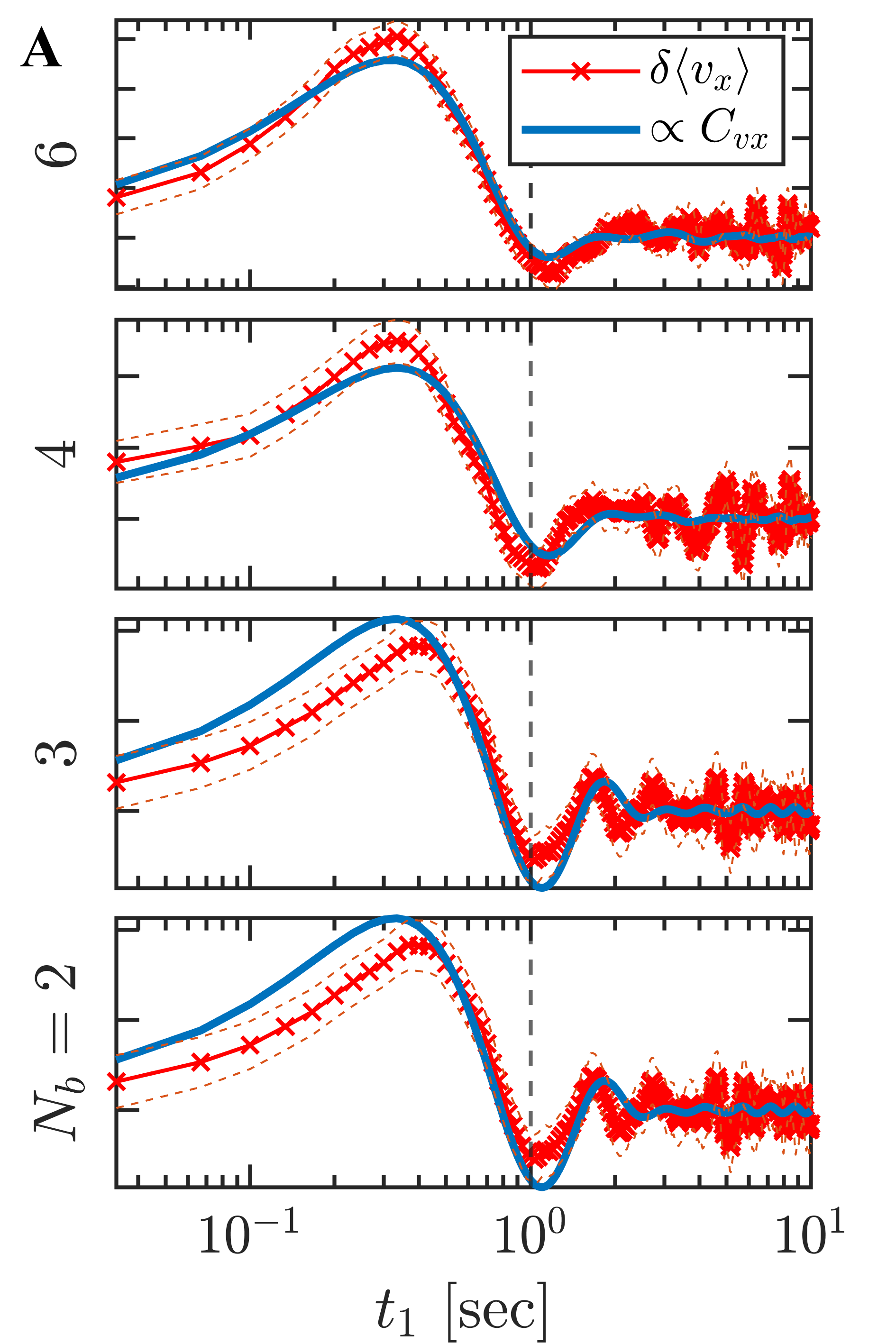}
    \includegraphics[width=0.23\textwidth]{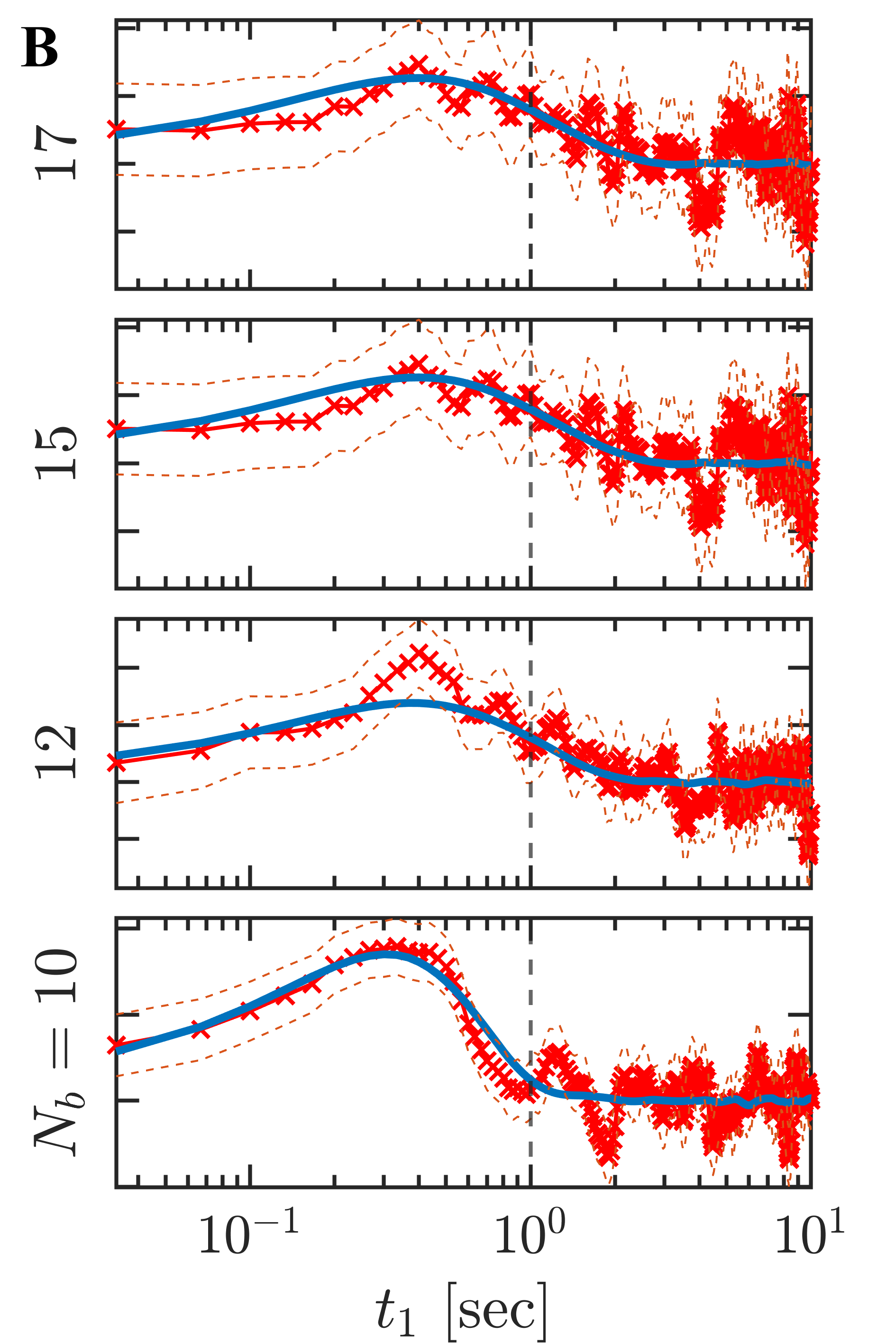}
    \caption{
    \textbf{Velocity ($v_x$) FDR results with different numbers of bbots.}
    \ \textbf{A,B.} FDRs with $T_{\text{eff}}\sim\langle \Delta x^2\rangle_0$ for $N_b = \{2,3,4,6,10,12,15,17\}$ bbot baths.
    Same details as in Fig.~\ref{fig:SIfig4}.
    }
    \label{fig:SIfig5}
\end{figure}

\begin{figure}[H] 
    \centering
    \includegraphics[width=0.23\textwidth]{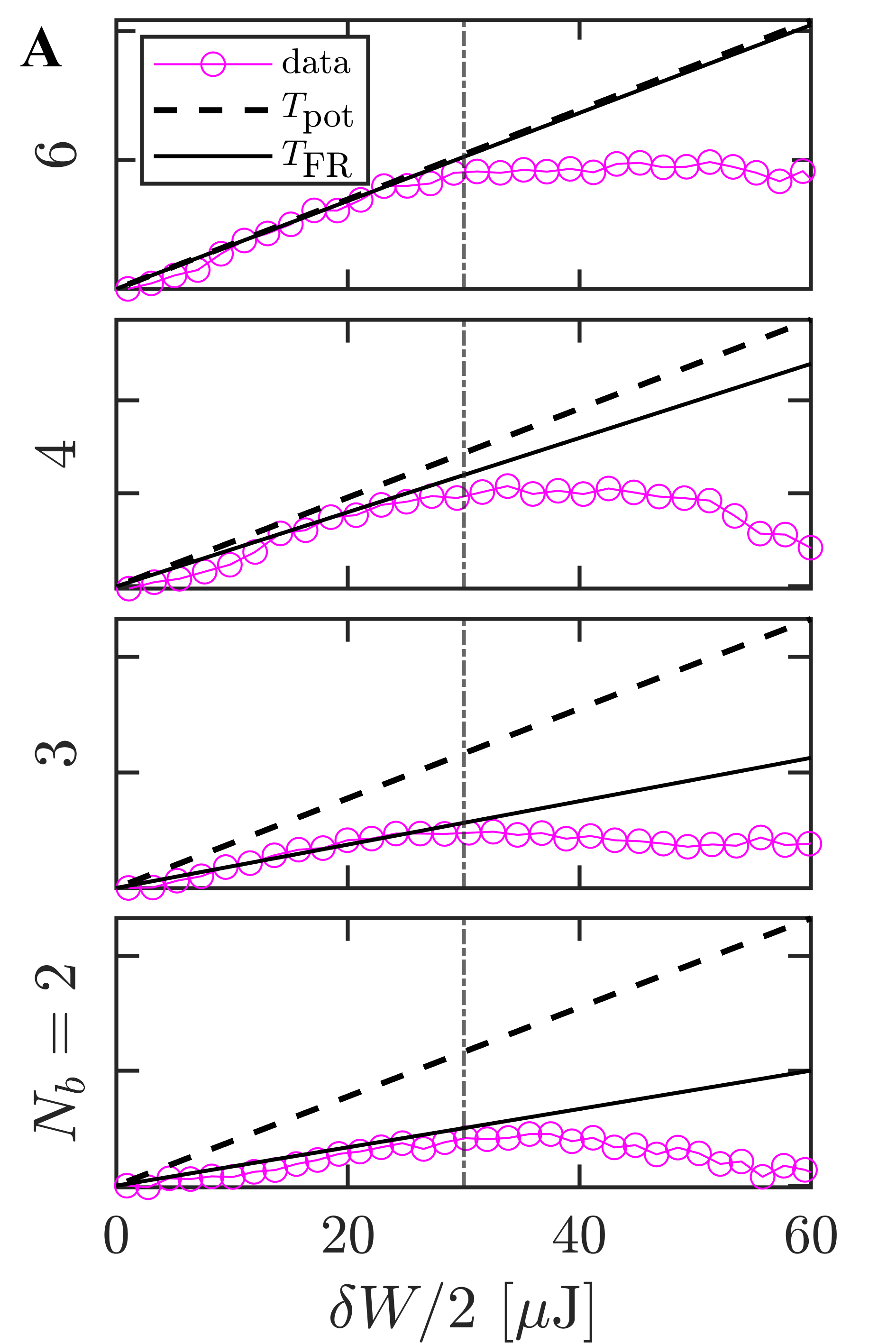}
    \includegraphics[width=0.23\textwidth]{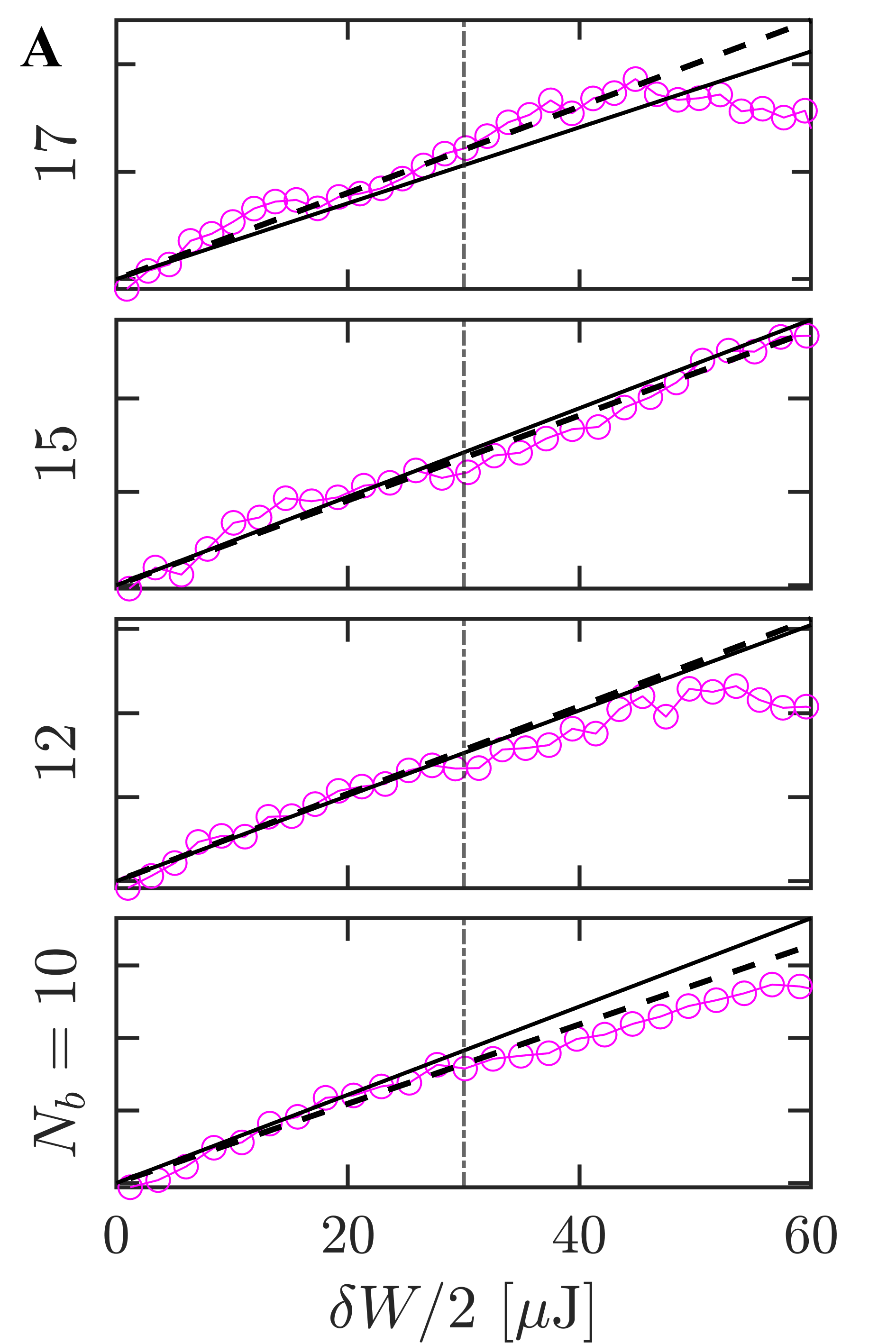}
    \caption{
    \textbf{Work FR results in (strongly) perturbed steady-states with different numbers of bbots.} 
    \ \textbf{A,B.} Work FRs with $N_b = \{2,3,4,6,10,12,15,17\}$. The solid line represent a linear fit with a slope of $\sim 1/T_{\text{FR}}$. The dashed line has a slope of $\sim 1/T_{\text{pot}}$.
    The results were obtained for an ensemble average of $M=375$ perturbation sequences, with the same tracer $m\approx 1$~g, gravitational stiffness $k\approx 28.2~\text{g/s}^{-2}$, and fan operating voltage of $13.5$~V.
    }
    \label{fig:SIfig6}
\end{figure}

We provide the following supplemental material: \ 
The average force exerted by the fan on the tracer was measured at different distances under a constant air stream (without bbots). 
The variance and skewness of the tracer’s stationary distributions (with $N_b=10$) are presented for different air stream amplitudes, corresponding to varying fan voltages (Fig.~\ref{fig:SIfig11}); \ 
FDR tests under constant $N_b$ and increasing perturbation amplitudes (Fig.~\ref{fig:SIfig3}); \ 
Individual FDR results for position and velocity tracer observables (Figs.~\ref{fig:SIfig4},\ref{fig:SIfig5}) and work FR results (Fig.~\ref{fig:SIfig6}), with different $N_b$. 
Particularly, these results were obtained with the \textit{standard} experimental setup, as detailed in the main paper.  

\begin{figure}[t] 
    \centering
    \includegraphics[width=0.20\textwidth]{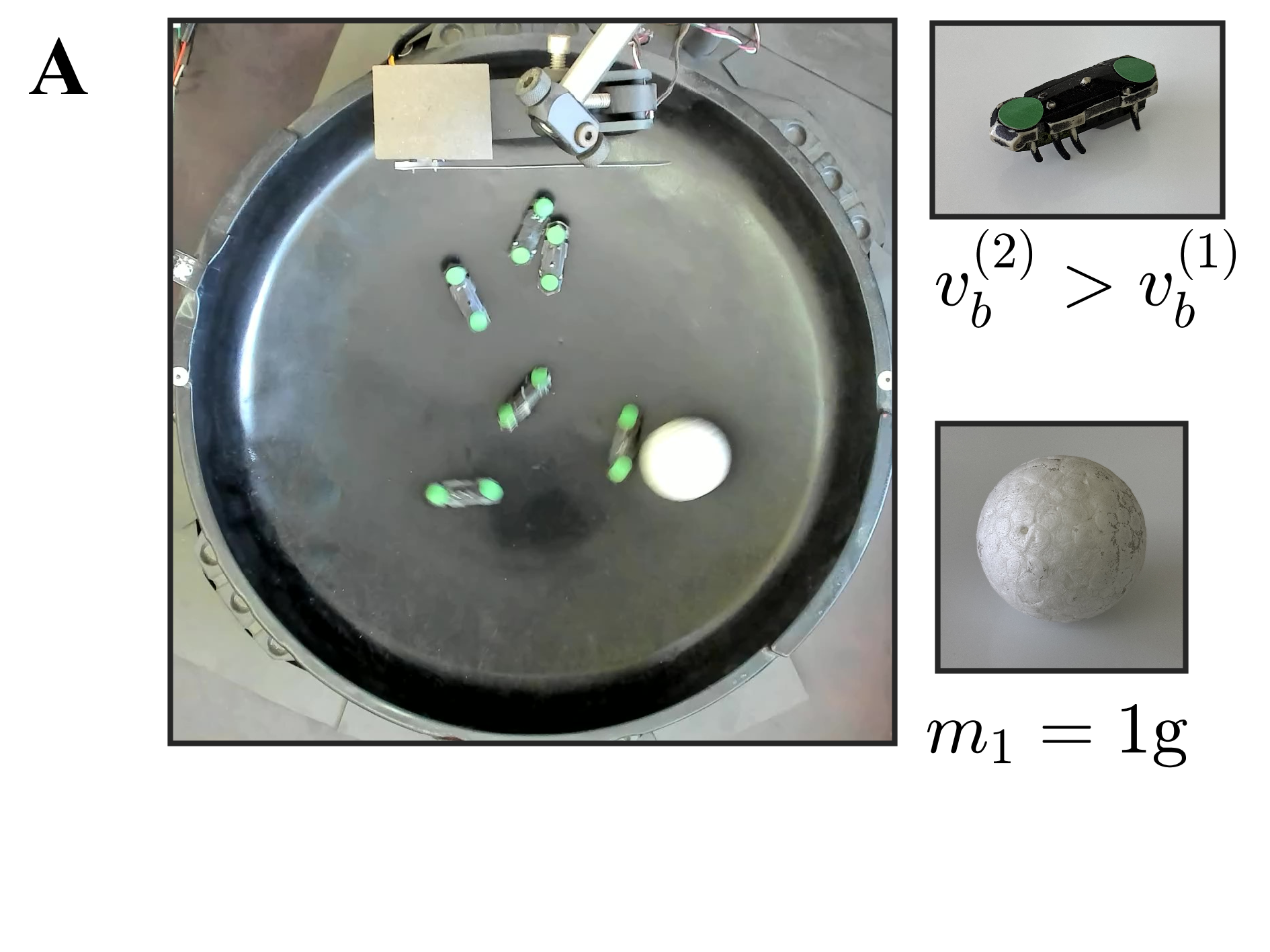}
    \includegraphics[width=0.20\textwidth]{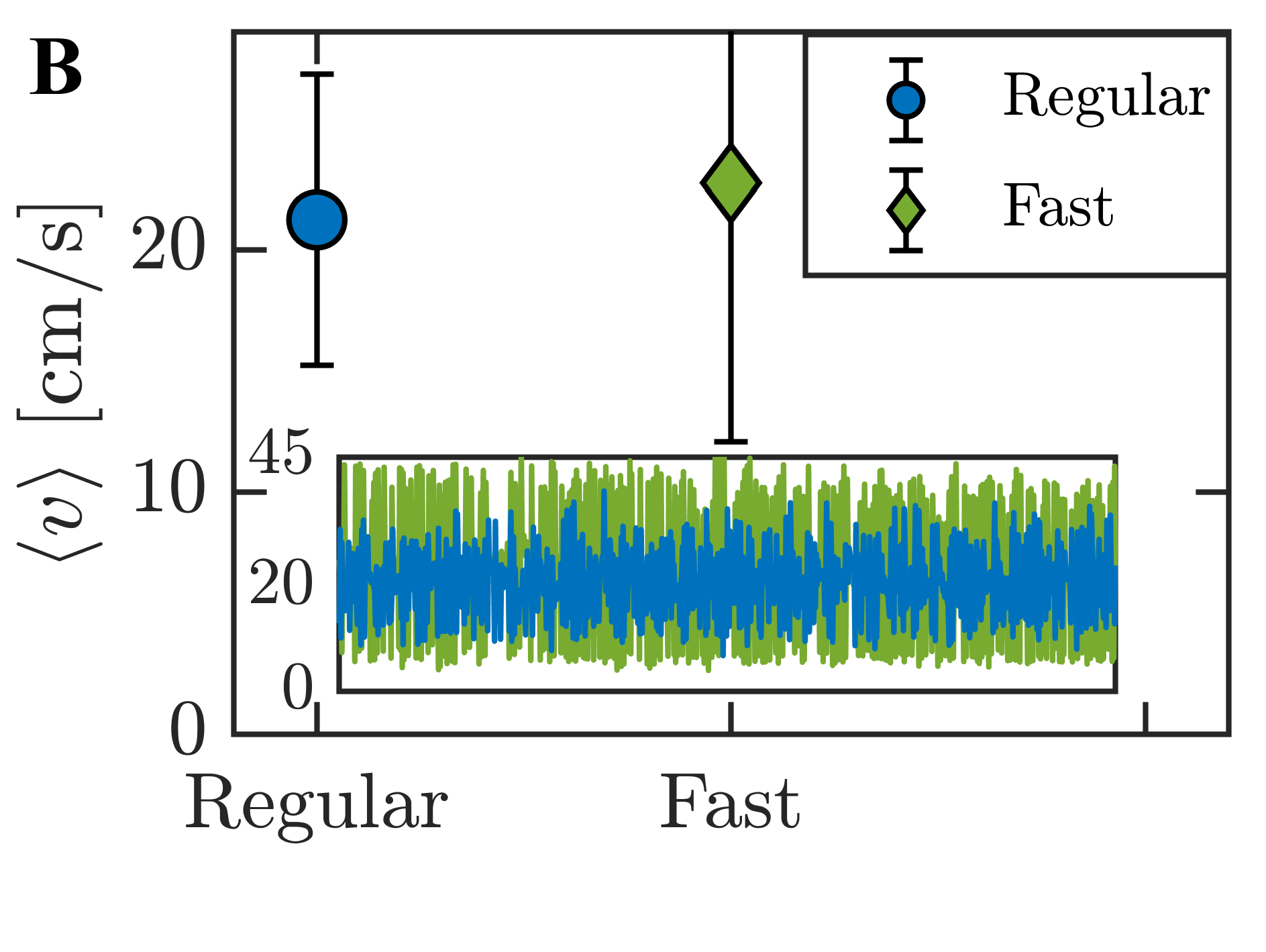}
    \includegraphics[width=0.20\textwidth]{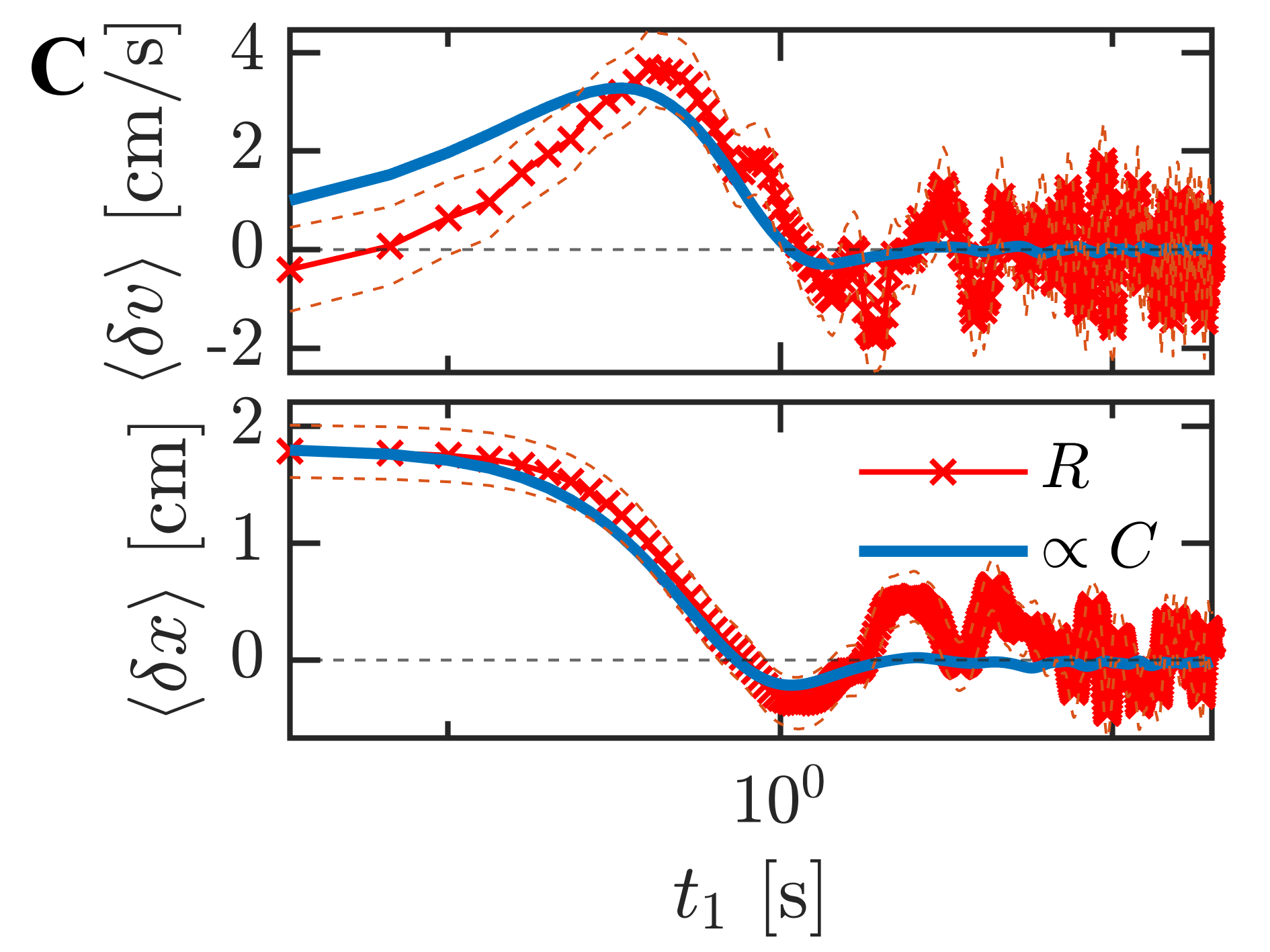}
    \includegraphics[width=0.20\textwidth]{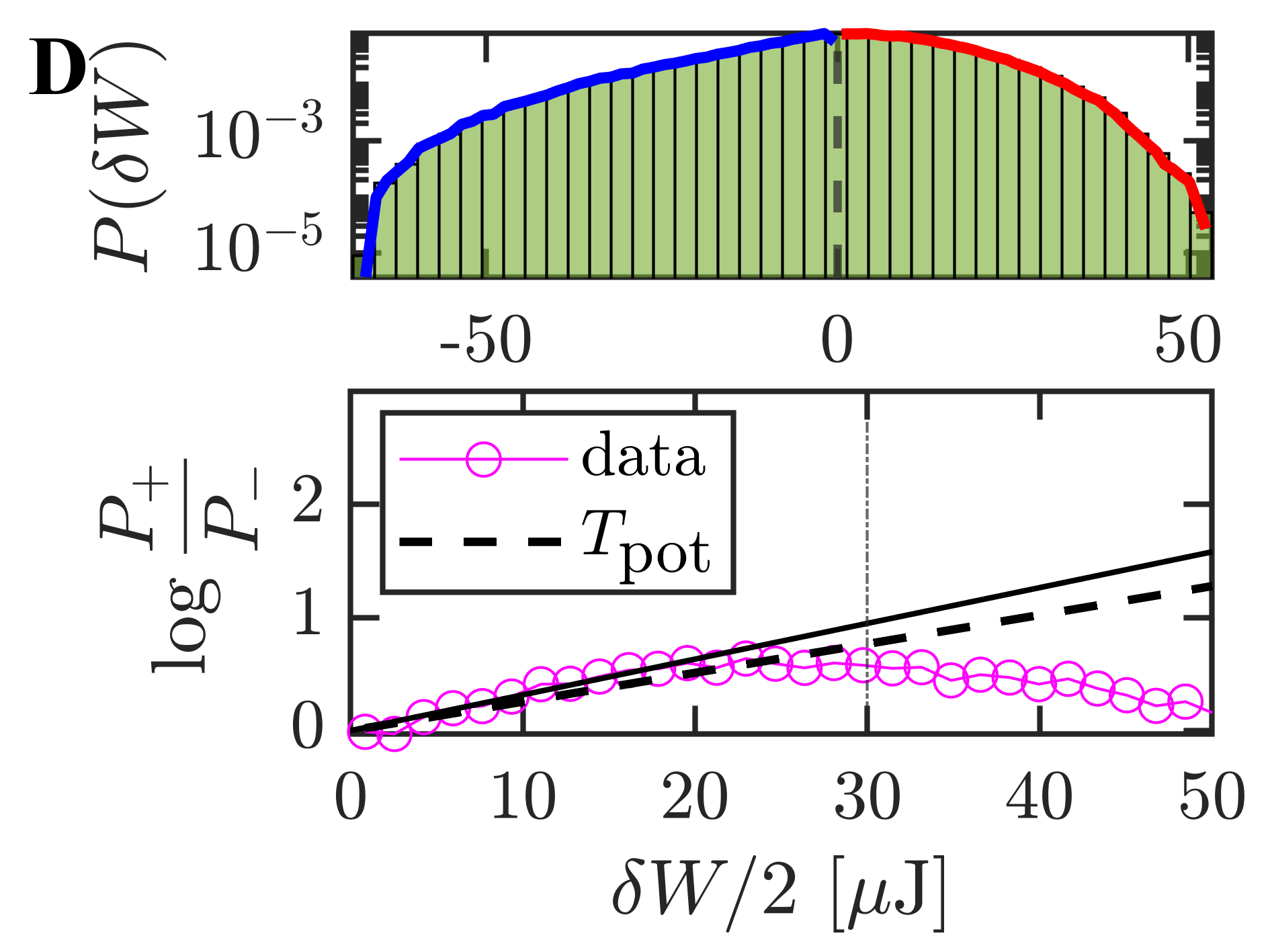}
    \caption{
    \textbf{Fast bbot setup.} \ 
    \textbf{A.} The active bath contains $N_b=6$ \textit{fast} bbots, the experimental setup is otherwise the same as in Figs.~\ref{fig:SIfig3}-\ref{fig:SIfig9}. \ 
    \textbf{B.} Average individual speed of a single bbot walking in a parabolic arena. Standard bbots (circle) and fast bbots (diamond), as referred to in the main paper. \ 
    \textbf{C.} FDR test for the tracer's position and velocity observables, under the same perturbation setup as detailed in Figs.~\ref{fig:SIfig4},\ref{fig:SIfig5}.
    The effective temperature is $T_{\text{eff}}=\kappa \langle x^2\rangle_0$. \ 
    \textbf{D.} Work FR test as detailed in Fig.~\ref{fig:SIfig6}. 
    }
    \label{fig:SIfig8}
\end{figure}

\begin{figure}[t] 
    \centering
    \includegraphics[width=0.20\textwidth]{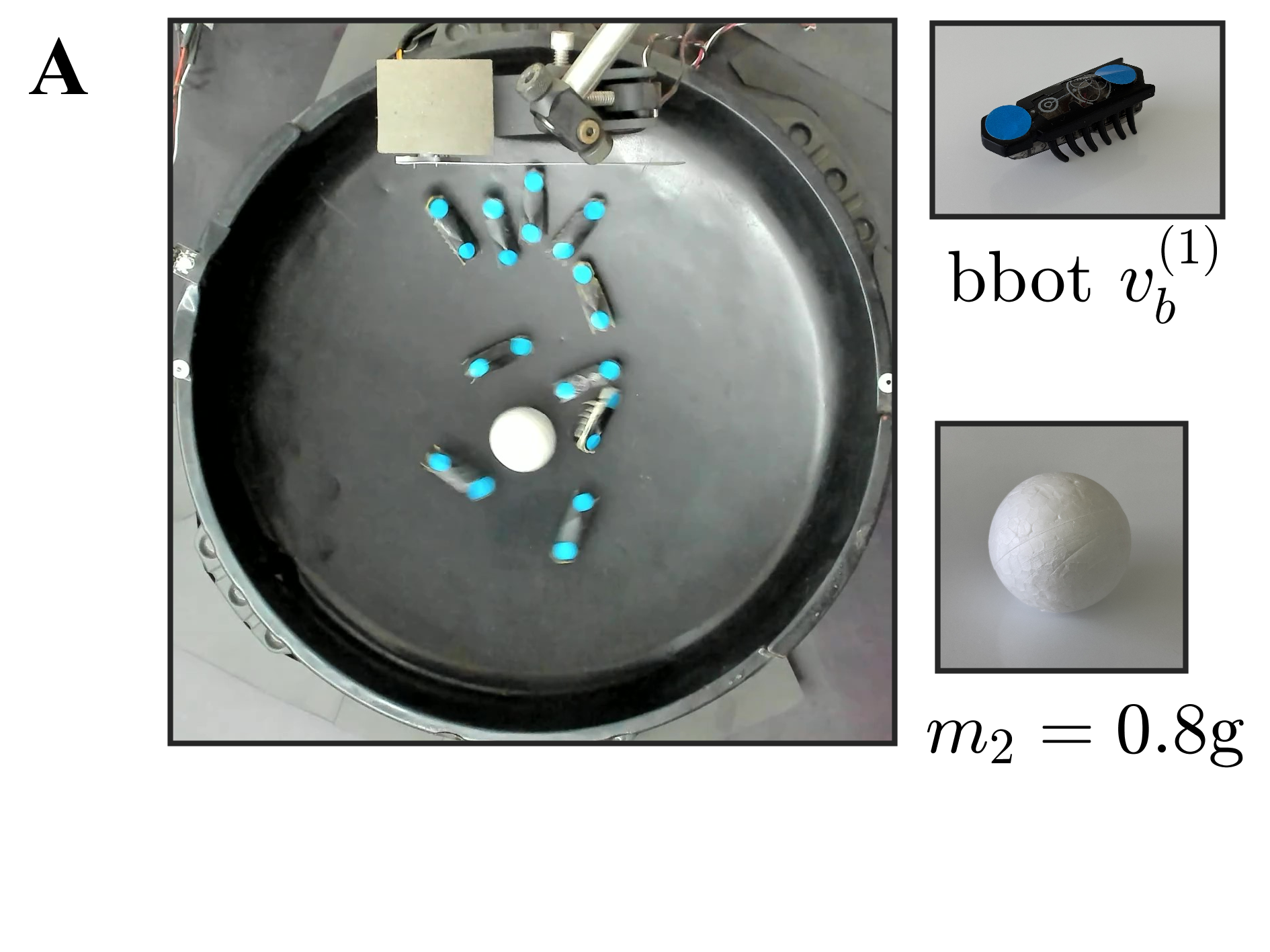}

    \includegraphics[width=0.20\textwidth]{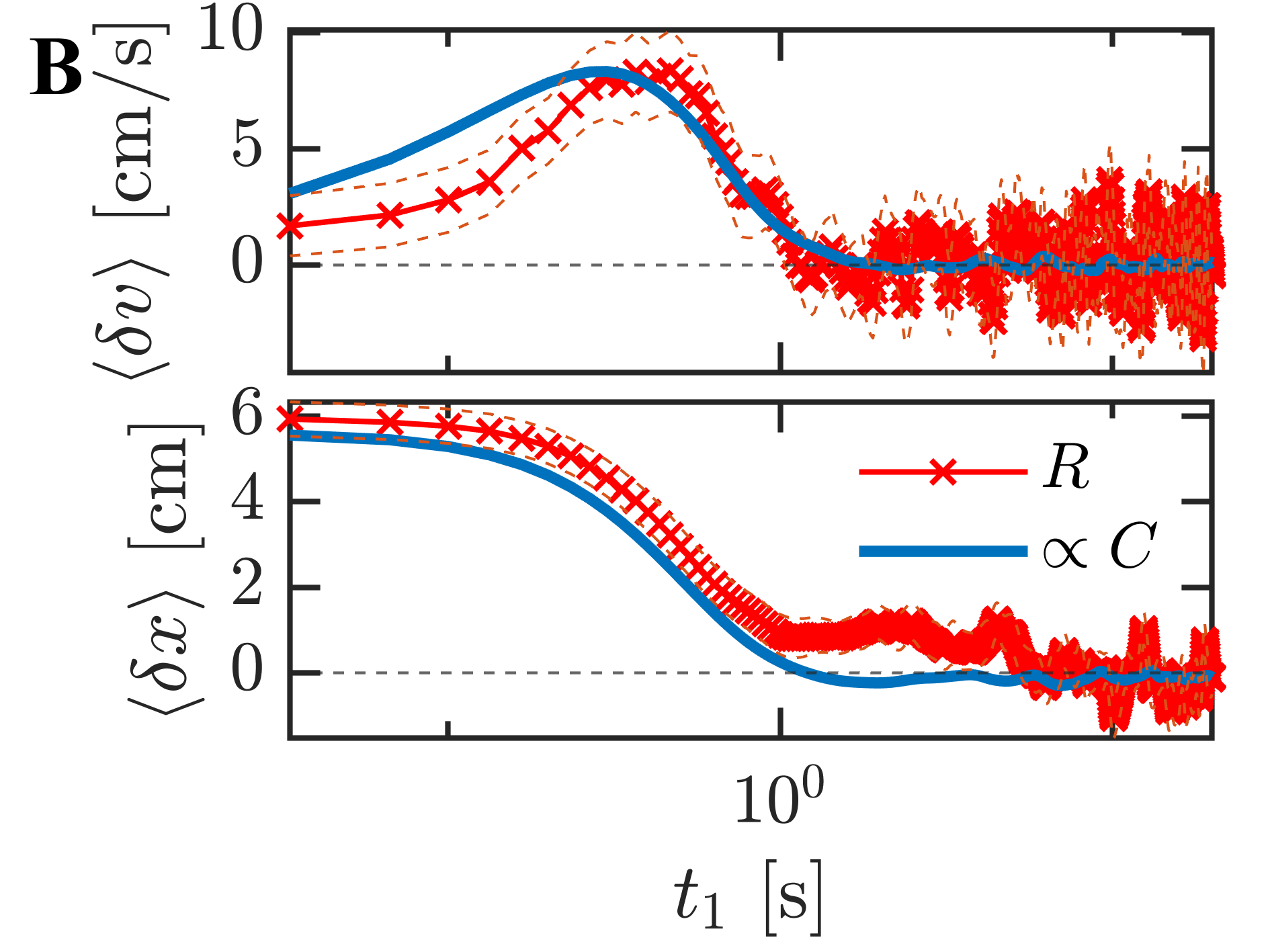}
    \includegraphics[width=0.20\textwidth]{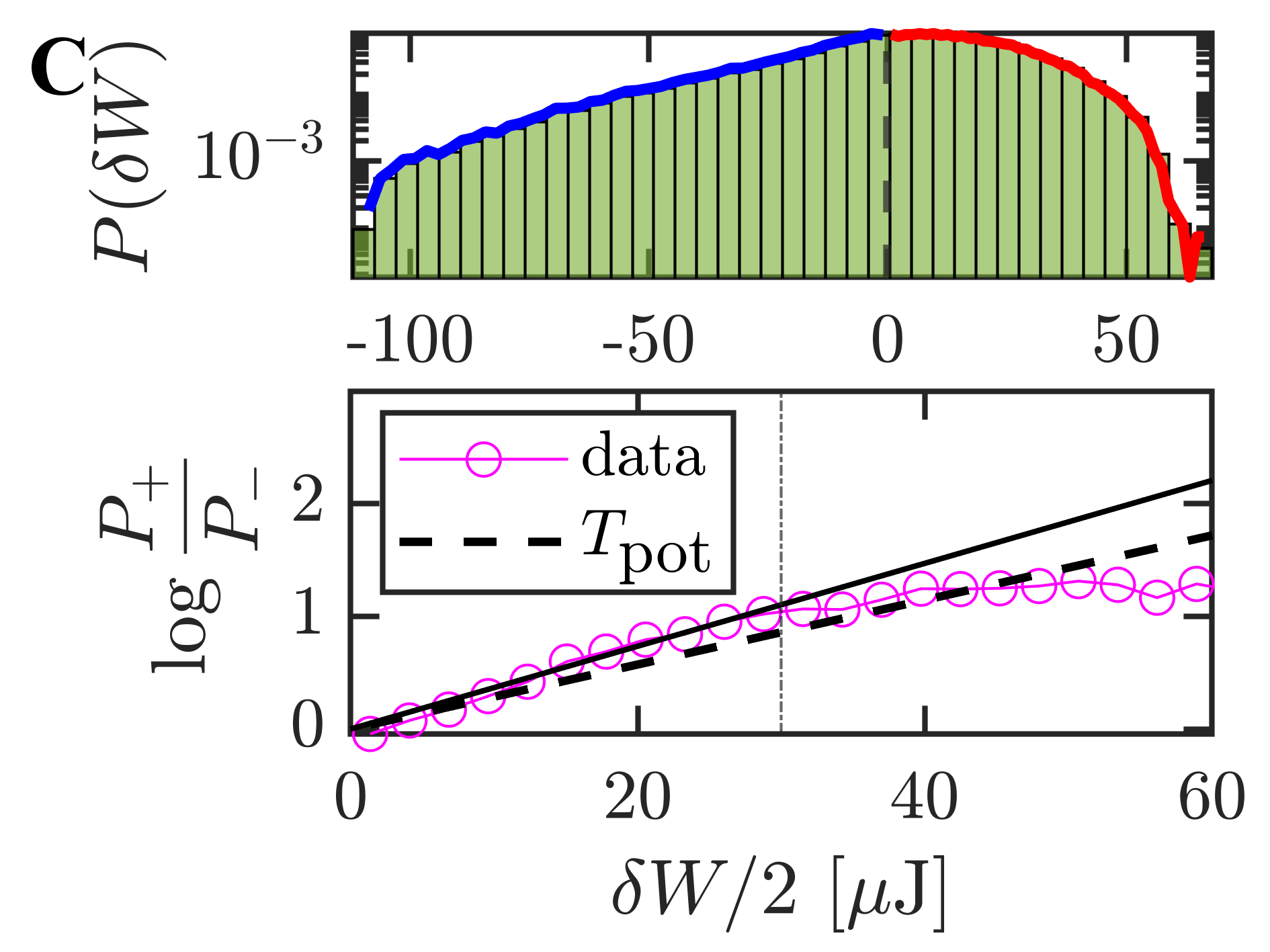}
    \caption{
    \textbf{Small tracer setup.} \ 
    \textbf{A.} A small styrofoam ball is used as a passive tracer ($3$~cm diameter, $0.8$~g). The active bath contains $N_b=10$ standard bbots. \ 
    \textbf{B.} FDR test for the tracer's position and velocity observables, under the same perturbation setup as detailed in Figs.~\ref{fig:SIfig4},\ref{fig:SIfig5}.
    The effective temperature is $T_{\text{eff}}=\kappa \langle x^2\rangle_0$. \ 
    \textbf{C.} Work FR test as detailed in Fig.~\ref{fig:SIfig6}. 
    }
    \label{fig:SIfig9}
\end{figure}

\begin{figure}[H] 
    \centering
    \includegraphics[width=0.20\textwidth]{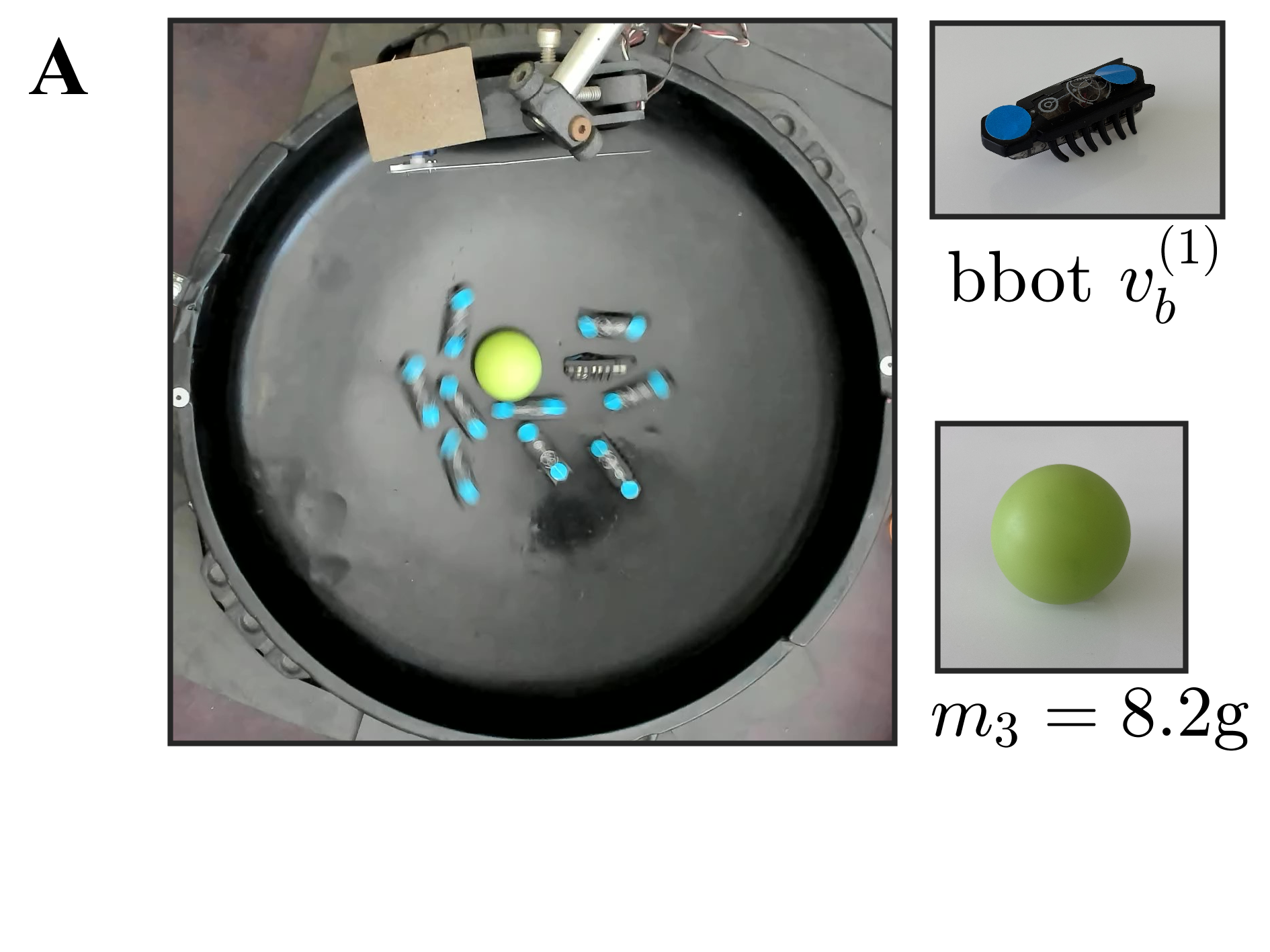}
    \includegraphics[width=0.20\textwidth]{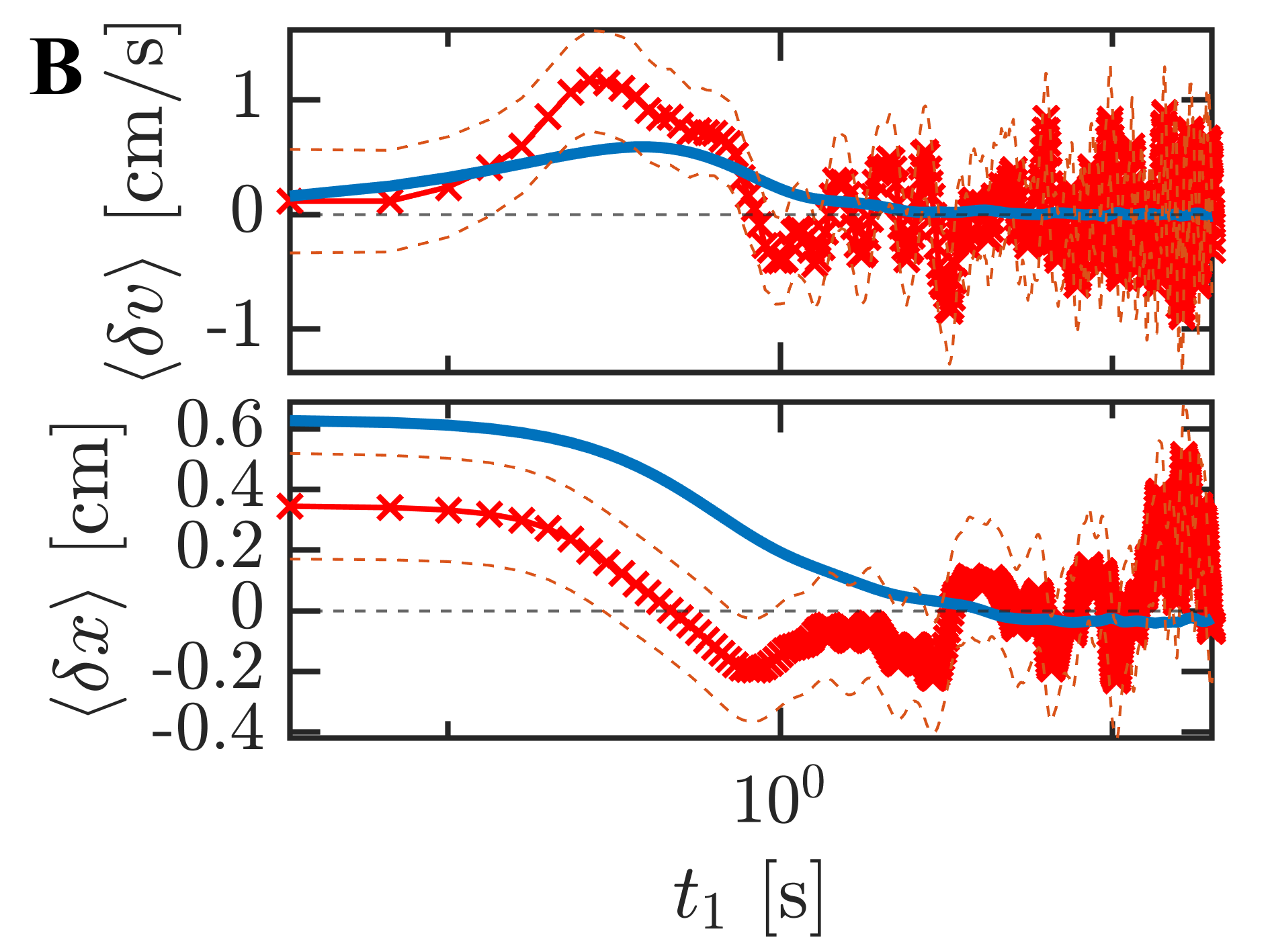}
    \includegraphics[width=0.20\textwidth]{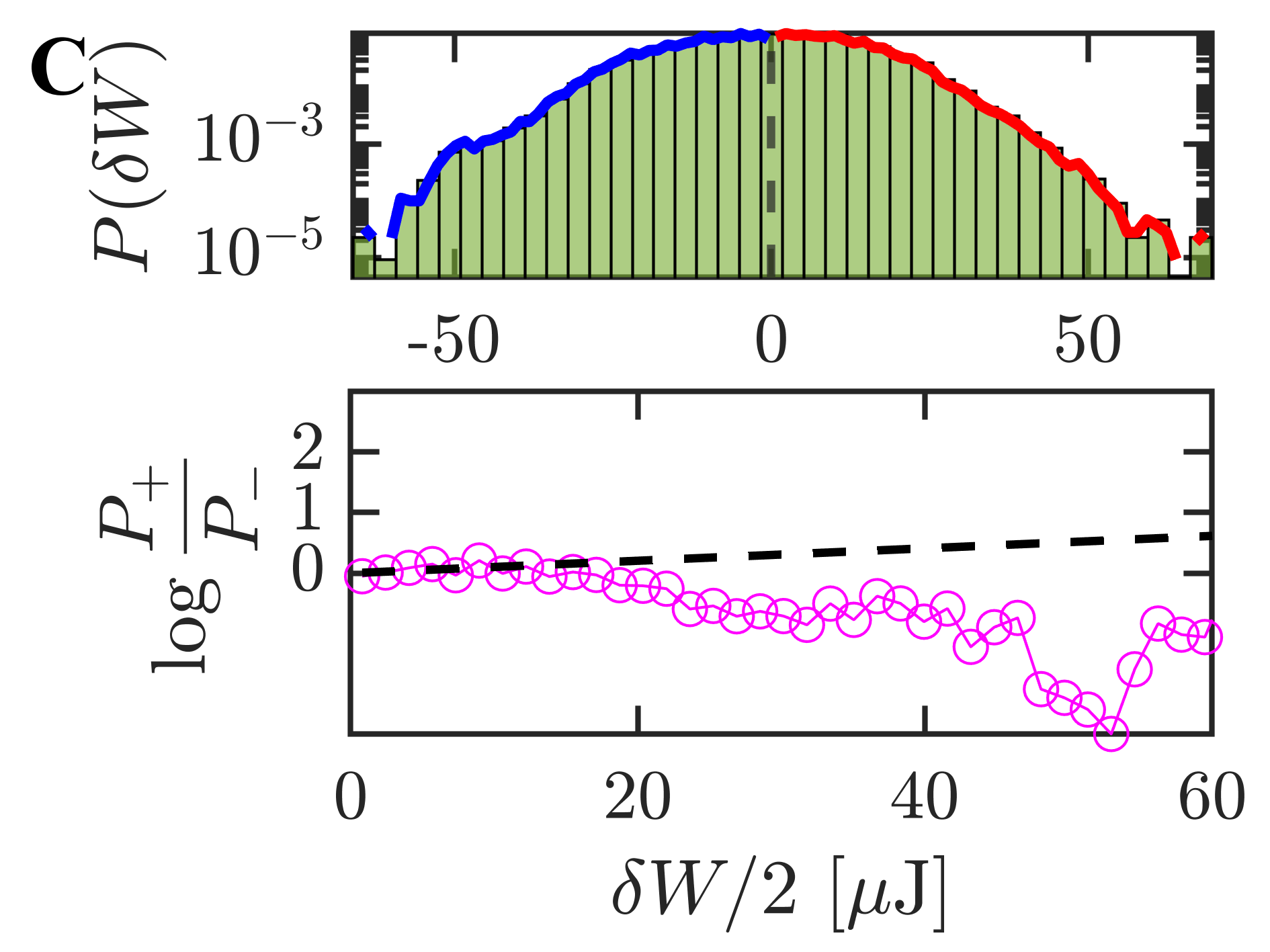}
    \caption{
    \textbf{Heavy tracer setup.} \ 
    \textbf{A.} A heavy plastic ball is used as a passive tracer ($3$~cm diameter, $8.2$~g). The active bath contains $N_b=10$ standard bbots. \ 
    \textbf{B.} FDR test for the tracer's position and velocity observables, under the same perturbation setup as detailed in Figs.~\ref{fig:SIfig4},\ref{fig:SIfig5}.
    A strong airflow ($13.5$~V fan operating voltage) was used as a step-perturbation.
    The effective temperature is $T_{\text{eff}}=\kappa \langle x^2\rangle_0$. The correlation functions fail to capture the response to the step-perturbation.
    \textbf{C.} Work FR test as detailed in Fig.~\ref{fig:SIfig6}, does not yield an effective temperature.
    }
    \label{fig:SIfig10}
\end{figure}

Fig.~\ref{fig:SIfig8} and \ref{fig:SIfig9} present FDR and FR tests with alternative system configurations.
Namely, we employ faster bbots and a smaller styrofoam ball as a passive tracer, respectively.
The position autocorrelation functions are approximately proportional to the responses with $T_{\text{eff}}=T_{\text{pot}}$, yet the FDR in the velocity observable show violations for short times.
The work FR is valid for sufficiently small values, from which a consistent effective temperature can be extracted with $T_{\text{FR}}\approx T_{\text{eff}}$.
These measurements fail with a heavy tracer setup (Fig.~\ref{fig:SIfig10}), as discussed in the main text.

We note that additional reliable $T_{\text{eff}}$ measurements across different $N_b$ would require additional tuning of the perturbation amplitude and the potential stiffness, to ensure stable dynamics in the system.

\clearpage
\newpage 

\clearpage
\newpage 


\clearpage
\newpage 

\subsection*{System technicalities} 
\begin{figure}[H] 
    \centering
    \includegraphics[width=0.20\textwidth]{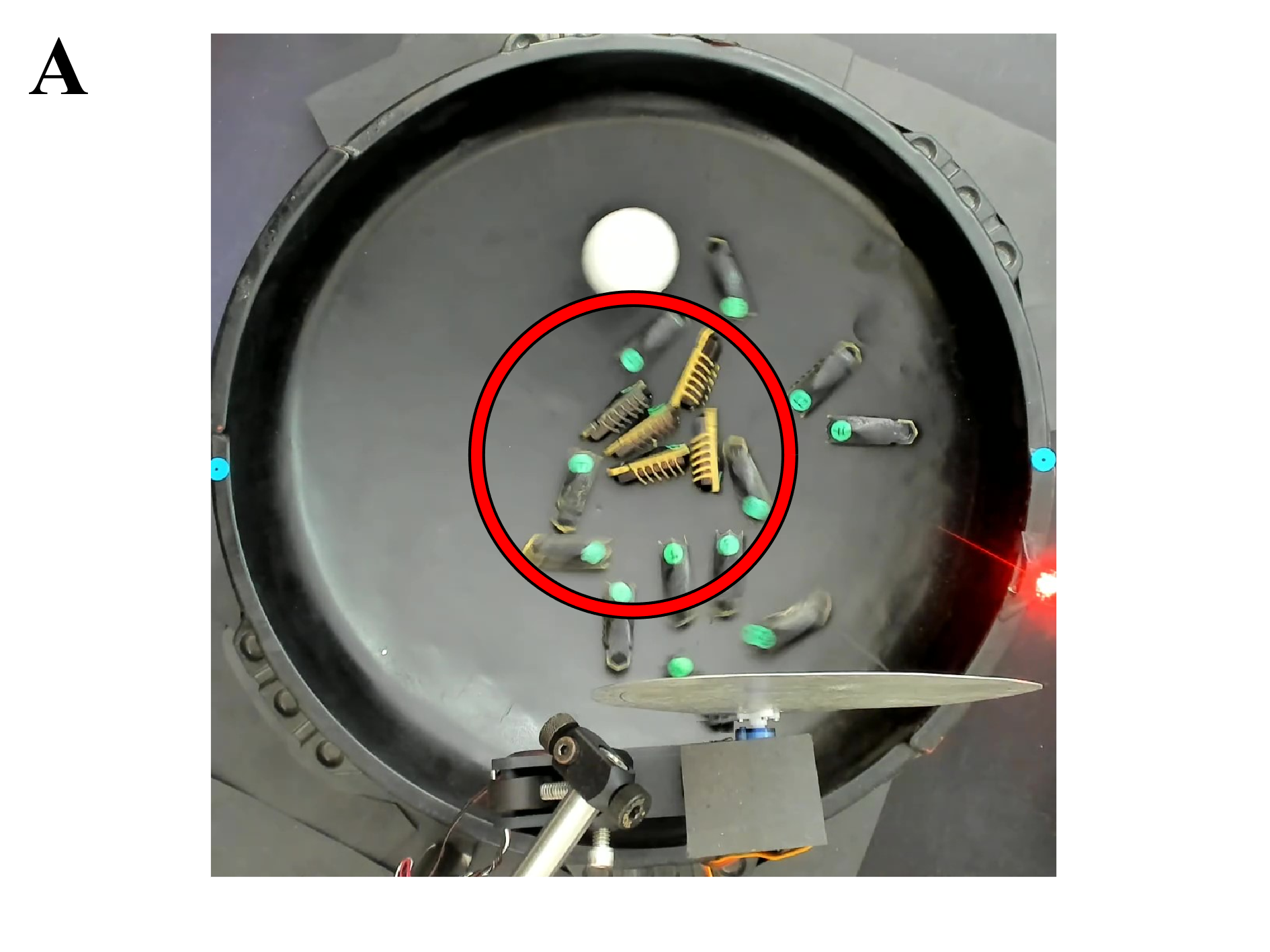}    
    \includegraphics[width=0.20\textwidth]{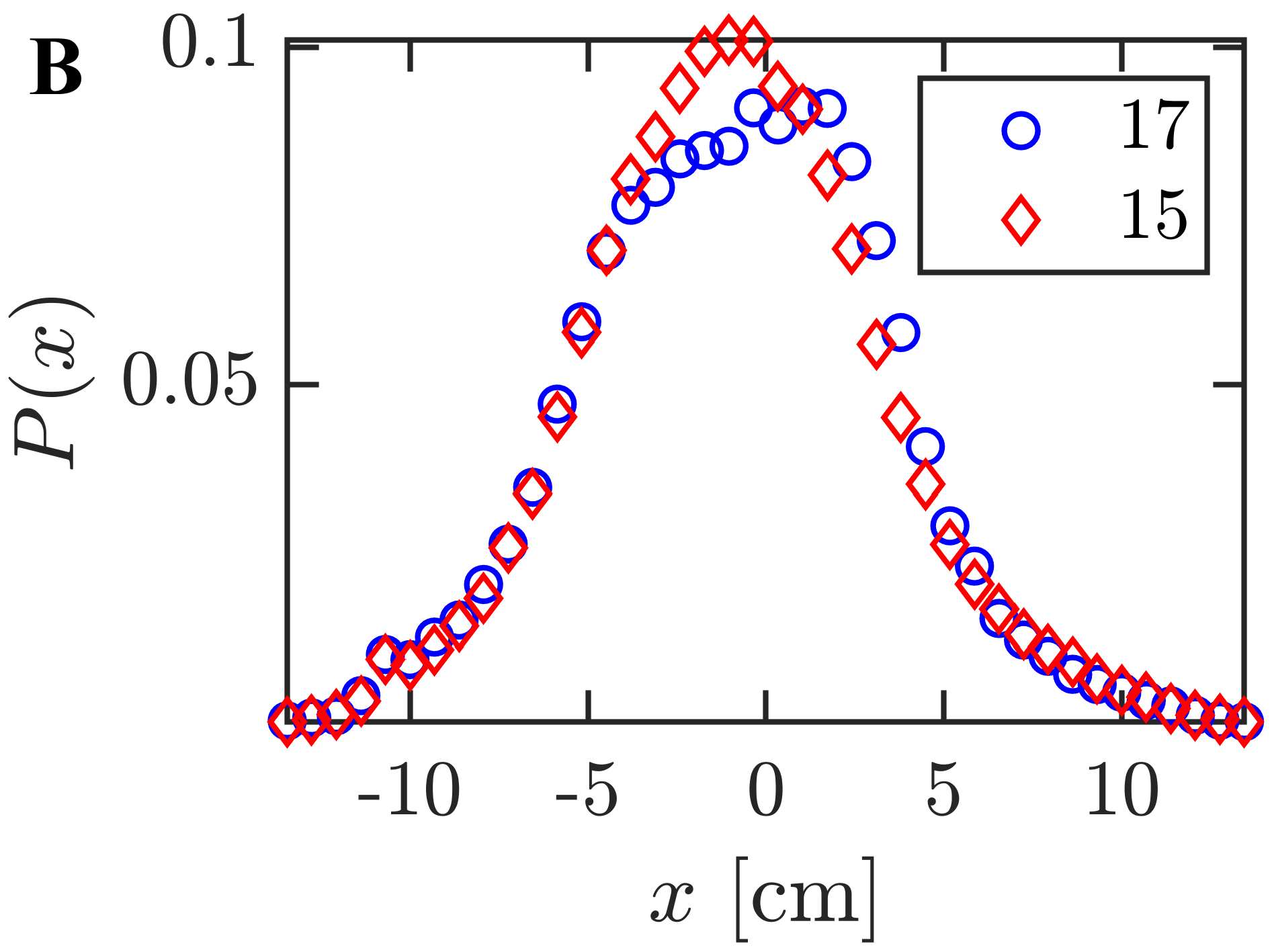}    
    \caption{
    \textbf{A.} \ Snapshot of an experiment with $N_b=17$, showing a temporal static cluster of tumbled bbots in the center trap region.
    \textbf{B.} \ The tracer's position probability distribution for $N_b=17$, showing a flat region in the center, reflecting the physical exclusion by static bbot clusters. }
    \label{fig:SIfig1}
\end{figure}

In our experimental setup, the number of bbots ($N_b$) is limited due to their tendency to occasionally tumble upon collisions with other bbots within the potential well. 
This tumbling effect becomes increasingly pronounced at high densities (and high propulsion speeds), leading to the formation of long-lived static clusters at the trap center.
Such clustering can prevent the tracer from accessing certain regions in the trap bulk, thus hindering reliable measurements of the its fluctuations and responses.
This effect can be partly observed in the stationary (tracer) probability distribution with $N_b=17$, as shown in Fig.~\ref{fig:SIfig1}.

\begin{figure}[t] 
    \centering
    \includegraphics[width=0.20\textwidth]{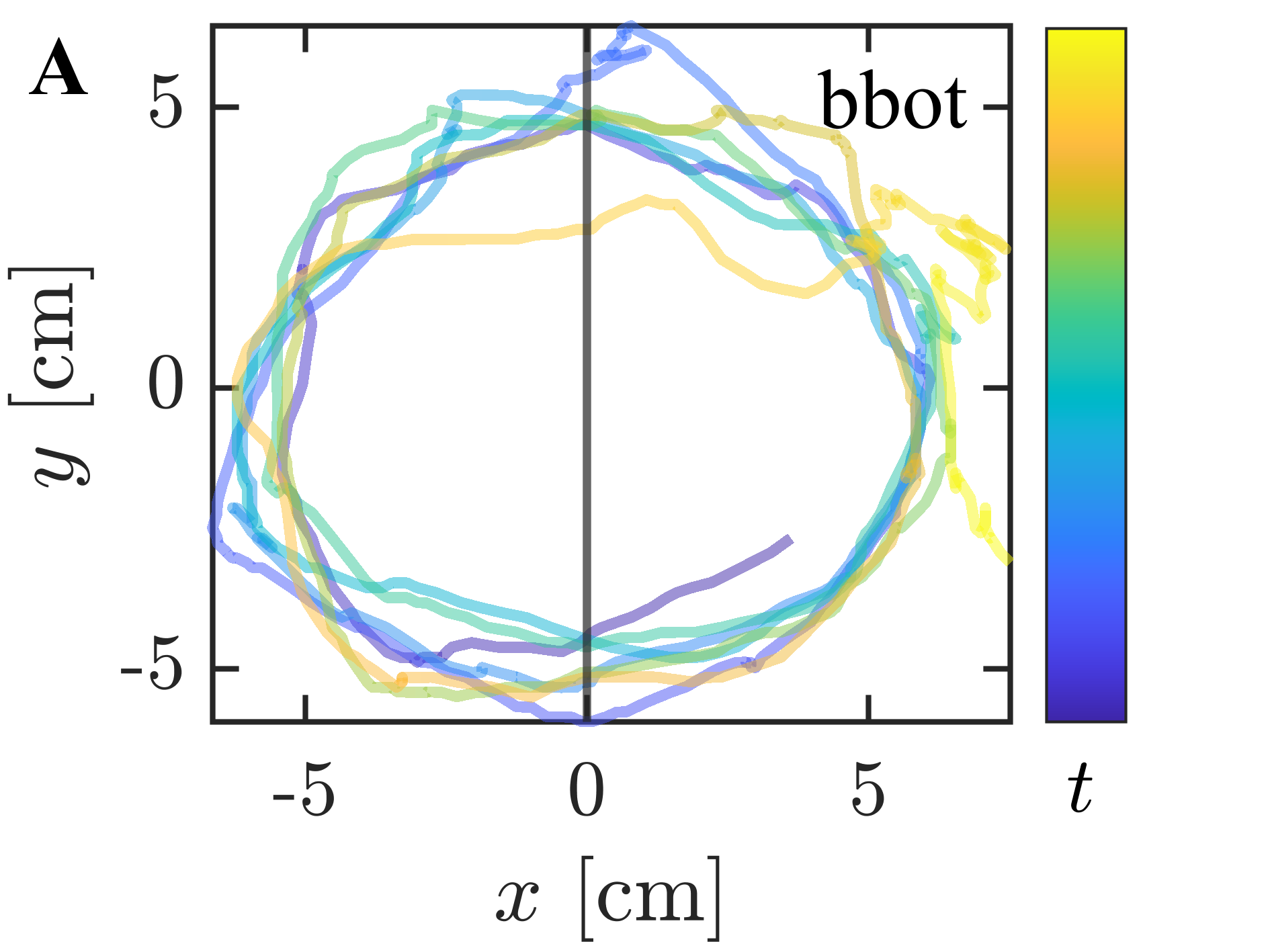}
    \includegraphics[width=0.20\textwidth]{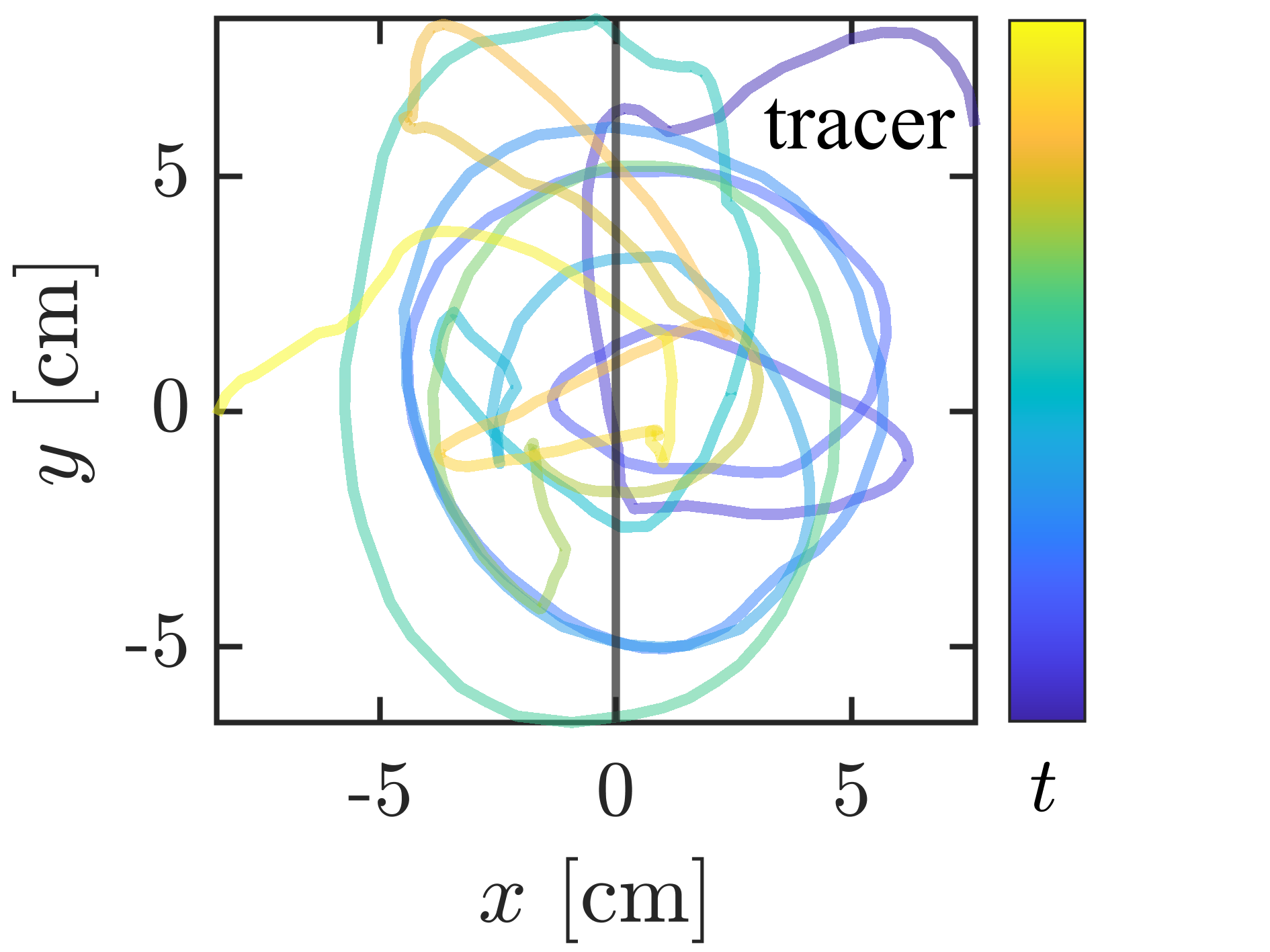}
    \includegraphics[width=0.20\textwidth]{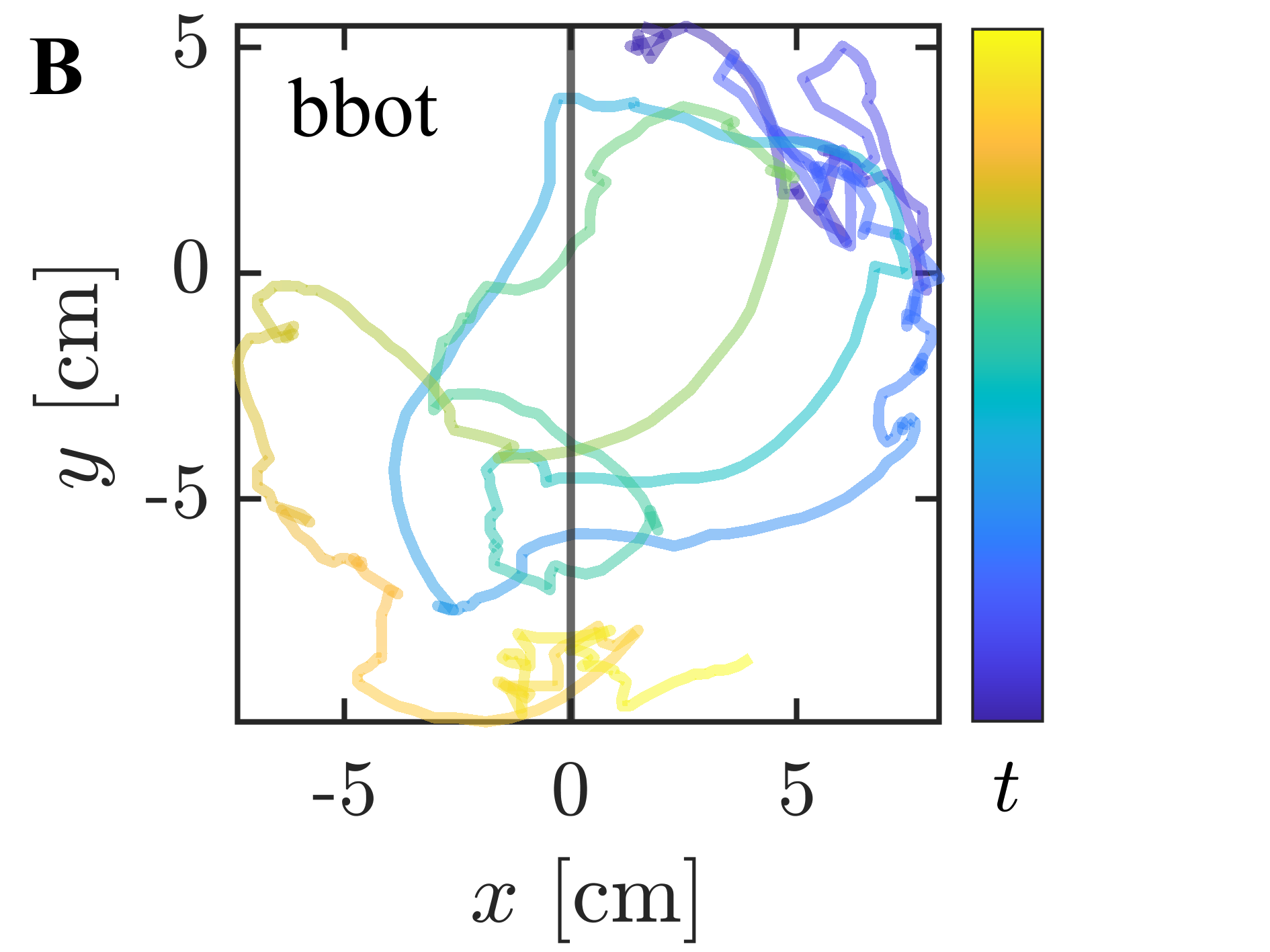} 
    \includegraphics[width=0.20\textwidth]{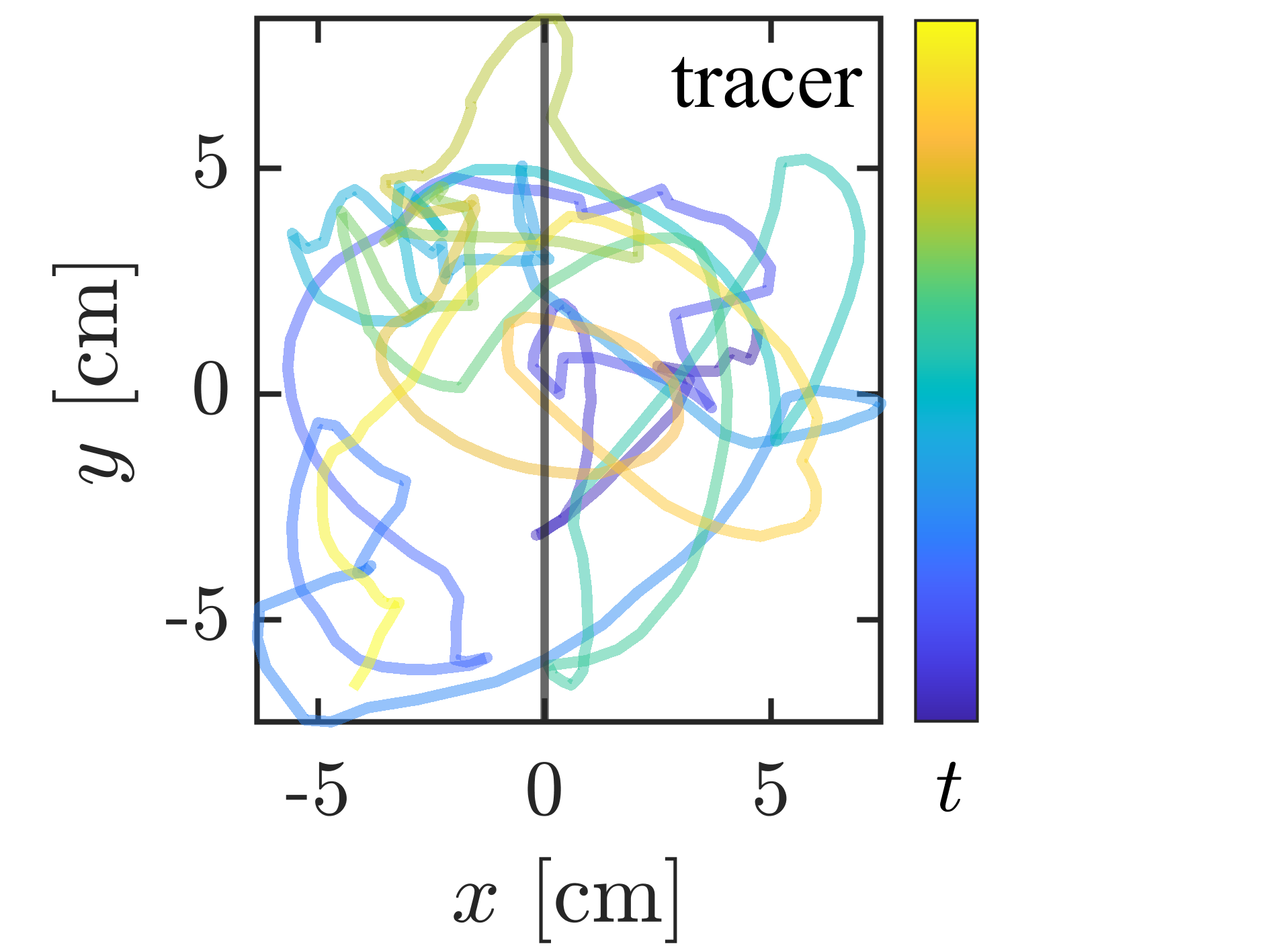}
    \includegraphics[width=0.20\textwidth]{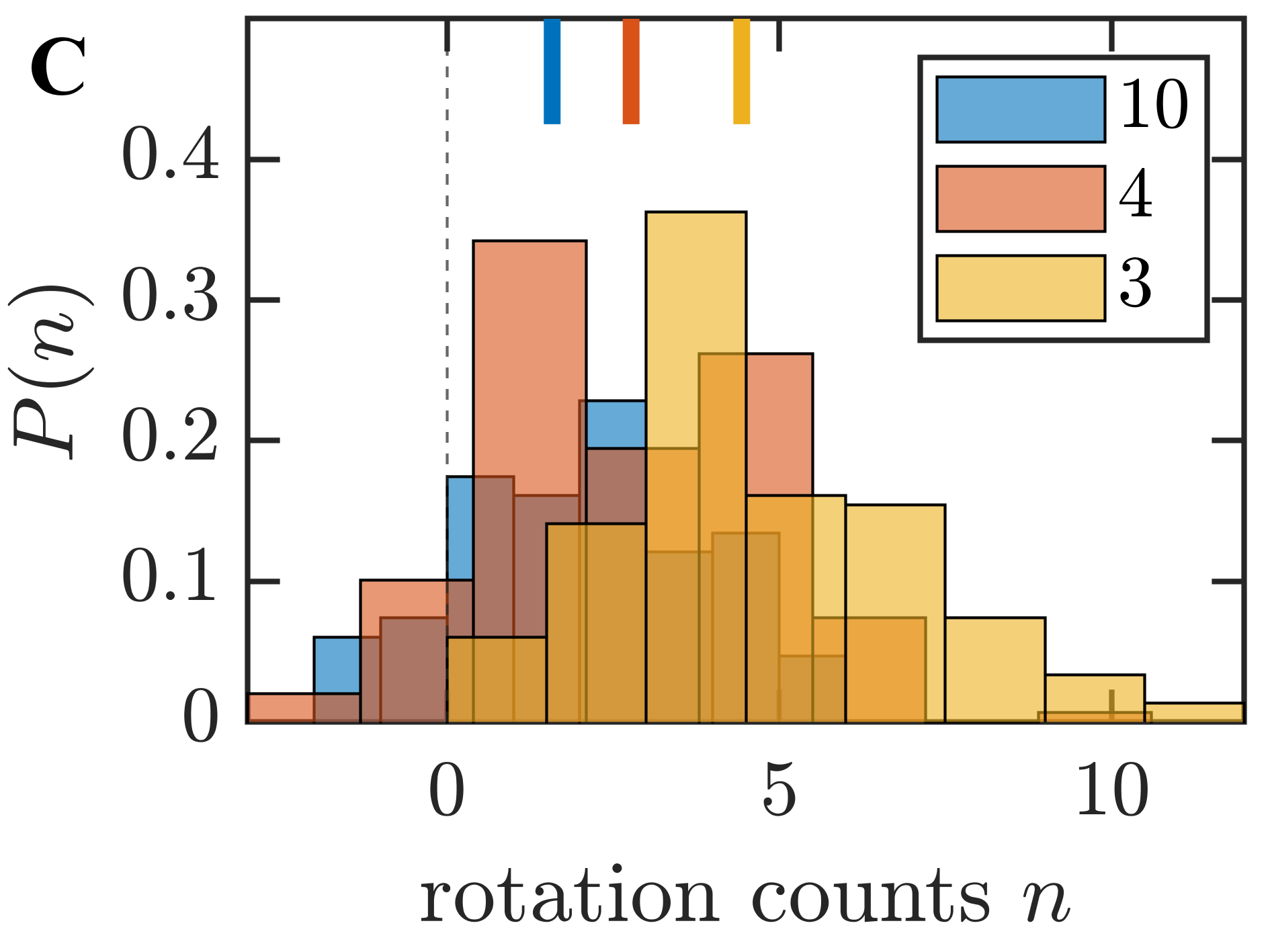}
    \includegraphics[width=0.20\textwidth]{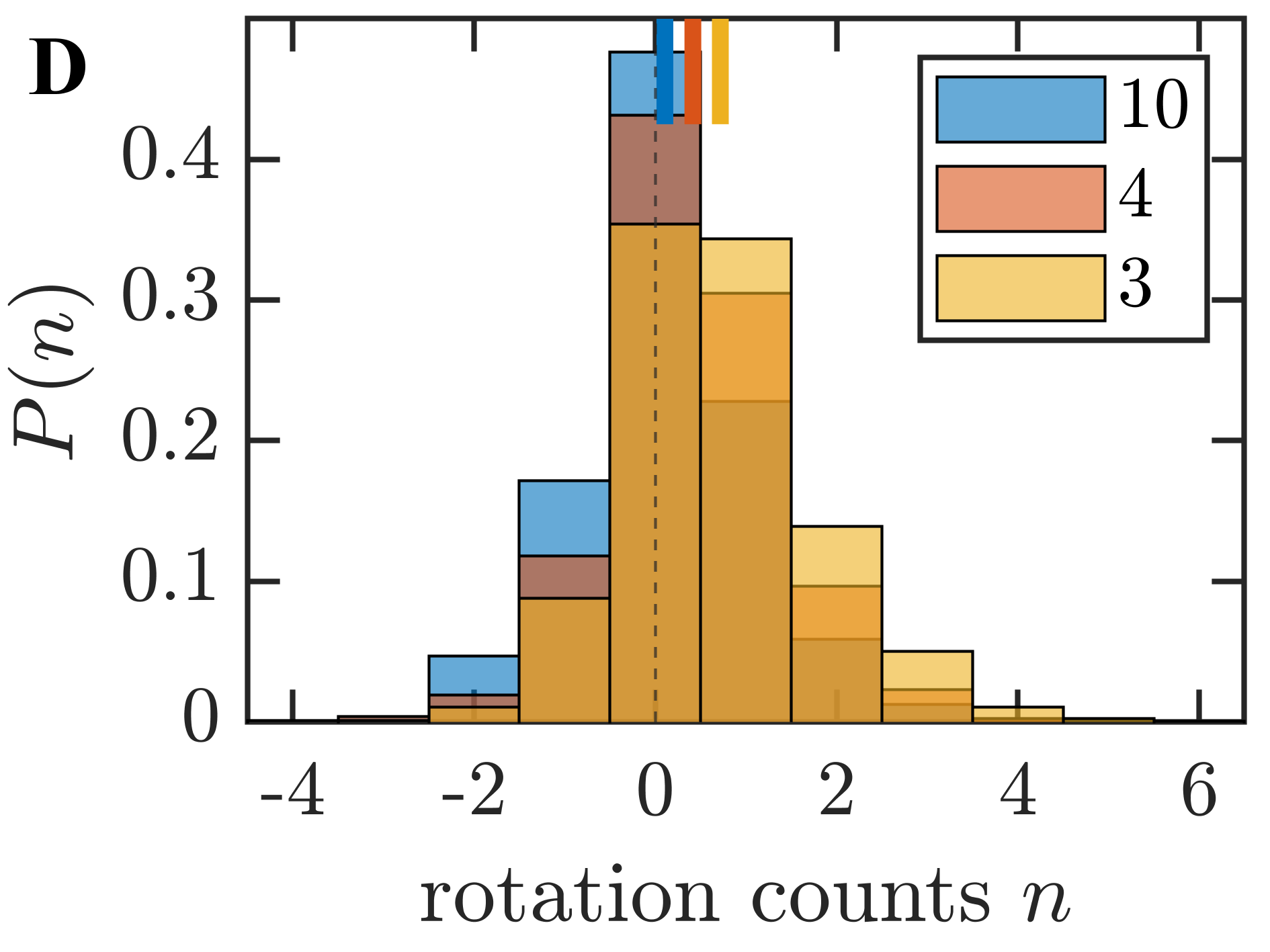}  
    \caption{
    \textbf{The system's chirality.} 
    Typical single bbot and tracer trajectories, in a system with $N_b=3$ (\textbf{A}) and $N_b=10$ (\textbf{B}).
    \textbf{C.} \ The distribution of bbot instantaneous rotations $\pm n$ across a central line at $x=0$, where $n>0$ indicates favored clockwise direction.
    Obtained by tracking 200 bbots trajectories ($t\in[0,5]s$).
    The upper lines are the means of $P(n)$, showing that net chirality is reduced by increasing $N_b$.
    \textbf{D.} \ The same chirality analysis for the passive tracer. }
    \label{fig:SIfig2}
\end{figure}

In Fig.~\ref{fig:SIfig2} we present measurements that quantify the chirality of a single bbot in an assembly composed of $N_b$ bbots (in the potential well).
We evaluate the net chirality of a system by counting the clockwise ($+1$) and anticlockwise ($-1$) crossings (across a central line at $x=0$).
The results were obtained by tracking 200 bbot trajectories ($t\in[0,5]$~s), and present a sum of $\pm n$ rotations per frame, and a probability distribution $P(n)$.
We find that the systems attain a net clockwise chirality that is reduced by increasing $N_b$, and therefore the collision frequency.
In comparison, the chirality of the passive tracer particle results in small values and vanish for sufficiently high $N_b$.

\vspace{1cm}

\clearpage

\end{document}